\documentclass[11pt,a4paper]{article}
\usepackage[english]{babel}
\usepackage[T1]{fontenc}
\usepackage{vmargin}
\usepackage{graphicx}
\usepackage{amsmath}
\usepackage{amssymb}
\usepackage{units}
\usepackage{subfigure}
\usepackage{etoolbox} 
\usepackage[hypertexnames=false]{hyperref}

\setmarginsrb
{2.5cm} 
{0.5cm} 
{2.5cm} 
{2.0cm} 
{0.5cm} 
{1.cm}   
{2.cm} 
{2.cm}   

\usepackage{authblk}

\usepackage{booktabs}
\newcommand{\botrule}{\hline}
\usepackage[font={small,sl}]{caption}

\newcommand{\backmatter}{}
\newcommand{\bmhead}[1]{\section{#1}}
\newcommand{\keywords}[1]{\paragraph{Keywords:}#1}

\newlength{\mytableboxsize}
\newtoggle{dist}
\togglefalse{dist}

\newcommand{\red}[1]{#1}
\newcommand{\sout}[1]{}

\newcommand{\comment}[1]{} 
\newcommand{\CompAng}{\Phi_\text{comp}}
\newcommand{\mmpermus}{mm/\mu s}
\def\myscale{0.7}

\def\etal{et al.} %

\newcommand{\IPHC}{1}
\newcommand{\UHA}{2}
\newcommand{\IFIC}{3}
\newcommand{\UPC}{4}
\newcommand{\UPV}{5}
\newcommand{\CPPM}{6}
\newcommand{\APC}{7}
\newcommand{\LAM}{8}
\newcommand{\CNESTEN}{9}
\newcommand{\Rabat}{10}
\newcommand{\Bologna}{11}
\newcommand{\BolognaUNI}{12}
\newcommand{\LNS}{13}
\newcommand{\LPMR}{14}
\newcommand{\NIKHEF}{15}
\newcommand{\ISS}{16}
\newcommand{\UvA}{17}
\newcommand{\Genova}{18}
\newcommand{\Roma}{19}
\newcommand{\RomaUNI}{20}
\newcommand{\Marrakech}{21}
\newcommand{\Bari}{22}
\newcommand{\UGRCITIC}{23}
\newcommand{\UPS}{24}
\newcommand{\Erlangen}{25}
\newcommand{\SalernoUNI}{26}
\newcommand{\ClermontFerrand}{27}
\newcommand{\LSIS}{28}
\newcommand{\GenovaUNI}{29}
\newcommand{\NIOZ}{30}
\newcommand{\GEOAZUR}{31}
\newcommand{\Leiden}{32}
\newcommand{\Wuerzburg}{33}
\newcommand{\Bamberg}{34}
\newcommand{\COM}{35}
\newcommand{\Catania}{36}
\newcommand{\IRFUSPP}{37}
\newcommand{\Napoli}{38}
\newcommand{\NapoliUNI}{39}
\newcommand{\UGRCAFPE}{40}
\newcommand{\IUF}{41}
\newcommand{\CasertaUNI}{42}

\begin{document}

\title{Acoustic Positioning for Deep Sea Neutrino Telescopes with a System of Piezo Sensors Integrated into Glass Spheres}

\author[\IPHC,\UHA]{A.~Albert}
\author[\IFIC]{S.~Alves}
\author[\UPC]{M.~Andr\'e}
\author[\UPV]{M.~Ardid}
\author[\UPV]{S.~Ardid}
\author[\CPPM]{J.-J.~Aubert}
\author[\APC]{J.~Aublin}
\author[\APC]{B.~Baret}
\author[\LAM]{S.~Basa}
\author[\APC]{Y.~Becherini}
% ...

\date{\normalsize \today}

\maketitle
\newpage
\begin{abstract}
Position calibration in the deep sea is typically done by means of acoustic multilateration using three or more acoustic emitters installed at known positions. Rather than using hydrophones as receivers that are exposed to the ambient pressure, the sound signals can be coupled to piezo ceramics glued to the inside of existing containers for electronics or measuring instruments of a deep sea infrastructure. The ANTARES neutrino telescope operated from 2006 until 2022 in the Mediterranean Sea at a depth exceeding $\unit[2000]{m}$. It comprised nearly 900 glass spheres with $\unit[432]{mm}$ diameter and $\unit[15]{mm}$ thickness, equipped with photomultiplier tubes to detect Cherenkov light from tracks of charged elementary particles. In an experimental setup within ANTARES, piezo sensors have been glued to the inside of such -- otherwise empty -- glass spheres. These sensors recorded signals from acoustic emitters with frequencies from $46545$ to $\unit[60235]{Hz}$. Two waves propagating through the glass sphere are found as a result of the excitation by the waves in the water. These can be qualitatively associated with symmetric and asymmetric Lamb-like waves of zeroth order: a fast (early) one with  $v_e \approx \unit[5]{mm/\mu s}$ and a slow (late) one with $v_\ell \approx \unit[2]{mm/\mu s}$. Taking these findings into account improves the accuracy of the position calibration. 
The results can be transferred to the KM3NeT neutrino telescope, currently under construction at multiple sites in the Mediterranean Sea, for which the concept of piezo sensors glued to the inside of glass spheres has been adapted for  monitoring the positions of the photomultiplier tubes.
\end{abstract}




\maketitle

\section{Introduction}
For deep sea neutrino telescopes, the Cherenkov light of charged particle tracks produced in neutrino interactions is detected by photomultiplier tubes (PMTs) which are typically mounted inside glass spheres, installed along vertical structures forming lines. To reduce the huge background from atmospheric muons, these glass spheres, commonly referred to as optical modules (OMs), are installed at great depth. Designing the components to withstand the huge pressure in the deep sea is one of the technical challenges of building a deep sea neutrino telescope. 
The lines are anchored with their bottom end on the sea floor and kept taut by submerged buoys at their top end. 
The directions of the charged particle tracks are reconstructed using the arrival times of the Cherenkov photons at the PMTs.
For this purpose it is important to know the positions of the OMs at any instance of time with a precision of $\lesssim \unit[20]{cm}$.
It is, however, technically impossible to keep the lines in a perfectly vertical alignment as they sway with the time-varying underwater currents. 
Therefore, deep sea neutrino telescopes contain a dynamic position reconstruction system with acoustic emitters installed at fixed positions at the sea floor and acoustic sensors (hydrophones) along the lines.

The ANTARES deep sea neutrino telescope~\cite{bib:ANTARES-paper} was operating in the Mediterranean Sea from 2006 until 2022. In its final phase of construction, a volume of $\unit[0.01]{km^3}$  was instrumented with 885 PMTs.
It was designed as a demonstrator to prove the feasibility of a cubic kilometre sized neutrino telescope in the Mediterranean Sea. For its successor, the multi-site telescope network KM3NeT~\cite{bib:KM3NeT-let-intent},
the first lines with PMTs
 were deployed in 2015.  

In ANTARES, three OMs together with an electronics container formed a so-called storey.  Some of these storeys were equipped with dedicated hydrophones of the position calibration system, exposed to the high ambient pressure. This setup pushes up the costs for the hydrophones and requires a cable that is connected through a penetrator with the pressure resistant electronics container. Such penetrators are single points of failure for a storey, as the failure of the seal can lead to the flooding of the electronics container with the subsequent failure of the complete storey.

The analysis presented here has been inspired by the position calibration for KM3NeT~\cite{bib:km3net-positioning-icrc2023,bib:km3net-positioning-sensors2020}, where the functionality of a complete storey of ANTARES has been integrated into a single OM, referred to as digital optical module (DOM)~\cite{bib:KM3NeT-DOMs}.
This design
makes hydrophones as external structures impractical for the deployment procedure. 
The solution for KM3NeT was the integration of  piezo-acoustic sensors into the DOMs, coupling to the sound signals in the sea through the glass sphere. In addition to the technically easier and less failure prone implementation compared to external hydrophones, this solution significantly reduces the costs of the acoustic sensors.
On the other hand, integrating the sensors into the glass spheres 
effectively creates an acoustic receiver with the dimensions of the sphere and in general, the sound wave from an acoustic emitter will not hit the sphere at the position of the acoustic sensor. Acoustic properties of the glass, such as its vibrations and the speed of sound in the glass, will affect the time of arrival of an acoustic signal and potentially introduce systematic errors to the position calibration. 
It is therefore crucial to study the acoustic properties of the combined system of acoustic sensor and glass sphere.
Such studies are conveniently done with the acoustic neutrino detection test system AMADEUS~\cite{bib:amadeus-2010} of the ANTARES detector that comprised three acoustic modules (AMs) with piezo ceramics glued to the inside of a glass sphere. 

In this paper, the propagation of sound waves inside the glass sphere of the AMs of the AMADEUS system is investigated. 
In Section~\ref{sec:antares+km3net}, the ANTARES 
neutrino telescope is introduced, in Section~\ref{sec:amadeus}, the acoustic system AMADEUS of the ANTARES detector is discussed before detailing the acoustic positioning with the AMADEUS system in Section~\ref{sec:pos-calibration}.
Subsequently, in Section~\ref{sec:measurement}, 
the propagation of the signal from an acoustic emitter to the sensors in the AM is investigated in detail, looking in particular into effects of the orientation of the AMs relative to the emitter.
For the signal to reach the sensor, two guided waves within the glass of the sphere are identified.
The speed of sound and its dependency on the frequency for each wave are investigated in Section~\ref{sec:sound-speed-vs-freq}.
In Section~\ref{sec:sys-errors}, systematic errors are addressed and 
the results of the investigation of the two waves are discussed in Section~\ref{sec:discussion-results}.
In Section~\ref{sec:interpret-2-waves}, the two waves are interpreted within the theoretical framework of guided Lamb waves and in Section \ref{sec:lamb-wave-prop}, Lamb waves are discussed in more detail.
In Section~\ref{sec:pos-reco-revisit}, it is shown that applying corrections to the position calibration procedure by taking into account the different speeds of sound in water and the glass of the AMs improves the precision of the reconstructed position.
Finally, in Section~\ref{sec:conclusions} the summary and conclusions are presented. The appendix contains three sections covering additional calculations for the propagation of sound in water, a detailed discussion of systematic errors, and a compilation of the elastic properties of the glass spheres.

\section{The ANTARES neutrino telescope}
\label{sec:antares+km3net}
The ANTARES deep sea neutrino telescope was located in the Mediterranean Sea at a water depth of 2475\,m, roughly 40 km South of the town of Toulon at the French coast at the geographic position of $42^\circ 48'$ N, $6^\circ 10'$ E.
Data taking with the complete detector started in 2008 and lasted until the telescope was turned off in February 2022.
It comprised 12 vertical structures, the detection lines.
Each detection line held up to 25 storeys that were arranged at equal distances of 14.5 m along the line, starting at about 100 m above the sea bed and interlinked by electro-optical cables. A standard storey consisted of a titanium support structure, holding three OMs (each one consisting of a PMT inside a water-tight pressure-resistant glass sphere) and one cylindrical electronics container. Horizontal distances between neighbouring lines were roughly $\unit[50]{m}$.

A 13th line, dubbed Instrumentation Line (IL), was equipped with instruments for monitoring the environment. It held six storeys, equipped with various instruments rather than OMs. For two pairs of consecutive storeys in the IL, the vertical distance was increased to 80 m.

Each of the 13 lines was fixed on the sea floor by an anchor equipped with electronics and held taut by an immersed buoy. An interlink cable connected each line via the anchor to the Junction Box from where the main electro-optical cable provided the connection to the shore station.

\begin{figure}[tbh]
\centering
\includegraphics[width=\columnwidth,angle=0]{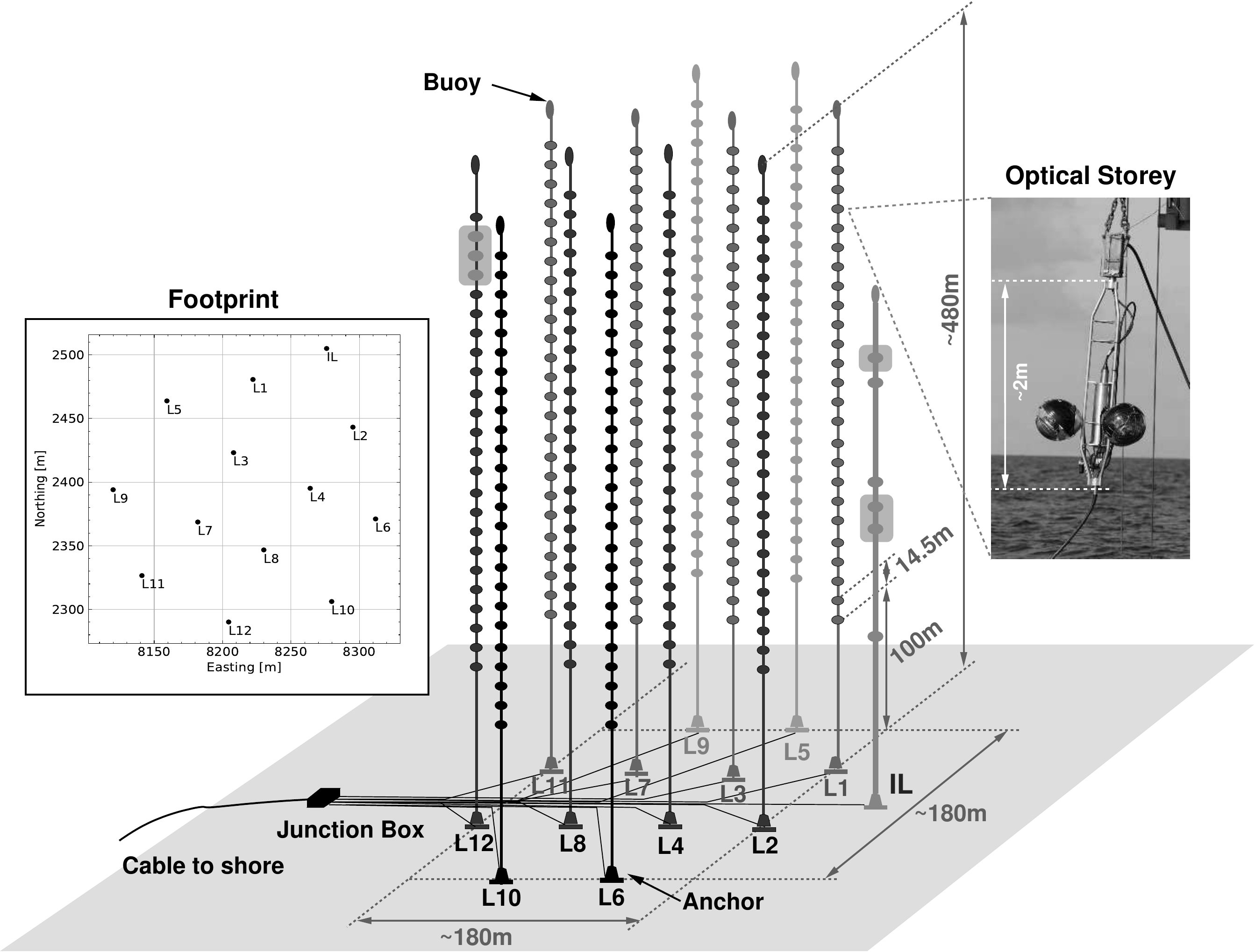}
\caption{
Schematic view of the ANTARES detector with the acoustic storeys of AMADEUS highlighted. Detection lines are labelled with the letter L followed by their number. Acoustic storeys are installed on the Instrumentation Line (IL) and line 12 (L12). 
The ``Footprint'' shows the positions of the anchors of ANTARES on the sea floor in UTM coordinates of UTM Zone 32.
}
\label{fig:antares-schematic}
\end{figure}

A schematic view of the ANTARES detector and a footprint of the anchors on the sea floor are shown in Figure~\ref{fig:antares-schematic}. The storeys of the AMADEUS acoustic test system that will be discussed in the next section are highlighted.

\section{The AMADEUS system}
\label{sec:amadeus}
\subsection{Design}
\label{subsec:amadeus-design}
\begin{figure}[htbp]
\centering
\subfigure[]{
\includegraphics[height=7.5cm,angle=0]{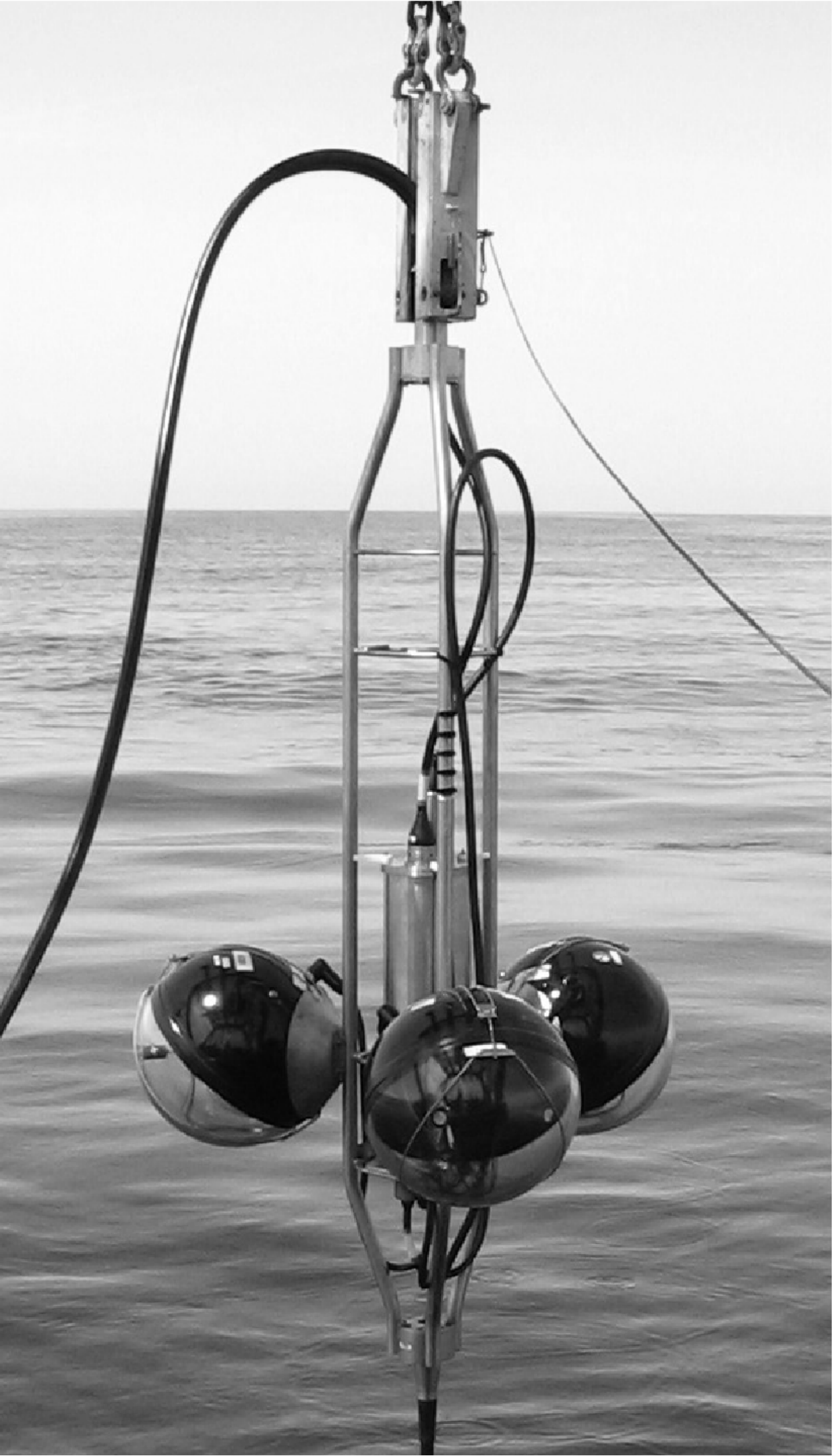}
\label{subfig:line12_storey21_AM_gray}
}
\subfigure[]{
\includegraphics[height=7.5cm,angle=0]{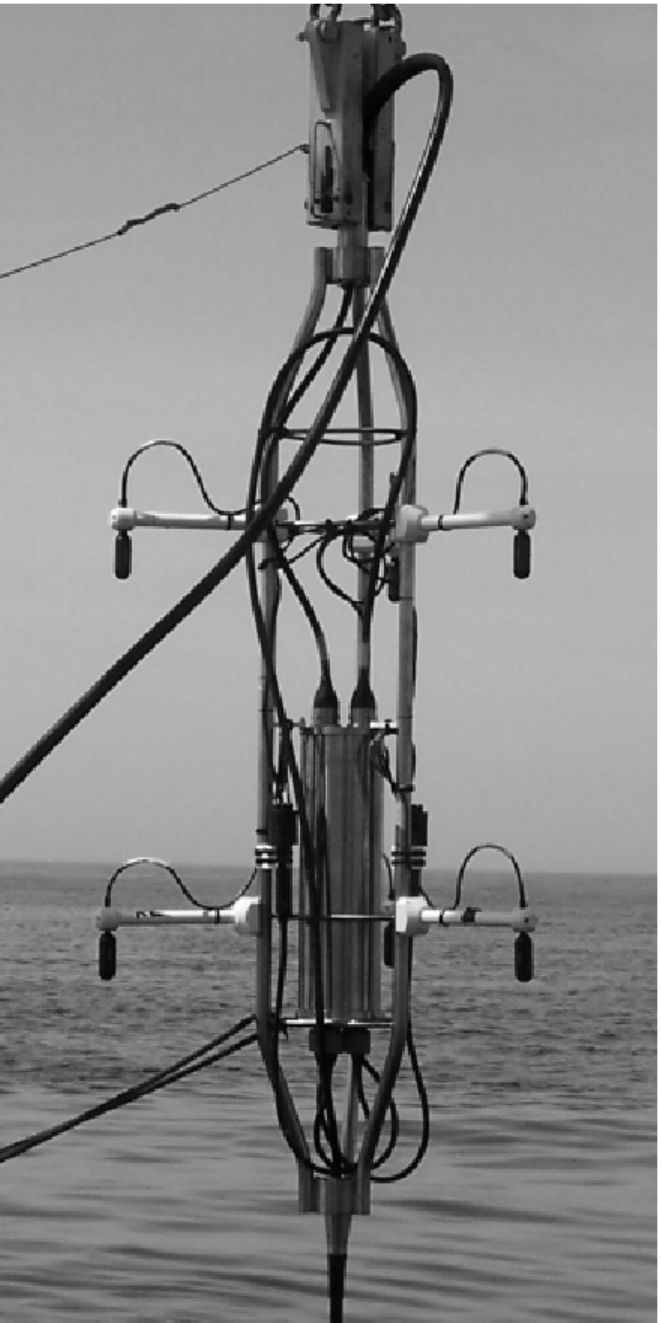}
\label{subfig:line12_storey22_HTI_gray}
}
\caption{
\subref{subfig:line12_storey21_AM_gray}
Storey with acoustic modules (AMS);
\subref{subfig:line12_storey22_HTI_gray}
Storey above the AMS holding standard hydrophones.
}
\label{fig:line12_AM_HTI_gray}
\end{figure}

Within the AMADEUS system~\cite{bib:amadeus-2010}, acoustic sensing was integrated in the form of acoustic storeys that were modified versions of standard ANTARES storeys, with the OMs replaced by custom-designed acoustic sensors. 
Dedicated electronics was used for the amplification, digitisation and pre-processing of the analogue signals. 
The AMADEUS system comprised a total of six acoustic storeys: three on the Instrumentation Line  and three on the detection line 12 (cf.\ Figure~\ref{fig:antares-schematic}). For each of the five standard acoustic storeys, six acoustic sensors were implemented, exposed to the ambient sea water and arranged at distances of roughly $\unit[1]{m}$ from each other.
Each of these sensors consisted of a piezo ceramic element with integrated pre-amplifier, coated in polyurethane.
The lowest acoustic storey on line 12 at a nominal height of about $\unit[400]{m}$ above the sea floor was equipped with alternative sensing devices: three AMs, each comprising a glass sphere -- the same type as used for the OMs -- into which two units of piezo ceramics with a custom-designed pre-amplifier were glued.
This storey will be referred to as the acoustic module storey (AMS).
Figure~\ref{fig:line12_AM_HTI_gray} shows the AMS and a standard acoustic storey, equipped with six standard hydrophones. 

The analogue signals from the acoustic sensors were routed to the Local Control Module (LCM) of the storey, a cylindrical titanium container housing the off-shore electronics. Here the analogue signal from each sensor was amplified by an adjustable gain that could be set to one of 12 values between 1 and 55\,dB.  The standard setting was a gain of 20\,dB, corresponding to a voltage amplification by a factor of 10.

Digitisation of the acoustic signals was done at a rate of $\unit[500]{ksps}$ and the sampled values are stored at $\unit[250]{ksps}$ after downsampling by a factor of 2. Prior to downsampling, a digital anti-alias FIR filter (128th order with Blackman window) with a corner frequency of about $\unit[100]{kHz}$ was applied in the electronics of the AMS. All resulting data from the 36 acoustic sensors was transmitted to the shore station. Here, an adjustable software filter selected events from the data stream for storage on disk and further offline analysis. Three filter schemes were in operation~\cite{bib:amadeus-2010}, of which the trigger on the signals from the acoustic emitters of the ANTARES position calibration system is the most relevant one for the analysis described in this article. 

\subsection{Acoustic modules of AMADEUS}
\label{sec:AMs}

The analysis described in this paper is based on the AMs of the AMADEUS system, see
Figure~\ref{subfig:line12_storey21_AM_gray}. 
The storey above the AMS, holding six standard hydrophones, will be used as a reference for signals observed with the AMS and
is shown in Figure~\ref{subfig:line12_storey22_HTI_gray}.

\begin{figure}[htb]

\centering
\subfigure[]{
 \begin{minipage}[b]{5cm}
\includegraphics[height=5cm,angle=0]{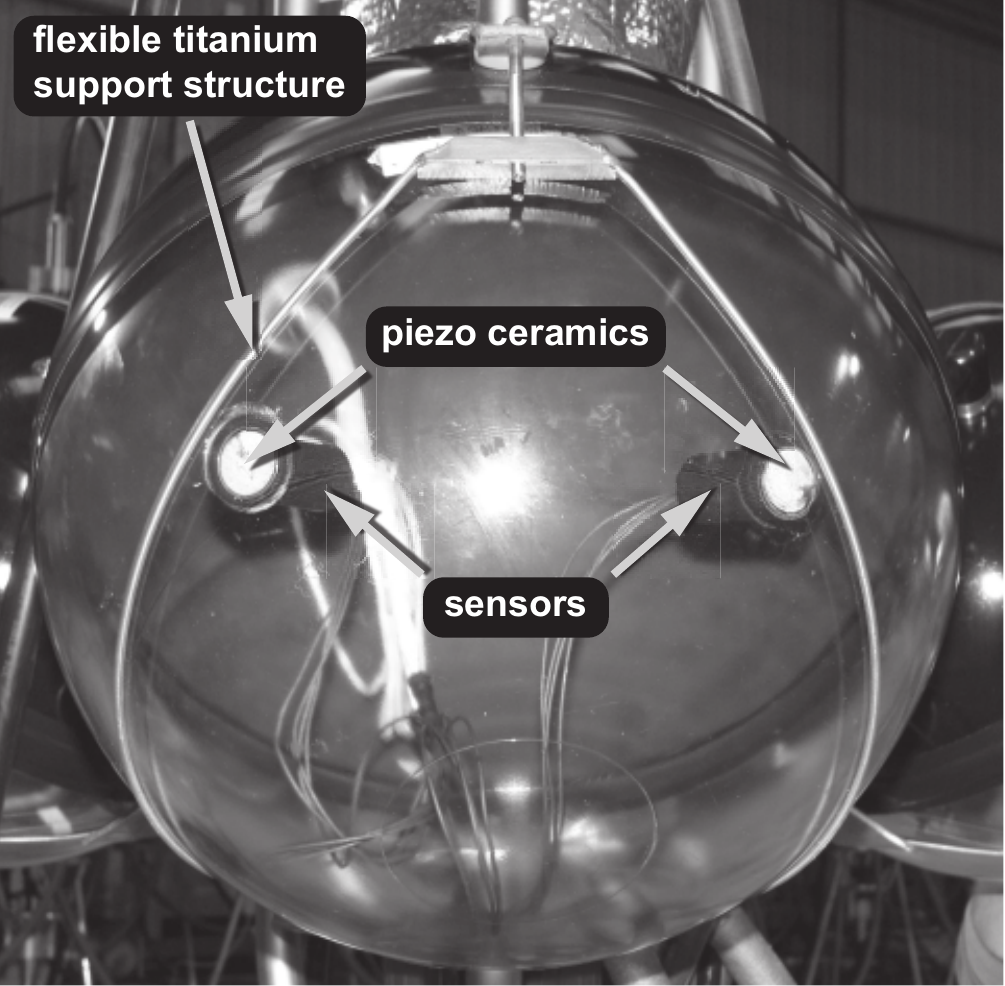}
  \end{minipage}
\label{subfig:AM_photo}
}
\hspace{5mm}
\subfigure[]{
 \begin{minipage}[b]{5cm}
\includegraphics[height=5cm,angle=0]{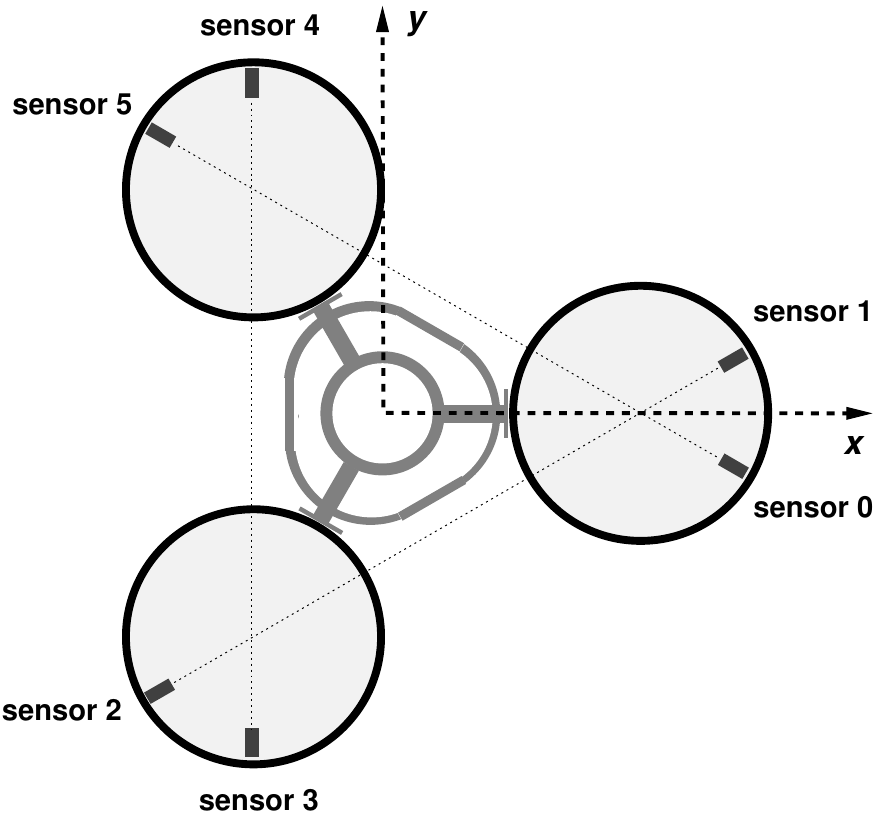}

  \end{minipage}
\label{subfig:AM_schematic_gray}
}
\caption{ \subref{subfig:AM_photo} Photograph of an acoustic module (AM) before deployment; 
  \subref{subfig:AM_schematic_gray} Cross-section of the acoustic module storey (AMS) in the plane of the sensors (i.e.\ the $xy$-plane of the local reference frame of the AMS), viewed from above. The fine dotted lines intersect at angles of 60$^\circ$ at the centres of the glass spheres. They are collinear with the
  longitudinal axes of the sensors and indicate the nominal arrangement of the sensor within the storey. 
  The bold broken lines indicate the $x$ and $y$ axes of the local reference frame of the AMS ($z$-axis pointing upwards).
   }
\label{fig:AM_gray}
\end{figure}

Figure~\ref{subfig:AM_photo} shows one of the AMs of the AMS. The piezo ceramics glued to the inside of the glass sphere have the shape of a solid cylinder with radius of $\unit[12.7]{mm}$ and thickness of about $\unit[1]{cm}$.
The glass spheres have an outer diameter of $2 r_o = \unit[432]{mm}$ and a thickness of at least  $\unit[15]{mm}$~\cite{bib:OMs}. 
The six sensors are located in the $xy$-plane of the local reference frame of the AMS that 
is defined by the nominal positions of the centres of the three spheres.
The intersection of this plane with the outer surface of each AM of the AMS will be denoted as horizontal orthodrome of the AM.
In order to obtain a $2\pi$ azimuthal coverage over the AMS,
the axes of the two sensors in an AM form an angle of $60^\circ$, see Figure~\ref{subfig:AM_schematic_gray}.
The positions of the sensors within the local reference frame of the AMS 
were measured before deployment. 
The position of a given sensor is defined as the intersection of the rotational symmetry axis of the piezo ceramic with the outer surface of the glass sphere.

The design of the ANTARES lines allows for the storeys not only to sway but also to rotate with the underwater currents. Each storey contains a compass board to record its bearing w.r.t.\ magnetic North at fixed time intervals in a central database. 
The direction North of the notional compass rose is nominally aligned with the $x$-axis of the local reference frame of the AMS. 

Unlike the conventional hydrophones, the AMs have no rotational symmetry around their vertical axis. Hence the recorded shape of a signal emitted from a given position is expected to depend on the rotation of the storey w.r.t.\ the acoustic source. 

\section{Position calibration of the acoustic module storey}
\label{sec:pos-calibration}
\subsection{Time of arrival determination}
\label{sec:TOA-determination}
For the acoustic positioning system of ANTARES, acoustic emitters were installed at known positions at the anchors of the lines~\cite{bib:antares-pos-2012,bib:antares-pos-old}). These emitters could also be used for the positioning of the acoustic storeys of the AMADEUS system. In intervals of 2~minutes, signals with a duration of $\unit[5]{ms}$ at a predefined single frequency were emitted from all emitters, following a predefined sequence.
Table~\ref{tab:emitter_frequencies} shows the frequencies of the emitters on the respective lines.
Note that the emitter on line 2 emits signals at two frequencies, which will be denoted by line 2(l) and line 2(h) for the low and high frequency emission, respectively.
 
The times of emission (TOEs) for all signals were logged in a central database. The travel time of a signal (time of flight, TOF) is the difference between its time of arrival (TOA) in a given sensor and its TOE from a given emitter.

The time when the signal in a given sensor exceeds a predefined threshold is defined as trigger time. The TOA is then derived by applying an offset to the trigger time to synchronise it with the phase of the signal for which the TOE was defined. 
Using the average speed of sound between the positions of the emitter and receiver, the distance between emitter and receiver can be calculated. The position of a receiver can then be inferred from its distances to at least three emitters.

\begin{table}[bht]
\centering
\caption{Frequencies used for the signals of the acoustic emitters on the respective lines. For line 2 (L2), signals at two frequencies are emitted, denoted by line 2(l) and line 2(h) for the low and high frequency emission, respectively.}
\begin{tabular}{ll}
\toprule
frequency [Hz] & line \\
\midrule
46545 & L3 \\
47000 & L5, L8, L11 \\
50000 & L1 \\
53895 & L7, L9 \\
56889 & L2(l) \\
60235 & L2(h), L4, L6, L10, L12 \\
\botrule
\end{tabular}
\label{tab:emitter_frequencies}
\end{table}

\begin{figure}[tb]
\centering
\includegraphics[width=\columnwidth,angle=0]{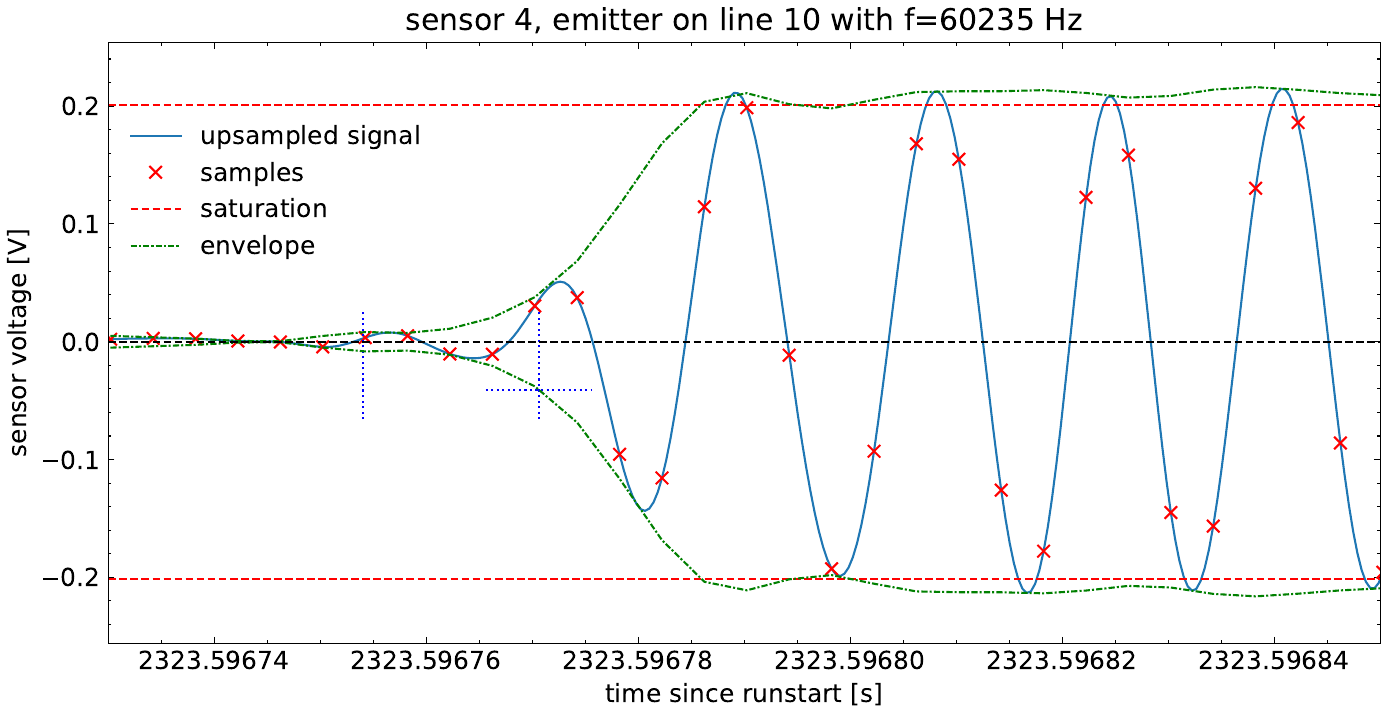}
\caption{Beginning of a signal emitted from the emitter on line 10 as recorded by sensor 4 of the AMS. 
Shown are the sampled values, the upsampled signal and the saturation of the ADC as explained in the text. The envelope on the upsampled signal is used to find the trigger time at which the signal exceeds a predefined threshold. The TOA used for the position reconstruction in this section is chosen to be one signal period before the trigger time.
The light dotted cross intersects at the trigger time and the trigger threshold, the light dotted vertical line indicates the time one period ($\unit[16.6]{\mu s}$ for the emitter on line 10) before the trigger time and is defined as TOA.
}
\label{fig:pinger_signal}
\end{figure}

Figure~\ref{fig:pinger_signal} shows the beginning of a signal from the emitter on line 10 as recorded by sensor 4 of the AMS. The sampling time is $\unit[4]{\mu s}$. 
On the $y$-axis of the plot, the output voltage of the pre-amplifier of the sensor is shown. This voltage is amplified by the standard gain of 20\,dB (cf.\ Section~\ref{sec:amadeus}) before it is digitised by a 16-bit analogue-to-digital converter (ADC). 
The ADC saturates at $\pm 2^{15} = \pm 32768$\, LSB\footnote{1 LSB (least significant bit), denotes the bit with the lowest significance and is typically used as unit for the digital output of an ADC.  }, which corresponds to an input voltage of the ADC of $\pm\unit[2.01]{V}$.
The gain of 20\,dB was chosen to record much weaker signals, hence,
as can be seen in the figure, the emitter signal saturates after the first few oscillations.
Note that the truncated signal is not reproduced in the region of saturation. This is due to the anti-alias filter applied to the digitised $\unit[500]{ksps}$ signal before downsampling, see Section~\ref{subsec:amadeus-design}.

The sensitivity of the sensors of the AM is roughly $\unit[-135]{dB\ re 1 V/\mu Pa}$ so that a sensor voltage of $\unit[0.1]{V}$ corresponds to a pressure change of roughly $\unit[0.5]{Pa}$ relative to the ambient pressure. The sensitivity, however, depends on the frequency and the orientation of the AM w.r.t.\ the emitter and the precision of the measurement of a few dB corresponds to a factor of about 2 or more on the linear voltage scale. To avoid such imprecision of the $y$-axis and because the analysis presented here does not depend on the normalisation of the sensor reading,
the output voltage of the sensors (i.e.\ after pre-amplification of the piezo signal but before applying the adjustable gain factor) was chosen to quantify the amplitude of the pressure signal.

For the standard determination of the TOA, a (negative) threshold on the envelope of the signal is used to define the trigger time. 
To improve the precision of this method and to facilitate the matching of the signals from different emission cycles (see Section~\ref{sec:signal-conditioning}), the signal is upsampled by a factor of 10. 

The threshold was chosen to be safely above the noise recorded by the sensor and scales with the sensitivity of the sensor and its distance to the emitter. 
The TOA is defined to be at one period of the emission frequency before the trigger time
to account for the fact that the threshold is crossed after the onset of the signal, when it already has built up some amplitude.

\subsection{Position reconstruction}
\label{sec:position-reco}
\begin{figure}[tbp]
\centering
\includegraphics[width=\myscale\columnwidth]{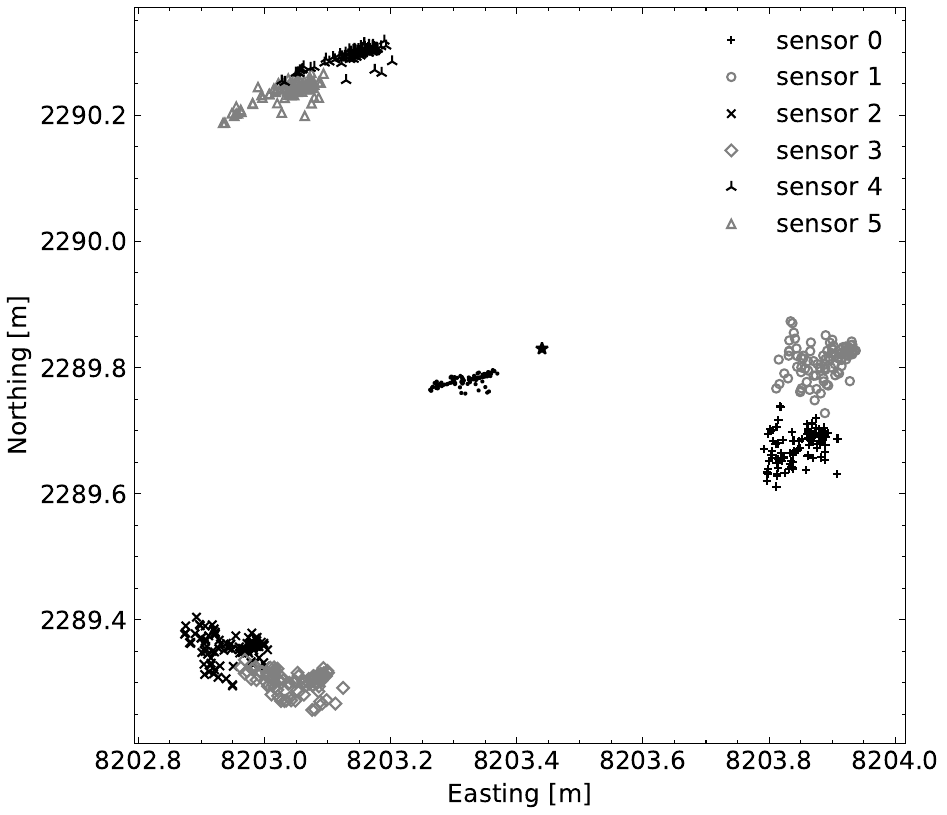}
\caption{
Demonstration of position reconstruction of the AMS for a 3 hour period with low sea currents. Shown are the individually reconstructed sensor positions and the reconstructed nominal centre of gravity of the three AMs in the $xy$-plane of the sea floor, corresponding to the origin of the local reference frame of the AMS (small black dots). The black star denotes the nominal position of the anchor of line 12. 
}
\label{fig:Sensor_reco_xy}
\end{figure}

To reconstruct the position of the AMS, at first the positions of the individual sensors were reconstructed in UTM coordinates. For each sensor it was required that it received signals from at least four emitters; with a minimum of three emitters needed for the position reconstruction, the signals from additional emitters provide redundancy. 
Then the position of the origin of the local reference system of the AMS in UTM coordinates and the orientation of the AMS (yaw angle $\Phi_\text{UTM}$) were obtained from a $\chi^2$-minimisation of the distances between the reconstructed sensor positions and those measured in the local reference frame of the AMS before deployment.  Roll and pitch of the AMS were assumed to be zero.
Reconstructed positions from at least four sensors were required.
The result of the position reconstruction of the AMS for a sample run\footnote{Data recording is organised in runs, which contain data collected typically over a time period determined by either a maximal duration or maximal data size. For AMADEUS, runs typically lasted 3 hours.} of 3 hours length is shown in Figure~\ref{fig:Sensor_reco_xy}.

\begin{figure}[htbp]
\centering
\includegraphics[width=\columnwidth]{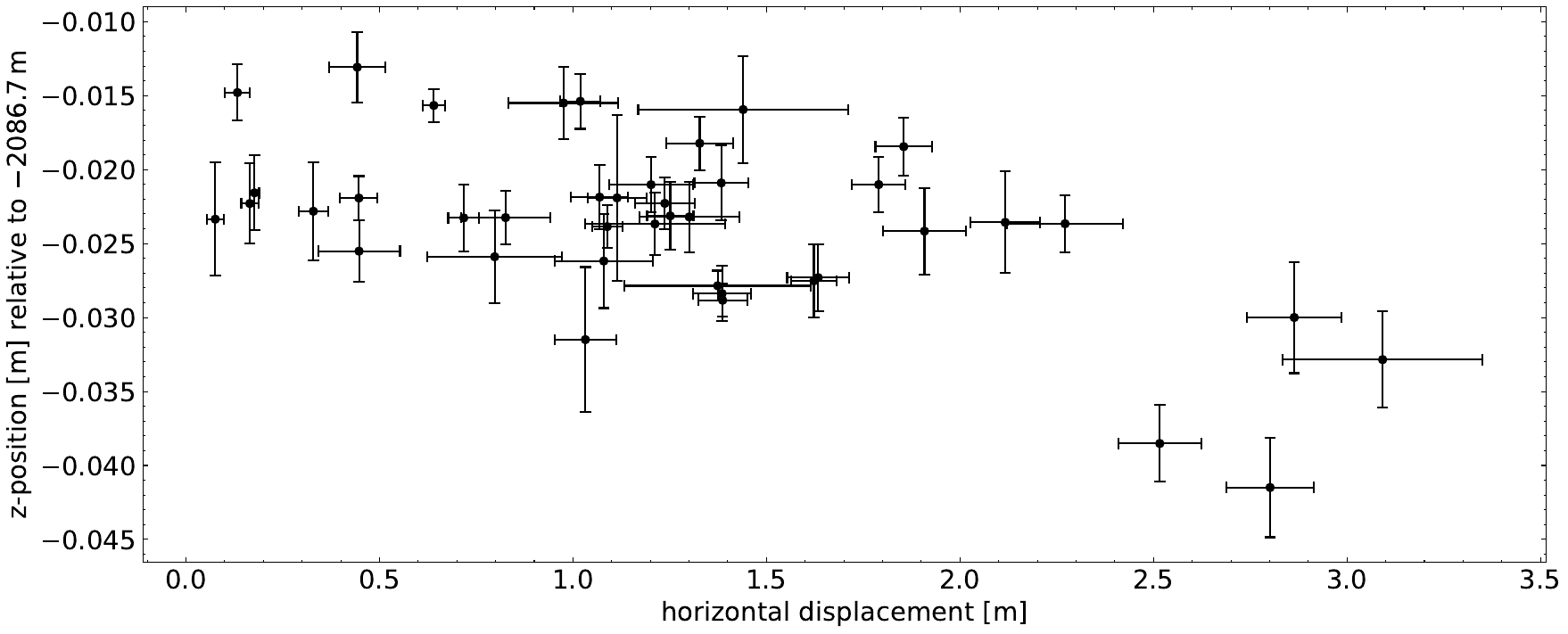}
\caption{
Vertical position of the AMS relative to its nominal position of $\unit[-2086.700]{m}$ as a function of the horizontal displacement of the AMS from the position straight above the anchor of line 12 on the sea floor. 
Each entry corresponds to one of the 41 runs used for the analysis.
The uncertainty on the horizontal displacement is dominated by the movement of the AMS during a run.
}
\label{fig:AM_z-position}
\end{figure}

Figure~\ref{fig:AM_z-position} shows the $z$-positions of the 41 runs 
used for the analysis described in this article, as a function of the horizontal displacement of the AMS from its nominal position. Note that the uncertainty on the horizontal displacement is mostly due to actual movements of the storey. The $z$-position is stable for all runs on a level of about $\unit[1]{cm}$. As expected for the largest displacements, the reconstructed $z$-positions move slightly towards the sea floor (cf.~\cite{bib:antares-pos-2012}).   

\begin{figure}[htbp]
\centering
\includegraphics[width=\myscale\columnwidth]{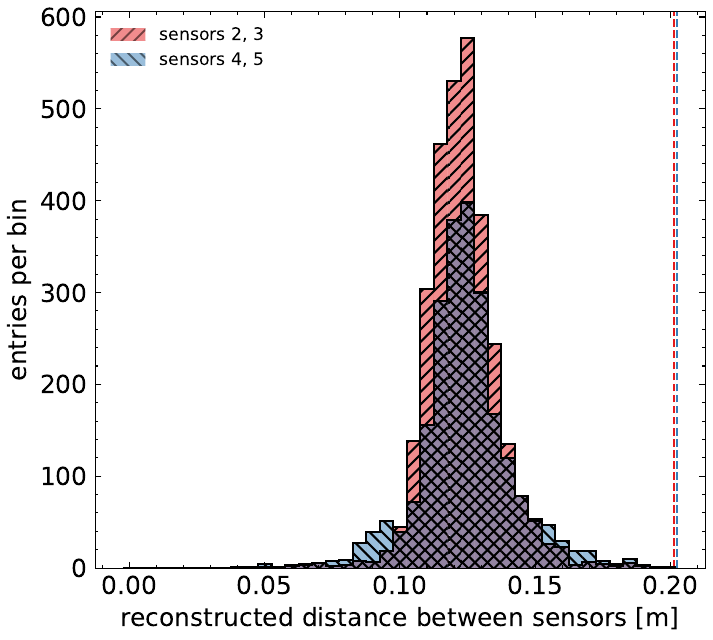}
\caption{
Reconstructed distances between the positions of sensors 2, 3 and 4, 5, where each pair is located in an individual AM. The broken lines show the nominal values of the distances calculated from the sensor positions measured before deployment, scaled to the inside of the glass sphere. The respective colours correspond to those of the sensor pairs. The cross hatched area indicates the overlap of the two distributions.
}
\label{fig:Sensor_distance}
\end{figure}

While the reconstruction of the AMS position yields a good statistical precision, a reconstruction of the fixed distance between the sensors in one AM yields systematic offsets, as shown in Figure~\ref{fig:Sensor_distance}. Only distances between sensor pairs (2, 3) and (4, 5) are shown, as sensor 1 provided data only for a fraction of the recorded runs due to a malfunction. 
The reconstructed distances  are
$123 \pm \unit[14]{mm}$
between sensors 2 and 3,
and
$125 \pm \unit[20]{mm}$
between sensors 4 and 5.
Compared to the distance between the sensors as deduced from the measurements before deployment of
$\unit[201]{mm}$ for sensors 2 and 3, and $\unit[203]{mm}$ for sensors 4 and 5,
the  reconstructed distances are about $\unit[8]{cm}$ too short. The positions of the sensors were measured at the outer perimeter of the glass sphere before deployment; the calculated distances given above were scaled to the inner perimeter for the comparison with the reconstructed distance. 
The dimensions of the piezo ceramics used in the AMs cannot account for the systematic offset. 

\begin{figure}[tbhp]
\centering
\includegraphics[width=\myscale\columnwidth]{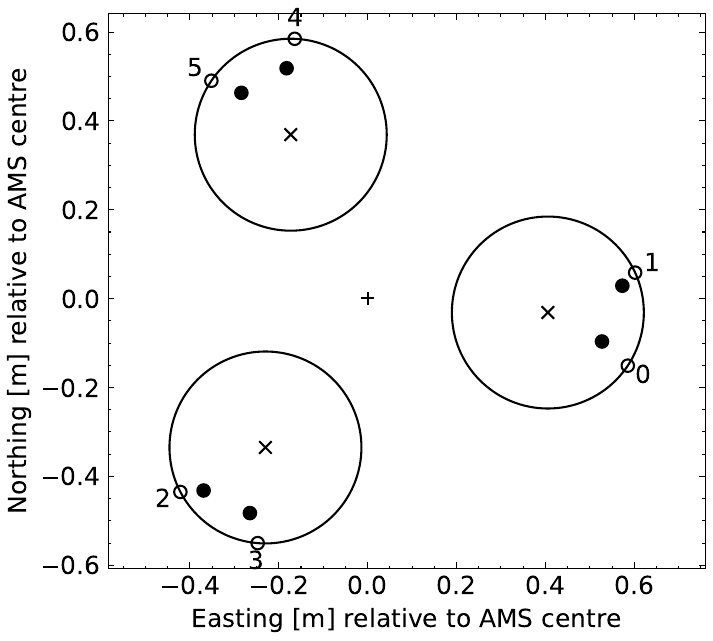}
\caption{
Sample position reconstruction for one of the reconstructed positions of the AMS from Figure~\ref{fig:Sensor_reco_xy}. Filled circles indicate the reconstructed sensor positions, the open circles indicate the  sensor positions measured before deployment, after moving the centre of the AMS (black plus) and rotating the AMS according to the outcome of the fit. The black crosses indicate the centres of the AMs and the black circles their circumferences.
}
\label{fig:AMS_reco_demo}
\end{figure}

Figure~\ref{fig:AMS_reco_demo} shows a sample position reconstruction as it entered Figure~\ref{fig:Sensor_reco_xy}. The reconstructed sensor positions can be seen to be moved inwards and towards each other in all AMs. These shifts cause the systematic offsets of the reconstructed distances between sensors in a given AM as shown in Figure~\ref{fig:Sensor_distance}.

Systematic effects on the TOA can easily stem from the procedure of estimating that the trigger time is delayed w.r.t.\ the TOA by one period of the emitter frequency (cf.~Figure~\ref{fig:pinger_signal}).
However, modifying the calculation of the TOA from the trigger time by setting the trigger time to the TOA or subtracting two periods of the emitter frequency from the trigger time does not significantly affect the difference between the sensor positions. The change in the TOF has mostly the effect of a coherent change of the $z$-position of the sensor positions.

For the reconstruction, it was assumed that the waves from the emitters propagate through the water until they are registered by the receivers. The results from Figure~\ref{fig:Sensor_distance} are an indication that this assumption needs some refinement. A closer investigation of the signal propagating in the glass sphere will be done in the following.
The determination of the TOF will be revisited in Section~\ref{sec:pos-reco-revisit}.

\section{Signal propagation to the receivers}
\label{sec:measurement}
\subsection{Analysis strategy}
\label{sec:analysis-strategy}

The propagation of the signal to the sensors was investigated in the following fashion:
For a number of orientations of the AMS spanning the full circle of $360^\circ$, runs were identified for which the orientation was stable, i.e.\ showed little variation as defined by the RMS value of the compass reading over the duration of the run, and the sea current was reasonably low. Then for a given orientation of the AMS and a given acoustic emitter, the point of impact (POI) of a plane wave from the emitter hitting an AM can be determined. 
The approximation of a plane wave is valid, because the distances between emitters and receivers are large ($>\unit[100]{m}$) compared to the distances between the receivers on the storey ($\sim\unit[1]{m}$).

Each of the two sensors of the corresponding AM has a specific distance from that POI (measured either along the glass sphere or as distance between two parallel planes of the wavefront, intersecting the POI and the sensor, respectively) that depends on the orientation of the AMS. Hence, from the measured difference of the respective TOA in the two sensors of a signal from a given emitter, and the calculated difference of the distances the signals propagates, the speed of sound of the signal can be deduced. 

The pairs of sensors (2, 3) and (4, 5), located in two different AMs, were used to derive the speed of sound of the emitter signal independently. 
Sensors 4 and 5 have very similar sensitivities and corresponding visualizations of the signals are 
more suitable for demonstration.
Hence the analysis is first explained and performed for this pair of sensors and then the results compared to those for sensors 2 and 3.

The analysis procedure will be demonstrated with signals from the acoustic emitter on line 10 -- the frequency $f_{10} = \unit[60235]{Hz}$ of the signals of this emitter is the highest one used in ANTARES and line 10 is the one that is closest to line 12 of all emitters operating at that frequency. The high frequency allows for the best resolution;
the effect of the distance of the emitter from the receiver -- or more precisely, of the angle $\vartheta_\text{POI}$ of the POI below the horizontal orthodrome of the AM -- will be discussed in Section~\ref{sec:effect-f-dist}.

\subsection{Orientation of the acoustic modules}
\label{sec:orientation-AM}

The storeys have certain preferred orientations, depending on recurring underwater currents, e.g.\ from the Coriolis force or tides, and on their $120^\circ$ rotational symmetry.
In particular, a southwards orientation (compass reading $\CompAng \approx 180^\circ$) is very rare.
Furthermore, periods with strong changes alternate with periods of low changes in orientation of the AMS. 

All angles will be defined as increasing in clockwise direction -- this corresponds to the convention of a compass while angles from coordinates in the $xy$-plane have to be adjusted.
The angle  $\Phi_\text{UTM}$ denotes the orientation of the $x$-axis of the local reference system of the AMS w.r.t. the Northing axis of the UTM grid. Then the difference to the compass reading is given by
\begin{equation*}
\CompAng - \Phi_\text{UTM} = \delta_\text{dec} + \delta_\text{cal} + \delta_\text{conv} =: \delta_\text{off} \ .
\end{equation*}
Here $\delta_\text{dec}$ is the magnetic declination;
the calibration error $\delta_\text{cal}$ combines magnetic deviation of the compass reading of $0^\circ$ from the direction magnetic North, and offsets of the $x$-axis of the local reference frame of the AMS w.r.t.\ the orientation magnetic North on the conceived compass rose; and $\delta_\text{conv}$ is the convergence angle.
The ANTARES detector is located on the edge of UTM zone 32 where the Northing axis has a convergence angle of $\delta_\text{conv} = +1.9^\circ$ w.r.t.\ the geographic North.

In practice it is most convenient to determine the combined offset $\delta_\text{off}$. 
As part of the position reconstruction described in Section~\ref{sec:position-reco}, 
the orientation of the AMS as obtained from the fit was compared to the compass reading.
The resulting offset of the compass reading is
$
\delta_\text{off} 
=  -3.0 \pm 2.3^\circ$. 
Hence, for a compass reading of $\CompAng = 357^\circ$ (or $-3^\circ$), the 
$x$-axis of the local coordinate system of the AMS is aligned with the Northing axis.

For the analysis, the orientation of a given sensor w.r.t.\ a given emitter needs to be determined. 
The positions of the emitters and the nominal position -- i.e.\ assuming a perfectly vertical alignment of the lines -- of each storey  are known as coordinates in the UTM grid. Displacements of the AMS from its nominal position due to the movement with the underwater currents have been kept small by selecting runs with low sea currents. This was demonstrated in Figure~\ref{fig:AM_z-position}, where the displacements for all runs used in the analysis are shown.

\begin{figure}[htb]
\centering
\includegraphics[width=\columnwidth,angle=0]{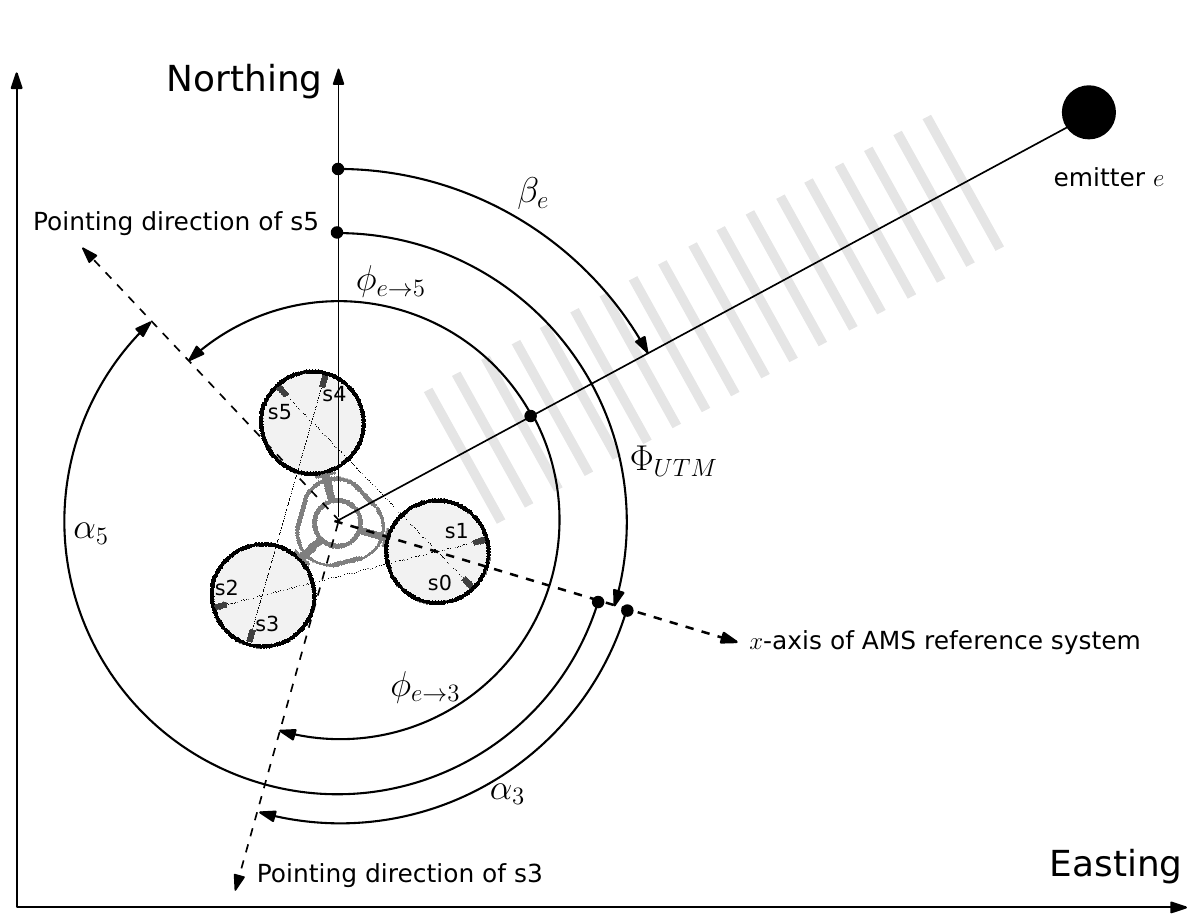}
\caption{
Various angles defined for the AMS as described in the text. Angles depending on the sensor are demonstrated at the example of sensors 3 and 5. The Northing axis and the pointing directions of sensors 3 and 5 have been shifted such that their origins coincide with the centre of the AMS for easier visualisation of the involved angles. 
Axes rotating with the AMS are shown as arrows with broken lines. 
This is indicated by the grey bars perpendicular to the line between the emitter and the AMS. 
The orientation of the storey, defined by the angle $\Phi_\text{UTM}$, changes with time due to time-varying underwater currents. 
}
\label{fig:definition_angle}
\end{figure}

The time-varying orientation of a given sensor $s$ relative to the Northing axis is  given by $\varphi_\text{s} = \Phi_\text{UTM} + \alpha_\text{s}$.
The fixed angle $\alpha_\text{s}$ of the orientation of sensor $s$ w.r.t.\ the $x$-axis of the local reference system of the AMS
is known from measurements before deployment. 
Figure~\ref{fig:definition_angle} demonstrates the 
geometry to find the time-varying angle $\phi_{e\rightarrow s}$ that  quantifies the relative orientation of sensor $s$  w.r.t.\ emitter $e$.
For $\phi_{e\rightarrow s}=0^\circ$, the horizontal component of the vector from the centre of the AM to  sensor $s$ points towards emitter $e$.

The fixed angle in the horizontal plane between the Northing axis of the UTM grid and
the line from a given emitter $e$ to the anchor of line 12, on which the AMS is located, is denoted by 
$\beta_\text{e}$. 
It is then
$\phi_{e\rightarrow s} =  \varphi_\text{s}  - \beta_\text{e} =
 \Phi_\text{UTM} + \alpha_\text{s} - \beta_\text{e} = \CompAng - \delta_\text{off} + \alpha_\text{s} - \beta_\text{e}$.

\subsection{Signal conditioning}
\label{sec:signal-conditioning}

\begin{figure}[htbp]
\centering
\includegraphics[width=0.95\columnwidth,angle=0]{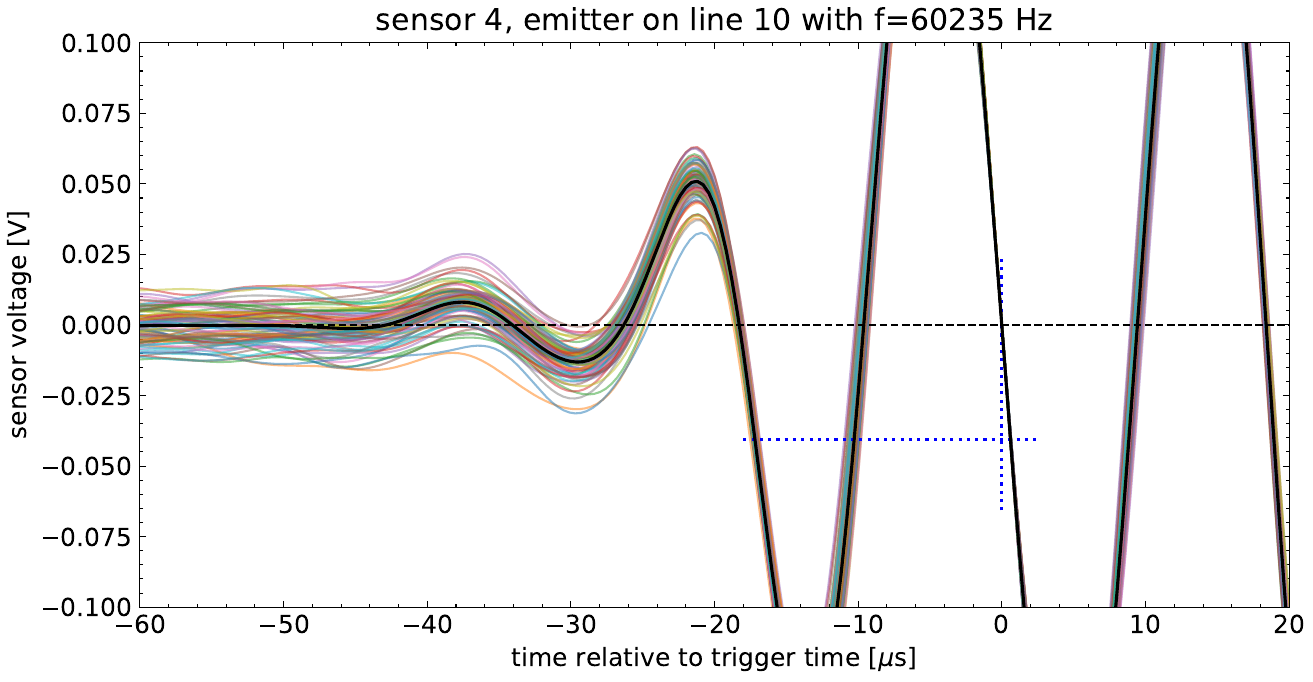} 
\caption{Total of 90 upsampled signals from the emitter on line 10 of a run overlaid (coloured graphs) for sensor 4 of the AMS. The black graph corresponds to the average. A zoom to the onset of the signal is shown.
The individual signals are aligned along the time axis to coincide at the trigger time (vertical blue dotted line) which has been defined as $t=0$ in the plot. The horizontal blue dotted line indicates the trigger threshold; the trigger time is the second zero crossing after the threshold was first passed.
}
\label{fig:pinger_signal_overlaid}
\end{figure}

In order to improve the signal to noise ratio at the onset of the emitter signal, 
all upsampled signals for a given emitter and sensor for a run were  overlaid such that the trigger times coincide, and then averaged. This is illustrated for 90 signals of a run in Figure~\ref{fig:pinger_signal_overlaid}.
For better matching of the individual signals,
the trigger conditions have been changed w.r.t.\ those described for the standard analysis in Section~\ref{sec:TOA-determination} by taking the time of the second zero crossing after exceeding the threshold as trigger time.
For the analysis presented here, the precise value of the trigger time is actually not important, as long as it allows for a precise matching of the signals from different emission cycles. This will be elaborated below.

\subsection{Signal shape vs.\ AMS heading}
\label{sec:sig-shape-vs-orientation}
For each of the combinations of all emitters with the two sensor pairs that were investigated, a set of runs was chosen for which the relative orientations $\phi_{e\rightarrow s}$ cover a range of angles which is as wide as possible. A total of 41 runs were used for the analysis (cf. Section~\ref{sec:position-reco}). 

A number of criteria had to be fulfilled in order for a run to be eligible:
The standard deviation of the compass values recorded during a run must not exceed $4^\circ$. 
Runs must have the standard length of 3 hours, resulting in about 90 emission cycles. 
In practice, 3-hour-runs frequently have less emission cycles recorded. This can be caused by high ambient noise, e.g.\ from precipitation or shipping traffic, that leads to high data rates and potential loss of data. Or signals are low when the AM is rotated away from the emitter and may not have a sufficiently high signal to noise ratio to be detected. As a minimum, 30 emission cycles, each one recorded in \emph{both} sensors of a given AM, were required. 
Furthermore, \sout{when possible, runs with low sea currents were chosen.}%
only runs with sea currents lower than $\unit[90]{mm/s}$ were considered,
\red{and from the runs eligible for a given heading of the AMS, those with the lowest prevalent sea currents were selected.}
The resulting displacements of the AMS were shown in Figure~\ref{fig:AM_z-position}. Using the position reconstruction described in Section~\ref{sec:pos-calibration}, it was confirmed that the centre of gravity of the AMS did not move more than $\unit[1]{m}$ during any of the runs.

Over the lifetime of AMADEUS, 
the headings of the AMS were not equally distributed over all directions on the compass rose, cf.~Section~\ref{sec:orientation-AM}. In addition, during some periods, not all emitters were functioning. 
And finally, for large relative orientation $\phi_{e \rightarrow s}$ between emitter and receiver, due to their specific properties, some combinations of emitters and receivers are more likely to record the emitter signal than others.
For these reasons, the range of relative orientations $\phi_{e \rightarrow s}$ for which a combination of emitter and receiver can record data is limited.  
For a given pair of sensors and a given emitter, at least 10 runs were required which cover a wide distribution of AMS headings $\Phi_\text{UTM}$.
As will be seen, the most relevant headings are those for which the differences of the TOAs between the two sensors are maximal.

For each waveform observed for a given sensor, resulting from the superposition of signals from emission cycles for a run, the time of the first discernible maximum
was defined as $t_0 = 0$ for that waveform. 
Then for each of the two sensors in an AM, the waveforms obtained for the set of runs corresponding to a given emitter-receiver combination can be aligned such that their $t_0$'s coincided. 

\begin{figure}
\centering
\includegraphics[width=0.95\columnwidth,angle=0]{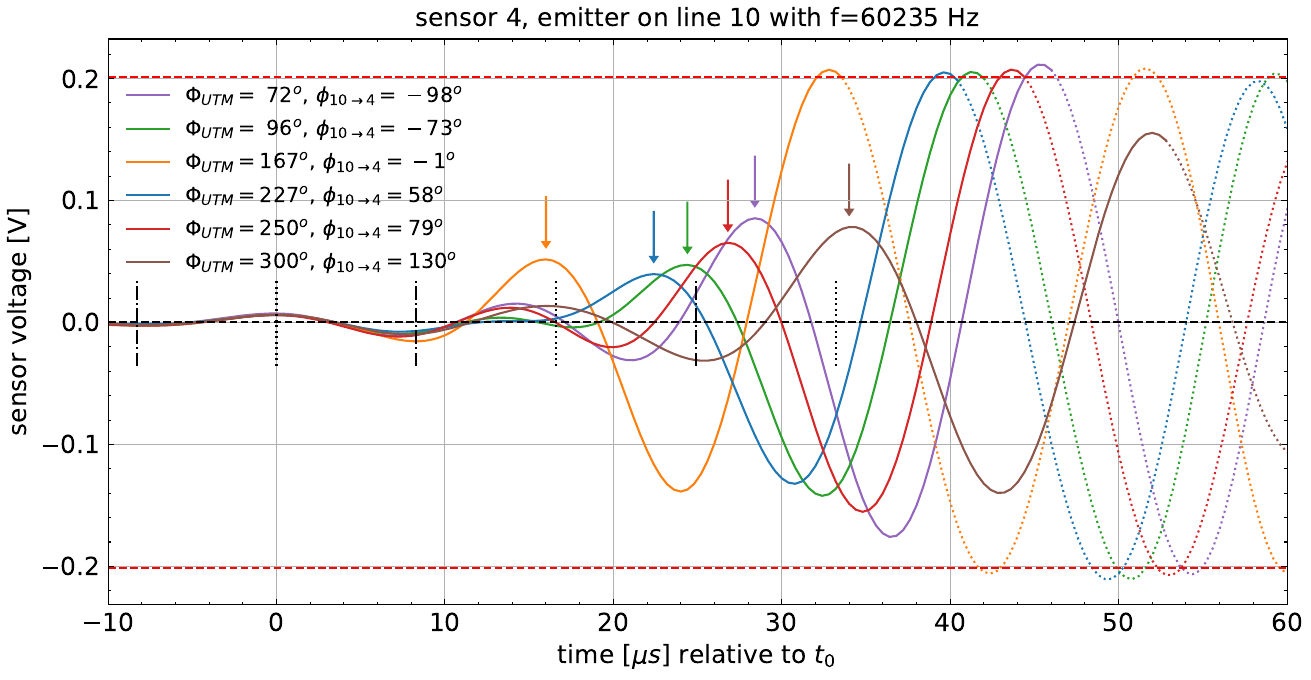}
\includegraphics[width=0.95\columnwidth,angle=0]{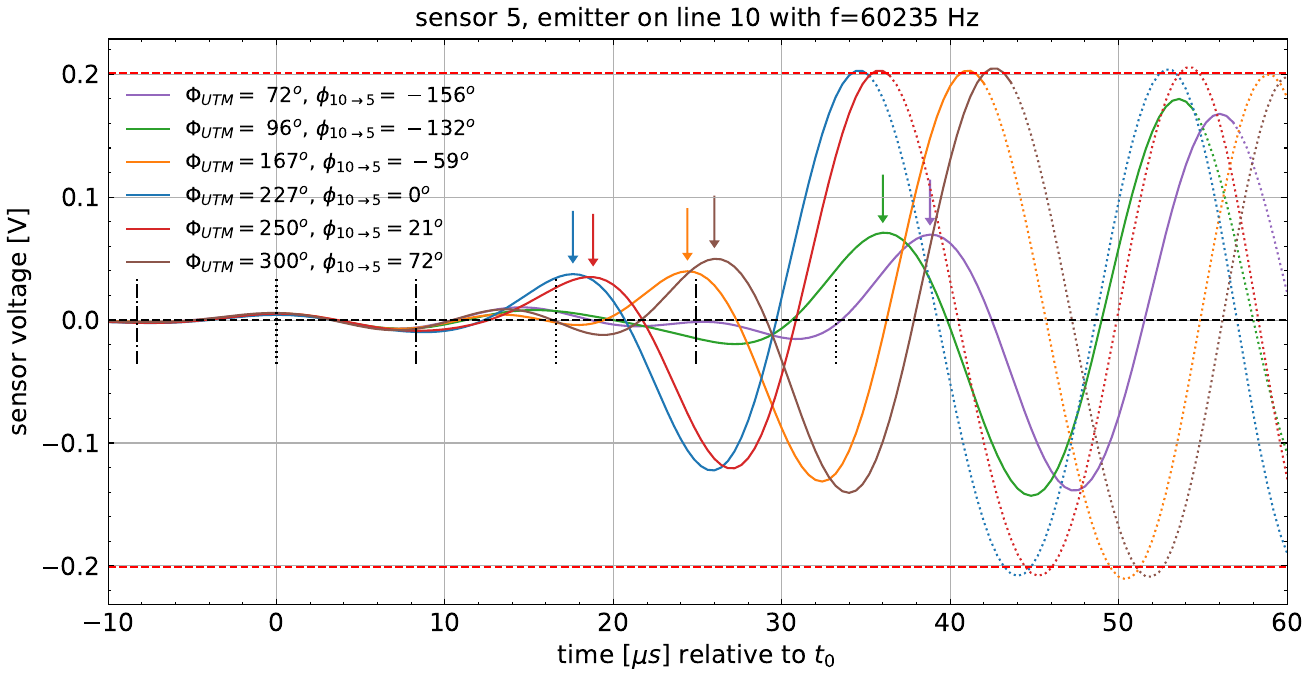}
\caption{Signals from the emitter on line 10 with frequency $\unit[60235]{Hz}$ for runs with various orientations of the sensors with respect to the emitter for 
sensor 4 (top) and 
sensor 5 (bottom). The first vertical dotted line indicates the maximum at the time of $t_0$, for subsequent dotted lines the time is incremented by one period of the emitter signal frequency. Dash-dotted lines indicate the nominal times of the minima of the early wave.
The arrows indicate the first maximum of the late wave for the respective signals.
After the signal has reached its maximum, the lines are continued as dotted lines for better clarity.
The red dashed line indicates the saturation of the ADC.
For the sake of a clearer representation in the figure, not all runs that were evaluated for the measurement of the speed of sound described in Section~\ref{sec:sound-two-waves} are shown.
  }
\label{fig:signal_vs_time_sens_4_5}
\end{figure}

The result of this procedure is shown for a subset of runs in Figure~\ref{fig:signal_vs_time_sens_4_5} for the acoustic emitter on line 10 and sensors 4 and 5.
%
Two waves can be observed:
An early wave with a first maximum that by definition is at $t_0$ with a low amplitude, and a late wave with a much higher amplitude that starts with a delay w.r.t.\ $t_0$. This delay increases as the AM is rotated further away from the emitter, i.e.\ the absolute value of the relative orientation $|\phi_{e \rightarrow s}|$ increases. 
For all waveforms, the first minimum after $t_0$ coincides well with the expected time of half the period of the signal after the maximum. \red{Small shifts will be discussed in Section~\ref{sec:model-superimposing-waves}.}
Negative relative orientations fit into the scheme as the delay is independent of the direction of the rotation.
For relative orientations with  $|\phi_{e\rightarrow s}| \gtrsim 100^\circ$,
the first maximum of the late wave rises so late that 
 the next maximum of the early wave after the first one at $t_0$ is observed very close to the expected time at $t_0 + 1/f_{10}$ with $1/f_{10} \approx \unit[16.6]{\mu s}$.
The start time of the late wave cannot be determined precisely, as it superimposes with the early wave while building up.

\begin{figure}[htbp]
\centering
\includegraphics[width=0.95\columnwidth,angle=0]{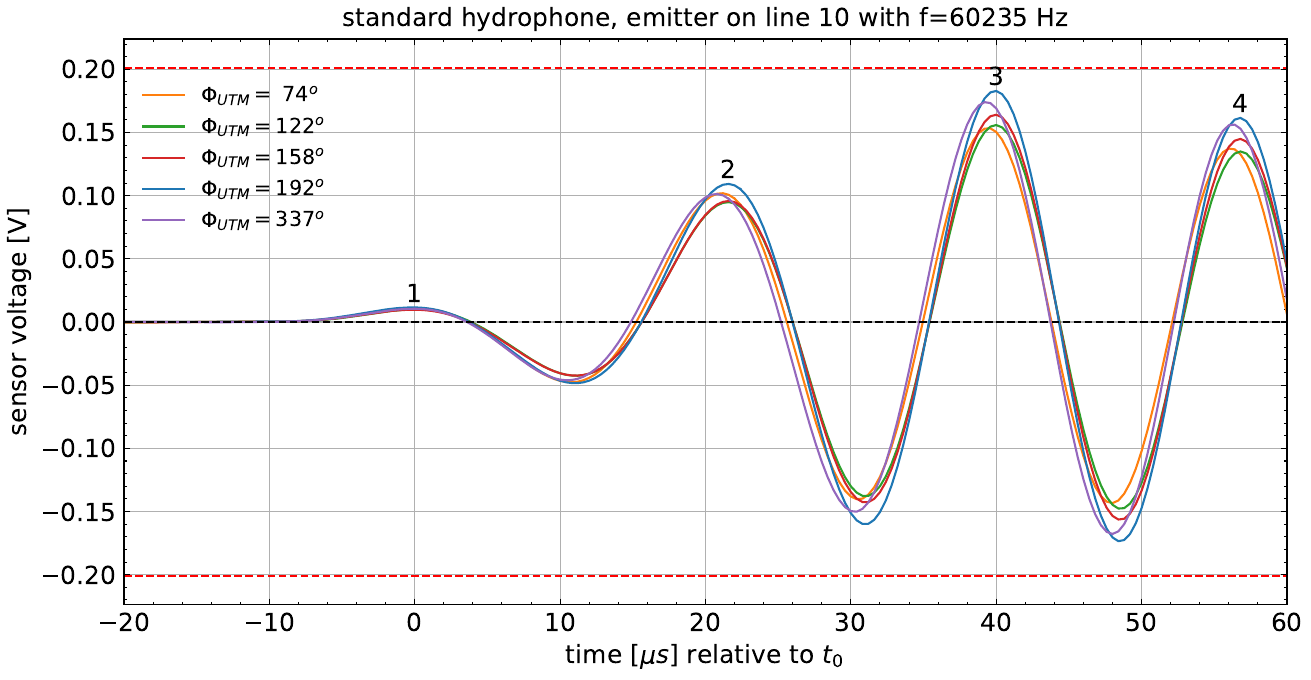}
\caption{Plot corresponding to Figure~\ref{fig:signal_vs_time_sens_4_5} for a commercial hydrophone in the floor directly above the AMS, recording signals from the emitter on line 10. No dependence of the signals on the orientation of the storey can be observed. Numbers indicate the count of maxima.}
\label{fig:signal_vs_time_sensor_13200}
\end{figure}

Figure~\ref{fig:signal_vs_time_sensor_13200} shows a corresponding plot for one of the commercial hydrophones installed on the storey above the AMS. 
There is no significant dependence on the rotation, so evidently the presence of the two superimposing waves is a feature of the AMs.  
As the commercial hydrophones have a lower sensitivity than the sensors of the AMS, their signals do not saturate in Figure~\ref{fig:signal_vs_time_sensor_13200}.
The principal shape of the signal as recorded by the commercial hydrophone is as follows: First a small increase of the pressure, followed by an oscillation to a negative amplitude with roughly three times the size of the first positive peak. 
The following maximum 2 has about twice the amplitude of the preceding minimum, and is already close to the 
maximal amplitude. From the third maximum onwards, the maximal amplitude is reached, which however shows some variations.

Each emitter has a specific pattern that may vary from that shown for the emitter on line 10 in Figure~\ref{fig:signal_vs_time_sensor_13200}, 
but the general pattern is a rather small first maximum, followed by a maximum with intermediate amplitude and the maximal amplitude reached with the third maximum.
We assume that the buildups of the early and late wave of the signal in the AM are qualitatively similar, i.e.\ two waves with very different maximal amplitude superimpose, each reaching their respective maximal amplitude with the third maximum.
Presumably, the first maximum that can be observed in Figure~\ref{fig:signal_vs_time_sens_4_5} for the early wave corresponds to maximum 1 in Figure~\ref{fig:signal_vs_time_sensor_13200}.
Note that -- to calculate the speed of sound of the early and late wave, as will be discussed in Section~\ref{sec:sound-two-waves} -- in principal it does not matter which maximum is used, as long as the same one is used for the waveforms observed in the two sensors of an AM. The amplitude of the early wave is only of relevance to estimate the effect of the superposition with the late wave, which will be discussed below.
To this end, assuming that the first maximum observed by a sensor in the AMS corresponds to maximum 1 -- rather than to maximum 2 with maximum 1 disappearing in the noise -- is the more conservative assumption.

\red{
\subsection{Model of superimposing waves}
\label{sec:model-superimposing-waves}
} 

\begin{figure}[tb]
\centering
\includegraphics[width=0.95\columnwidth,angle=0]{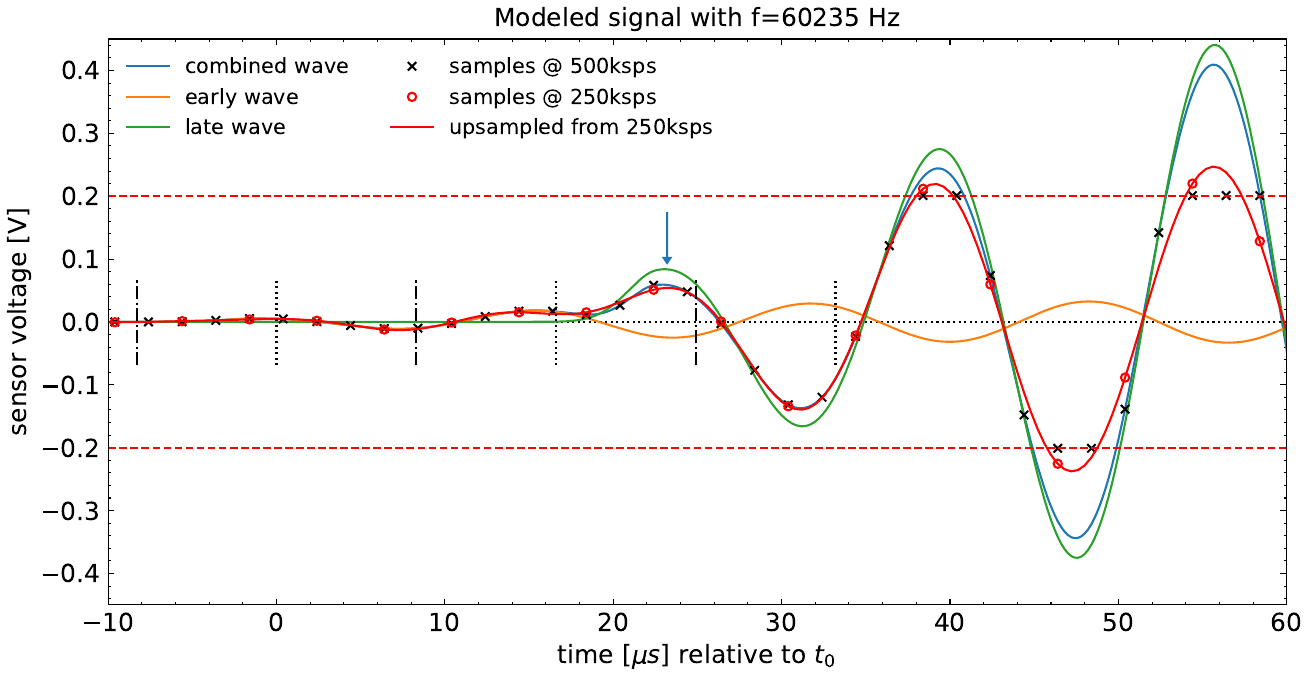}

\caption{Model of received signal in a sensor of an AM as superposition (combined wave) of early and late wave as described in the text.
The definition of $t_0$ (zero point on the $x$-axis) and indications of the saturation of the ADC (red dashed line), first maximum of the combined wave (arrow), and nominal times of maxima and minima of the early wave (dotted and dash-dotted vertical lines, respectively) correspond to those in Fig.~\ref{fig:signal_vs_time_sens_4_5}. 
}
\label{fig:superimposing_waves}
\end{figure}

\red{
A simple model for the superposition of early and late wave was devised, see Figure~\ref{fig:superimposing_waves}. 
To approximate an analog emitter signal as received by a sensor, first the time series of a sine function with frequency $f_{10} = \unit[60235]{Hz}$, and a time resolution of $\unit[0.2]{\mu s}$, corresponding to 10 times the sampling  rate of the acoustic signals of 500\,ksps, see Section~\ref{subsec:amadeus-design}, was produced. The rising part to the first maximum of the sine function, i.e.\ its first quarter period, including some zero padding for the time before the first maximum, was replaced by the rising half of a Gaussian. This way the smooth onset of the signal as seen in Figure~\ref{fig:signal_vs_time_sensor_13200} is approximated. The standard deviation $\sigma$ of the Gaussian was chosen such that its FWHM corresponds to the FWHM of a half period of the sine function, approximated well by $\sigma = 1/(7 f_{10})$.
The late wave (green curve in Figure~\ref{fig:superimposing_waves}) was produced by taking the amplitude of this function as $\unit[0.5]{mV}$ and multiplying it with a Fermi function of the form $1 - 1/(\exp((t-t_c)\lambda)+1)$, with $\lambda$ and $t_c$ tuned to resemble the observed signals. 
The early wave  (orange curve) was derived from the late wave by reducing its amplitude by a factor of $1/15$ and assuming it reaches the sensor a certain time $\tau$ before the late wave. 
The combined wave (blue curve) was then created by adding the two contributions. 
The time delay was taken as
 $\tau = 1.45/f_{10}$, as measured\footnote{
The time difference $\tau$ between the first maximum of the early and late wave as a function of AMS heading $\Phi_\text{UTM}$ will be discussed in Section~\ref{sec:discussion-results}. The $\tau$-value for $\Phi_\text{UTM} = 167^\circ$ can be deduced from Figure~\ref{fig:late_minus_early_peak_sensors_l10_11}, bottom right plot, for signals from the emitter on line 10, received by sensor 5.} for the signal received by sensor 5 for the run with AMS heading $\Phi_\text{UTM} = 167^\circ$ in 
Figure~\ref{fig:signal_vs_time_sens_4_5}, also shown in Figure~\ref{fig:Signal_timedelay}.
Clearly, the model is limited in that e.g.\ the emission pattern is simplified, noise and averaging effects are not considered (cf.\ Figure~\ref{fig:pinger_signal_overlaid}) and the transfer function of the sensors is not applied.

The combined wave was sampled at 500\,ksps and the saturation of the ADC, cf.\ Section~\ref{subsec:amadeus-design}, taken into account by applying a cutoff at $\unit[2.01]{mV}$ (black crosses in Figure~\ref{fig:superimposing_waves}). On these sampled values, the low pass filter for downsampling by a factor of 2 was applied and consequently every second sample was selected (red circles). To these samples, an upsampling by a factor of 10 was applied (red line in Figure~\ref{fig:superimposing_waves}), as was done with the sampled data used in the analysis.

For the value of $\tau$ used for the model shown in Figure~\ref{fig:superimposing_waves}, the separation between the  early and late wave is large enough for the first maximum of the combined wave to be in good agreement with the first maximum of the early wave. 
The times of the maxima and minima of the early wave as extrapolated from $t_0$ (indicated by the dotted and dash-dotted vertical lines, respectively, in Figure~\ref{fig:superimposing_waves}) are delayed w.r.t.\ the actual maxima and minima of the early wave. This is due to the transient behaviour of the signal (multiplication with the Fermi function in the model) that shifts the first maximum of the early wave, and  therefore $t_0$, towards a later time. This effect can also be observed in Figure~\ref{fig:signal_vs_time_sens_4_5} for the data recorded with the AMs.
} 

While the amplitude of the early wave is clearly smaller than that of the late wave, it is difficult to  determine their actual ratio from the data.
\red{In the model, a ratio of 1:15 has been assumed, as it resulted in a combined wave that shows reasonable resemblance with the data.
For increasing absolute values of the relative orientation of the sensors, $|\phi_{e\rightarrow s}|$, the distance propagated from the POI to the sensor increases and the amplitude of the recorded wave decreases, see Section~\ref{sec:lamb-wave-prop} for further discussion.}
For the runs with $|\phi_{e\rightarrow s}| \gtrsim 100^\circ$ from Figure~\ref{fig:signal_vs_time_sens_4_5}, the late wave does not saturate the ADC and the early and late wave have a better temporal separation.
For these runs, maximum 2 of the early wave at $t = 1/f_{10} \approx \unit[16.6]{\mu s}$ is not strongly affected by the late wave. 
From these runs, the ratio of the respective maxima 2 for early and late wave is roughly estimated as  $\sim$1:10.
For the maxima 1 of the early and late wave, the ratio is affected by the low amplitude of the early wave and the interference of the early wave with maximum 1 of the late wave, but the ratio from above can be roughly confirmed.
\red{
Note that the early and late wave do not necessarily have the same attenuation coefficient. The nature of the waves will be further discussed in Sections~\ref{sec:interpret-2-waves} and \ref{sec:lamb-wave-prop}.

In the analysis, the zero crossings of the signal upsampled from the 250 ksps data are measured. These correspond to the zero crossings of the combined wave in the model, which are then assumed to be identical to the zero crossings of the late wave.  In Figure~\ref{fig:superimposing_waves}, the differences can be seen to be very small, which however depends on both the amplitude of the early wave and its time shift w.r.t.\ the late wave. 
Taking the worst case where the early and late wave have a phase difference of $\pi/2$, the combined wave -- ignoring the transient behaviour at the beginning --  can be expressed as
$A(t) = A_0\sin\alpha \sin (2\pi f_{10} t + \pi/2) + A_0\cos\alpha \sin(2\pi f_{10} t) = A_0\sin\alpha \cos(2\pi f_{10} t) + A_0\cos\alpha \sin(2\pi f_{10} t) = A_0\sin(2\pi f_{10} t + \alpha) $
where $A_0\sin\alpha $ and $A_0\cos\alpha $ are the amplitudes of the early and late wave, respectively.
The ratio of the amplitudes corresponds to $\tan\alpha$, and assuming a maximal ratio of $\tan\alpha = 1/10$, it is 
$\alpha \approx 6^\circ$. This corresponds to only about $\unit[0.35]{\mu s}$ for the lowest frequency of $\unit[46545]{Hz}$
and has no significant effect on the systematic error of the measurement, as will be discussed in Section~\ref{sec:appendix_sys-errors} of the appendix.
} 

\sout{The effect will be quantified in Section~\ref{sec:sys-errors}, where
sources of systematic errors will be summarised.}

\subsection{Speed of sound of the two waves}
\label{sec:sound-two-waves}
In liquids, only one speed of sound exists, whereas in solids, multiple modes of oscillation and correspondingly different speeds of sound exist.
It is plausible to assume that the early (faster) wave is a guided wave inside the glass sphere. 
The late (slower) wave could reach the sensor directly through the water, as the corresponding speed of sound is slower than that of any wave propagating through the glass.
To better interpret the waves, their speeds will be derived in this section. 
For the late wave, the speed of sound will be derived first under the assumption of propagation in glass. Subsequently, it will be tested if the result is consistent with the assumption of the wave propagating through the water.

\subsubsection{Propagation time}
\label{sec:propagation-time}

\begin{figure}[tbhp]
\centering
\includegraphics[width=\columnwidth,angle=0]{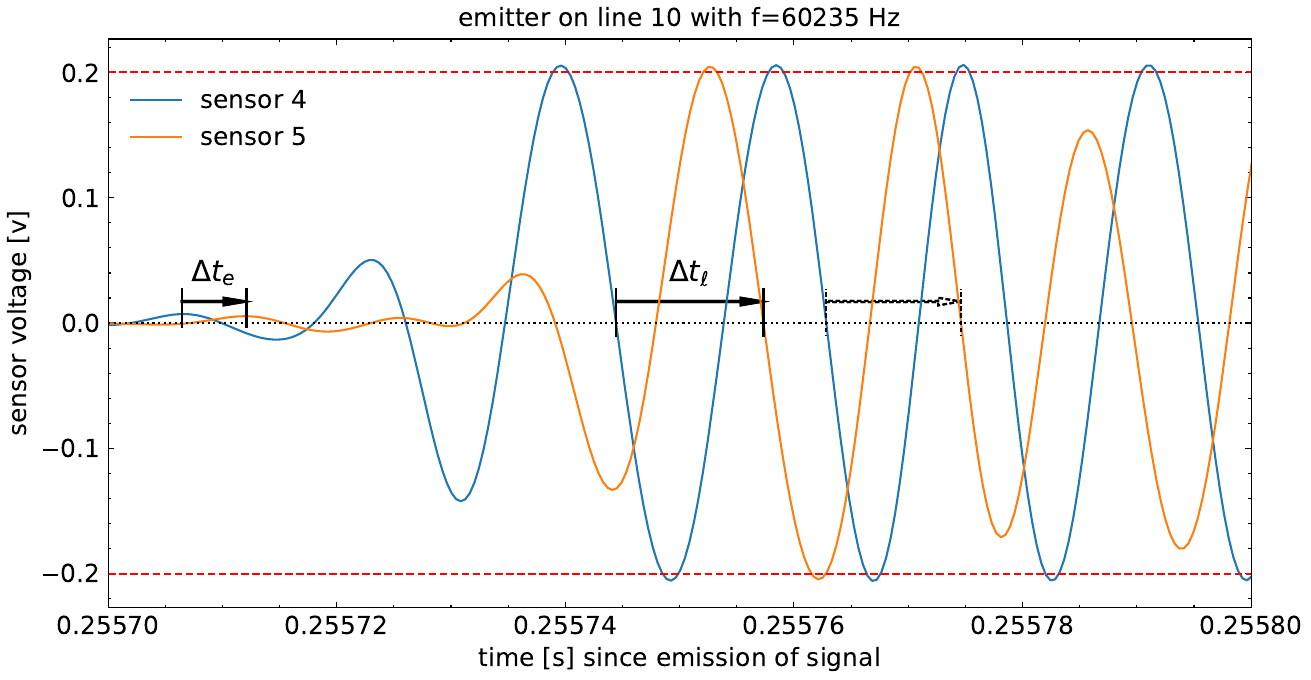}
\caption{Signal amplitude (upsampled and averaged for one run) for sensors 4 and 5 as a function of time since emission of the signal from the emitter on line 10. The time differences between the two waves as discussed in the text are indicated in the plot. 
}
\label{fig:Signal_timedelay}
\end{figure}

In Figure~\ref{fig:Signal_timedelay},  a particular run/orientation was chosen from those shown in Figure~\ref{fig:signal_vs_time_sens_4_5} and the time profiles of the two signals were overlaid starting from the TOE of the signal. This time will be referred to as absolute time as the signals recorded in the two sensors now have a common reference time. This is in contrast to setting the time of the first maximum of the early wave to $t_0 = 0$ for all runs recorded with a given sensor as done in Section~\ref{sec:sig-shape-vs-orientation}.

For the early wave, the time difference $\Delta t_e$ (index $e$ for early) between the first discernible maxima in the signals recorded by the two sensors in an AM is calculated. This corresponds to the time difference between the two $t_0$'s of the sensors from Section~\ref{sec:sig-shape-vs-orientation} in absolute time.
For the late wave, to reduce potential effects of the signal saturation, the time differences were calculated from the zero crossings rather than the positions of the maxima. 

In principle, the time difference between any corresponding zero crossings of the waves recorded in the two sensors should yield the same result. In practice, the early zero crossings, where the late wave has not yet reached full amplitude, could be affected by the superposition with the early wave.
For the later zero crossings, effects such as superpositions from reflected waves could affect the waveform.  
Based on these considerations, the zero crossing after maximum 2 of the late wave was chosen to determine the time difference $\Delta t_\ell$ between the late waves recorded in the two sensors, as indicated in Figure~\ref{fig:Signal_timedelay}.

\red{The dashed arrow in Figure~\ref{fig:Signal_timedelay} indicates the time difference as derived from the zero crossing after maximum 3 of the late wave in the respective sensors. This time difference is slightly smaller than  $\Delta t_\ell$ shown in the same figure, derived from the zero crossings one period earlier.}
\sout{Looking closely at the zero crossings of Figure~\ref{fig:Signal_timedelay}, there is a hint of time-dependant differences between zero crossings.}%
Errors from the choice of zero crossing that is used to calculate the speed of sound of the late wave will be discussed in Section~\ref{sec:sys-errors}.

\subsubsection{Propagation distance in the glass sphere}

The time differences $\Delta t_{e}$ and $\Delta t_\ell$, introduced above, depend on the heading of the AMS. The corresponding difference between the distances propagated by the signals to the respective sensors must be found. 

The POI is the intersection of the glass sphere with the line from the emitter to the centre of the AM. The shortest distance from the POI to the position of the receiver along the glass sphere -- taken at the middle of the glass  -- is taken as the distance propagated by the signal in the glass. 
The unit vector pointing from the centre of the AM to the POI shall be denoted as $\vec{n}_\text{POI}$ and the unit vector from the centre of the AM to a given sensor for its AMS heading in the UTM reference frame as $\vec{n}_{s}$. Then the distance within the glass between the two points as a function of the heading of the AMS is given by 
\begin{equation}
d = r_{c}\arccos(\vec{n}_\text{POI}\cdot\vec{n}_{s})\ . 
\label{eq:dist_in_glass}
\end{equation}
Here $r_{c}$ denotes the radius of the sphere, taken at the centre of the thickness of the glass. 
The differences of the two distances from the POI to the receivers $s_1$ and $s_2$ in an AM is then calculated according to 
\begin{equation}
\Delta d  = r_{c}\arccos(\vec{n}_\text{POI}\cdot\vec{n}_{s_2}) 
 - r_{c}\arccos(\vec{n}_\text{POI}\cdot\vec{n}_{s_1}) \ .
\label{eq:delta_d}
\end{equation}

\subsubsection{Calculation of the speed of sound in the glass sphere}
\label{sec:speed_in_glass_sphere}
The speed of sound of the early wave is given by $\Delta d/\Delta t_e$ and that of the late wave by
$\Delta d/\Delta t_{\ell}$. 
To reduce the effects of potential offsets of the POI (e.g.\ from 
offsets of the compass reading) or of $\vec{n}_s$ (e.g.\ due to systematic errors of the measured sensor positions within the AM), $\Delta d$ will be plotted as a function of $\Delta t$ and the speed of sound derived from the slope of a linear fit.
Here and in the following, the omission of the index for $\Delta t$ indicates either of the time differences $\Delta t_e$ or $\Delta t_{\ell}$ that correspond to a given $\Delta d$.

\begin{figure}[tb]
\centering
\includegraphics[width=\columnwidth,angle=0]{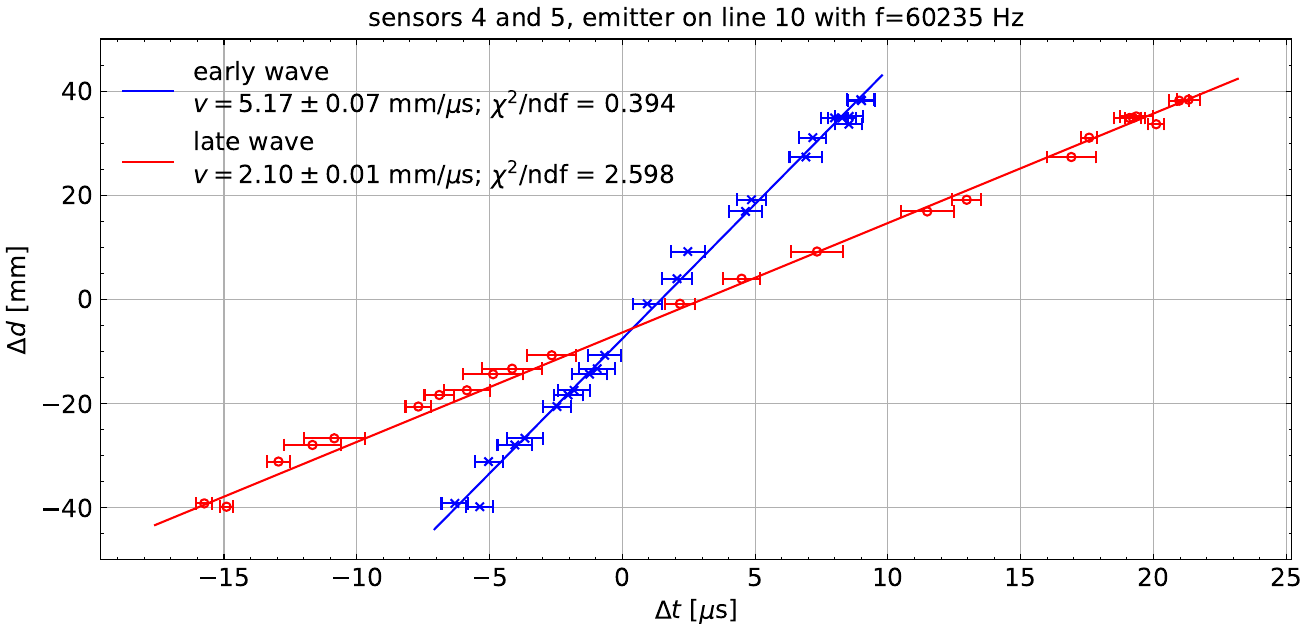}
\caption{Speed of sound derived from a linear fit for the early and late wave for the emitter on line 10, as described in the text. For each run, the average orientation was determined, which corresponds to a particular value of $\Delta d$. 
Each run contributes a value for $\Delta t_e$ and $\Delta t_{\ell}$, which correspond to the two entries on the $x$-axis for the $y$-value of a given run.
}
\label{fig:dist_vs_time_cycle10}
\end{figure}

In Figure~\ref{fig:dist_vs_time_cycle10} the results are shown 
for the emitter on line 10. The values of $\Delta d$ for the time differences $\Delta t_e$ and  $\Delta t_{\ell}$ show a good linear dependence as indicated by the linear fit.  
The errors are statistical only. They were calculated from the individual signals that were superimposed to yield the combined waveform (cf.\ Figure~\ref{fig:pinger_signal_overlaid}) in the following manner:
For each individual signal that reaches both sensors, the difference between the absolute trigger times 
were calculated; the standard deviation of this time difference for all individual pairs of trigger times was assigned as statistical uncertainty for a given value of $\Delta t_\ell$. 
For the early wave, 2/5th of this value was assumed as statistical uncertainty. This is motivated by the ratio of the respective speeds of sound that will be calculated below, which implies a propagation time of the early wave that is smaller by that factor.

In addition, an uncertainty for the determination of the first visible maximum for the early wave and for the zero crossing of the late wave was assumed:
for the early wave this uncertainty was estimated to be $\unit[0.5]{\mu s}$ 
(combined for the time difference of the waveforms from the two sensors),
for the late wave it was calculated from the standard deviation of the times of the zero crossings of the individual waveforms before averaging (cf.\ Figure~\ref{fig:pinger_signal_overlaid}).
This uncertainty varied for the individual runs and was typically in the range $\unit[0.1\sim 0.3]{\mu s}$, again combined for the time difference for the two sensors. The statistical uncertainties shown in Figure~\ref{fig:dist_vs_time_cycle10} and used for the $\chi^2$-minimisation were calculated by adding in quadrature the
two underlying uncertainties discussed above.

The statistical error from the difference of the trigger times in the two sensors is mostly due to small rotations of the sphere. The effect of such movements is larger for a time difference near zero, where one sensor will move towards and the other away from the POI. For orientations with maximal time differences, small rotations will move both sensors away from or towards the POI. Hence, the statistical uncertainties are smaller (larger) for large (small) values of $|\Delta d|$. This can be observed in Figure~\ref{fig:dist_vs_time_cycle10}. 

A statistical error on $\Delta d$ would result from the uncertainty of the compass measurement. This uncertainty however propagates to uncertainties on $\Delta t$ so that no explicit uncertainties for $\Delta d$ were used for the fit.

Fits were done by $\chi^2$-minimisation with the iminuit\footnote{\url{https://iminuit.readthedocs.io/en/stable/}, version 2.21.2} package of python. As the procedure is simplified by having an error on the $y$-value rather than the $x$-value, the fit was done to $\Delta t$ vs. $\Delta d$, giving  the reciprocal of the speed of sound as fit result. The speed of sound and its error can then be easily deduced.

Clearly, the process of formation of the wave in the glass is not trivial and it is an oversimplification to assume that the POI would be the point source of the waves in the glass. More likely, they are formed in some area -- a formation region -- around the POI. However, for symmetry reasons, any effect of the formation process on the distance propagated by the waves cancels out when taking the difference $\Delta d$ for the two sensors. 
Recall that even when a sensor is oriented into the direction of an emitter, the distance between the POI and the sensor is greater than zero because the sensors are located near the horizontal orthodrome of the AM whereas the emitters send their signals from below.
The shortest distance between the POI and the horizontal orthodrome of a given AM, measured along the centre of the glass of the sphere, varies from about 
$\unit[230]{mm}$
to   
$\unit[295]{mm}$ for the emitter with the largest (line 1) to shortest (line 8) distance from line 12.
\sout{This corresponds to several wavelengths, even for the late (slow) wave of the lowest frequency and it is hence reasonable to assume that the formation region does not extend from the POI all the way to the sensors, regardless of the rotation of the AMS.}

\red{Assuming that early and late wave are forming in the same region, a clear temporal separation between the first maxima of the two waves should indicate that the sensor is positioned at some distance from the formation region. In Section~\ref{sec:effect-f-dist} the situation of the first maxima of early and late wave reaching the sensor with a rather short delay will be discussed.}
The formation of the signal will be discussed in Section~\ref{sec:discussion-results}.

\subsubsection{Hypothesis of the late wave propagating through the water}
\label{sec:late-wave-in-water}

The speed of sound in water at the ANTARES site has been derived from the Chen-Millero formula~\cite{bib:Chen-Millero-1977} using sets of temperature and salinity profiles measured at the site.
The speed of sound of the late wave of about $\unit[2]{\mmpermus}$ from the fit shown in Figure~\ref{fig:dist_vs_time_cycle10} is not compatible with the speed of sound in water at the depth of the AMS of  $v_\text{water} = \unit[1.54]{\mmpermus}$.
However, the difference $\Delta d$ of propagation distances from the POI to the respective sensors has been based on the assumption of the waves propagating through the glass.
For a plane wave propagating through the water, it would be more accurate to assume that the plane wave hits the sphere at the POI and then continues as a plane wave to the receiver. 
In principle, if the sensor is located in the shadow zone of the emitter, diffraction has to be taken into account so that the wave reaching the sensor cannot be a pure plane wave. 
A sensor is considered to be in the shadow zone of a given emitter if the AM is rotated such that the direct line between the emitter and the sensor position -- extrapolated to the outer surface of the sphere -- intersects the glass sphere. 
In this case, the distance travelled by a notional plane wave can be seen as the shortest possible distance to be propagated by a wave in the water from the POI to the sensor, therefore yielding a lower limit on the speed of sound.  
The distance propagated by the plane wave in water then is
\begin{equation}
d_w = r_{o}(1 - \vec{n}_\text{POI}\cdot\vec{n}_{s}) \ ,
\label{eq:dist-in-water}
\end{equation}
where $r_{o}$ is the outer radius of the sphere. 

In practice, however, the difference between distances $\Delta d$ according to Eq.~\ref{eq:delta_d} and the difference $\Delta d_w$ that follows from the definition of $d_w$ above is minimal. 
Recalculating the speed of sound of the late wave with $\Delta d_w$ rather than $\Delta d$ yields
$v_{\ell w} = 2.17 \pm \unit[0.01]{\mmpermus}$ with $\chi^2/\text{ndf} = 2.84$ for the emitter on line 10. 
This has to be compared to the fit results of the late wave given in Figure~\ref{fig:dist_vs_time_cycle10}.
The reason that the speed of sound \emph{increases} w.r.t.\ the assumption of the late wave propagating in the glass is due to the fact that the outer radius of the sphere was used in Eq.~\ref{eq:dist-in-water} whereas the central radius of the glass was used in Eq.~\ref{eq:delta_d}.
This corresponds to a factor of $r_{o}/r_{c} = 1.036$.

To reduce the speed of sound $v_{\ell w}$ to the known speed of sound in water of  $\unit[1.54]{\mmpermus}$ at the given depth, a systematic error of more than 40\% would have to be present. 
The total systematic error does not reach that size and
the conclusion is that indeed both waves travel in the glass of the sphere.
The speeds of sound as calculated for the assumption of the late wave propagating through the glass and the water, respectively, are compared in Section~\ref{sec:appendix-late-wave-in-water} of the appendix.

\subsection{Effect of signal frequency and distance of emitter}
\label{sec:effect-f-dist}

In Figure~\ref{fig:signal_vs_time_sens_4_5}, a clear separation between the first visible maximum of the early wave
and maximum 1 of the late wave is visible. For a relative orientation of $\phi_{e\rightarrow s} = 0^\circ$ for a given sensor $s$ and emitter $e$, the delay between the two waves is smallest. Even in this case, the separation between maximum 1 of the late wave and the first visible maximum of the early wave -- on which the calculation of the speed of sound is based -- is about one period of the frequency $f_{10} = \unit[60235]{Hz}$ which should prevent any effect of the late wave on the early wave -- see Figure~\ref{fig:signal_vs_time_sensor_13200} where the signal at $1/(2f)\approx \unit[8]{\mu s}$ before maximum 1 is essentially zero.

Apart from the frequency, the separation of the two waves also depends on the location of the emitter. For a signal emitted from directly below an AM, the path length of the signal in the AM would be about a quarter of the circumference of the AM to each sensor near the horizontal orthodrome. Alas, such an emitter position would not allow for a measurement of propagation times, as $\Delta t$ and $\Delta d$ would be zero for any orientation of the AM.
Still, for a shorter horizontal distance between emitter and line 12 holding the receiver -- and therefore the POI being located closer to the lowest point of the AM --
the propagation path from the POI to the sensor will be longer. This results in
a better separation of the early and late wave recorded by a given sensor.
 
\begin{figure}[phbt]
\centering
\includegraphics[width=0.95\columnwidth,angle=0]{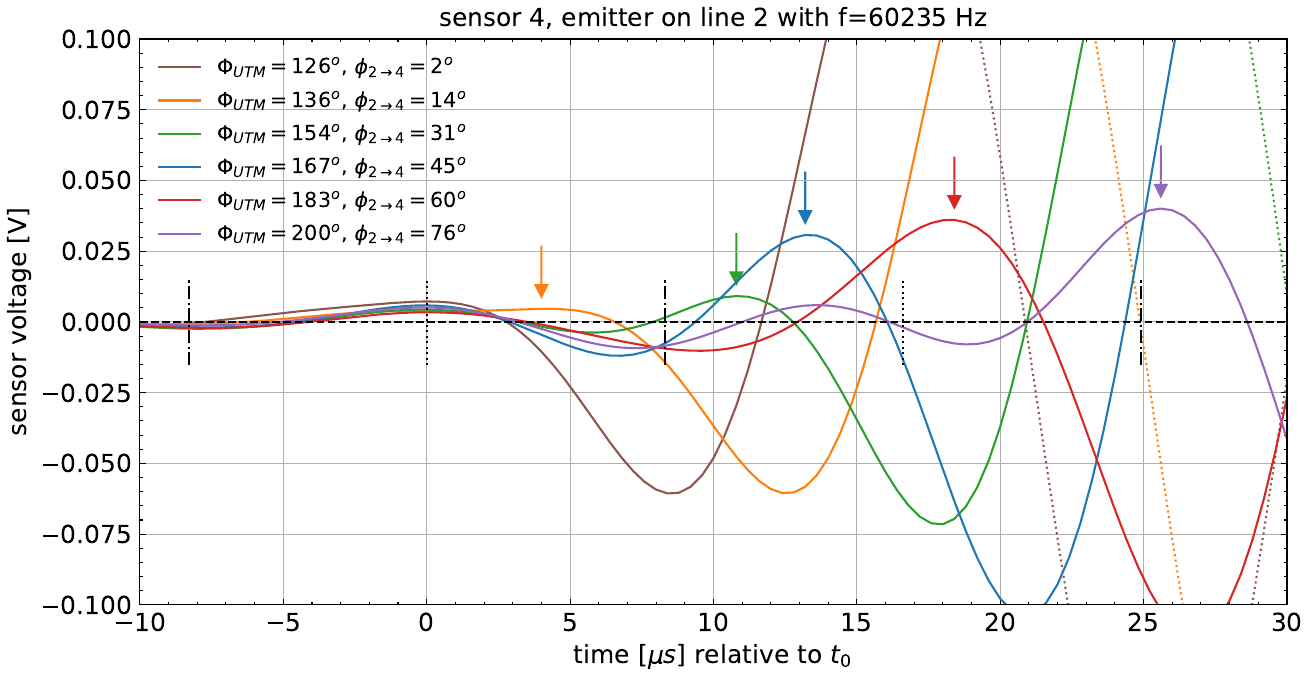}
\includegraphics[width=0.95\columnwidth,angle=0]{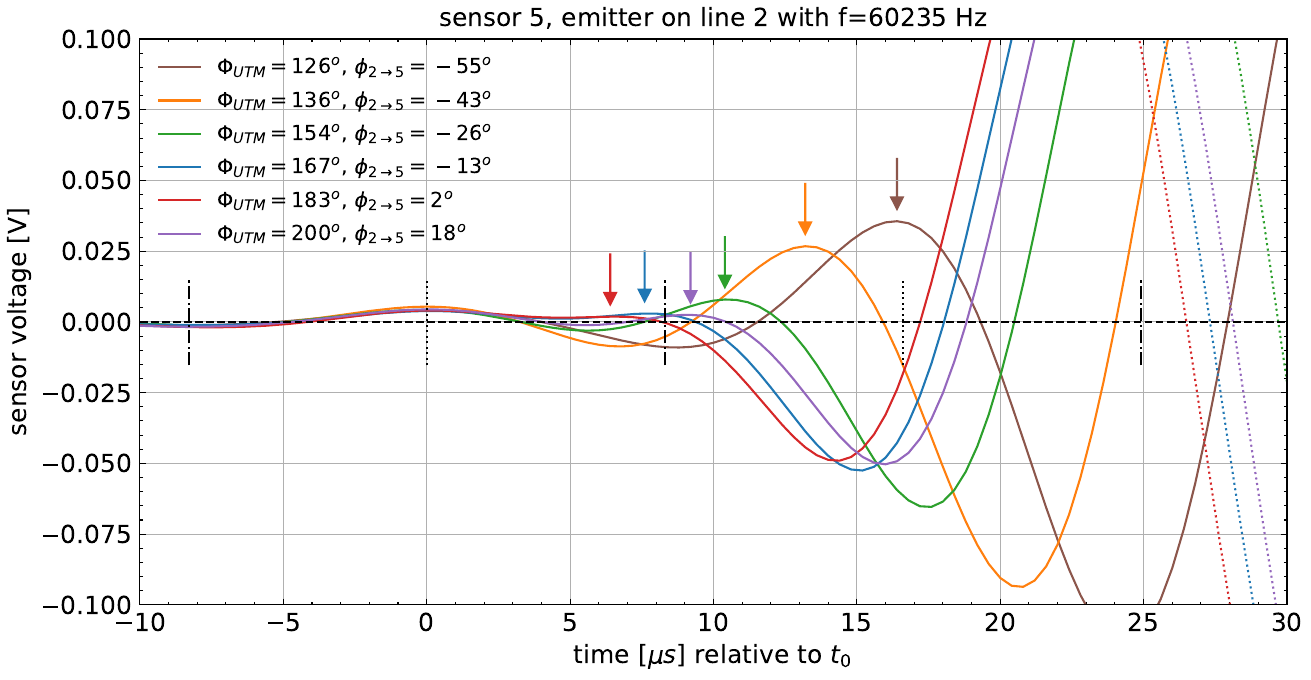}
\caption{
Signals from the emitter on line 2 with frequency $\unit[60235]{Hz}$ for a selection of runs with various orientations of the sensors with respect to the emitter for 
sensor 4 (top) and 
sensor 5 (bottom). 
Shown is a zoom to the region of the onset of the late wave.
The arrows indicate the first maximum of the late wave for the respective signals.
  }
\label{fig:signal_vs_time_sensor_4_5_cycle_2_zoom}
\end{figure}

The effects are demonstrated for the emitter on line 2 in Figure~\ref{fig:signal_vs_time_sensor_4_5_cycle_2_zoom}. This emitter has the same frequency as the emitter on line 10 (cf.\ Figure~\ref{fig:signal_vs_time_sens_4_5}) but is located at a greater distance. 
Comparing the positions of the first maximum of the early waves (at $t_0 =0$) with that of the late waves (indicated by the arrows)  in Figures~\ref{fig:signal_vs_time_sens_4_5} and \ref{fig:signal_vs_time_sensor_4_5_cycle_2_zoom}, the waves originating from the emitter on line 2 show interferences as the relative orientations $\phi_{2\rightarrow s}$ approach zero.
For the signals from the emitter on line 2 recorded with sensor 4, shown in the top plot of Figure~\ref{fig:signal_vs_time_sensor_4_5_cycle_2_zoom}, 
it can be observed that a clear separation between the first visible maximum of the early wave and maximum 1 of the late wave is not reached until the relative orientation of the sensor w.r.t.\ the emitter exceeds $|\phi_{2\rightarrow s}| \sim30^\circ$. 
%
Consequently, for the emitter on line 2, relative orientations $|\phi_{2\rightarrow s}| < 30^\circ$ (for sensors $s = 4, 5$) are excluded from the calculation of the speed of sound of the early wave.
With a separation of the two sensors of $\sim60^\circ$, a total range of $\sim 120^\circ$ has to be excluded. 
For a given distance of the emitter, the separation of waves gets worse for lower frequencies.

From the properties of the emitted signal, a time separation of the first visible maximum of the early wave and maximum 1 of the late wave of at least $0.6/f$ for signal frequency $f$ is expected to provide sufficient separation in order for the first visible maximum of the early wave -- which is used to calculate the speed of sound -- not to be affected by the late wave.
\red{The value of $0.6/f$ was deduced from the data, but from the model introduced in Section~\ref{sec:model-superimposing-waves}, it is easy to understand that a delay of less than $\tau = 0.5/f$, i.e.\ of half a period, will lead to interferences of the first maxima of early and late wave. 
Note that in  Figure~\ref{fig:signal_vs_time_sensor_4_5_cycle_2_zoom} another limitation of the model becomes visible: As the delay $\tau$ is decreasing for $|\phi_{2\rightarrow 4,5}|$ approaching zero, the first maximum of the late wave disappears. This cannot be explained by superposition of early and late wave alone. Presumably, this is an effect of the formation of the wave, cf.\ Section~\ref{sec:discussion-results}. 
} 

For sensors 4 and 5, the requirement on the temporal separation of the maxima is met for the emitters on lines 4, 6, 7, 8, 10, and 11. For lines 3 and 9, a cut on the compass reading as discussed above removes runs for which the requirement of the aforementioned separation of at least $0.6/f$ was not fulfilled, while leaving enough measurements of $\Delta t_e$  for a fit of the early wave. For lines 1, 2 and 5, such a cut removes too large a portion of the measured range of $\Delta t_e$, so that $v_e$ will not be derived for these lines.

\begin{figure}[tb]
\centering
\includegraphics[width=\columnwidth,angle=0]{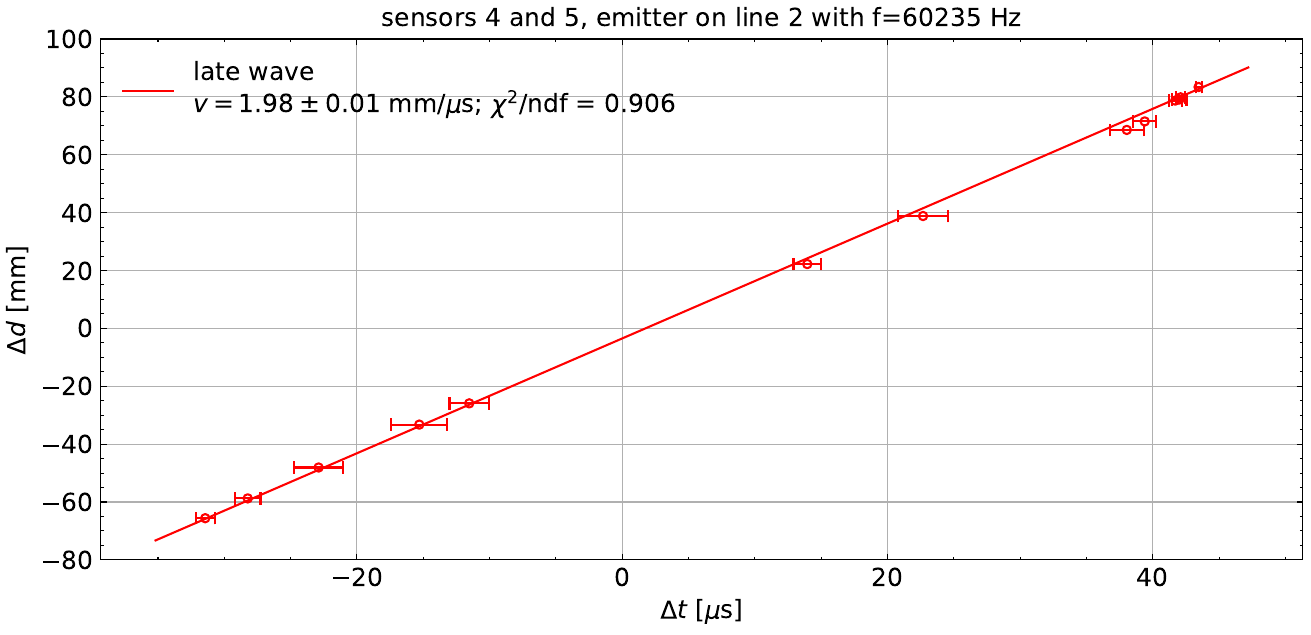}
\caption{
Speed of sound derived from a linear fit for the late wave for the emitter on line 2, as described in the text. For each run, the average orientation was determined, which corresponds to a particular value of $\Delta d$. 
For the emitter on line 2, only the speed of sound of the late wave was determined from the values of $\Delta t_{\ell}$.
}
\label{fig:dist_vs_time_cycle2}
\end{figure}

The restrictions discussed above apply to the early wave only; 
the fit for the speed of sound of the late wave from the emitter on line 2 is shown in Figure~\ref{fig:dist_vs_time_cycle2}. Compared to Figure~\ref{fig:dist_vs_time_cycle10}, the range of $\Delta t_\ell$ and consequently $\Delta d$ is about twice as large. This is expected, 
as for the POI closer to the lowest point of the AM in the case of the emitter on line 10, the difference  between the signal paths from the POI to the respective sensors is smaller. 

\subsection{Speed of sound for sensors 2 and 3}
\label{sec:v-for-sens-23}
The analysis described above was repeated for sensors 2 and 3, located in a different AM, in exactly the same manner as for sensors 4 and 5. A notable difference is that for sensors 2 and 3, the distances over which the signals propagate inside the glass sphere apparently are longer, resulting in a better separation of the early and late wave in the sensors. 
Presumably this is due to the sensors 2 and 3 being positioned slightly above the horizontal orthodrome of their AM, as will be discussed in Section~\ref{sec:discussion-results}.
The condition from Section~\ref{sec:effect-f-dist} that 
the first visible maximum of the early wave and maximum 1 of the late wave shall be separated by at least $0.6/f$ for signal frequency $f$, for sensor pair (2, 3) is fulfilled for all emitters. Hence the speed of sound can be calculated for the early wave for all emitters with the exception of 
the emitter on line 5: the first maximum of this signal is relatively low and seems to be superimposed by some intrinsic noise which distorts the first maximum of the early wave.
For this reason, for emitter 5, only the speed of sound of the late wave will be calculated for sensor pair (2, 3). As a reminder, for the sensor pair (4, 5), the signal of emitter 5 shows an insufficient separation of the early and late wave, so that the speed of sound of the early wave is not calculated. Hence in this case any distortion of the first maximum of the early wave is of no concern.

\section{The speed of sound as a function of frequency}
\label{sec:sound-speed-vs-freq}

\begin{figure}
\centering
\includegraphics[width=\columnwidth,angle=0]{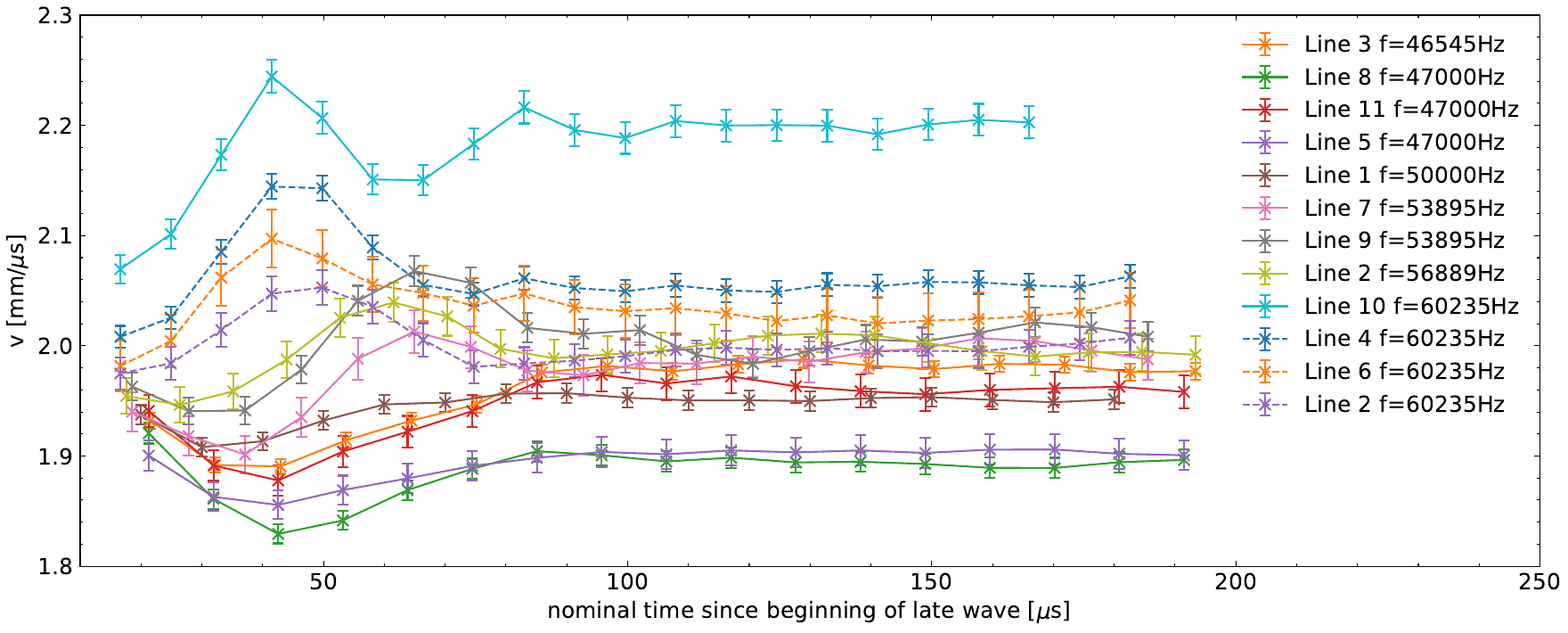}
\\
\includegraphics[width=\columnwidth,angle=0]{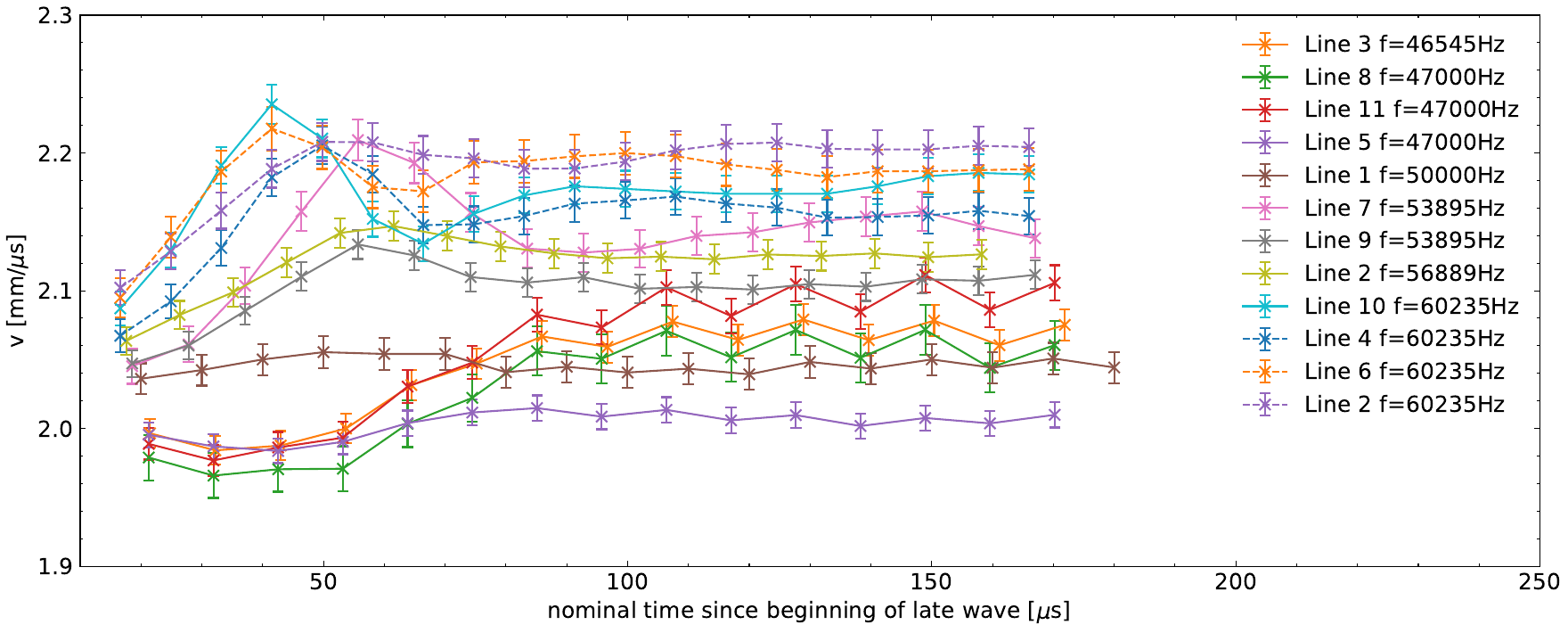}
\caption{Speed of sound of the late wave for sensors 4 and 5 (top), 2 and 3 (bottom) as derived for subsequent zero crossings. See text for explanations. 
}
\label{fig:speed_of_sound_vs_Zero_X_sens_4_5}
\label{fig:speed_of_sound_vs_Zero_X_sens_2_3}
\label{fig:speed_of_sound_vs_Zero_X}
\end{figure}

Figure~\ref{fig:speed_of_sound_vs_Zero_X_sens_4_5} shows the speed of sound of the late wave as calculated from the sensor pairs (4, 5) and (2, 3), respectively, for the subsequent zero crossings of the late wave.
Each entry was obtained
 from a speed of sound fit using the values of $\Delta t_\ell$ as resulting from corresponding zero crossings of the waveforms recorded by the two sensors of an AM, cf.~Figure~\ref{fig:Signal_timedelay}. 
Along the $x$-axis, 
subsequent entries are plotted at intervals of half a period for the respective nominal emitter frequency. 
The first entry for each emitter shows the speed of sound as derived for the zero crossing of the rising amplitude of maximum 2 of the late wave at a time corresponding to one period. 
The  beginning of the signal at $t=0$ is hence assigned to the nominal time where the amplitude starts to rise to maximum 1.
The speed of sound resulting from the fit for the late wave shown in Figure~\ref{fig:dist_vs_time_cycle10} for sensors 4 and 5 corresponds to the second entry of line 10 at $\unit[24.9]{\mu s}$ (1.5 periods for a frequency of $\unit[60235]{Hz}$).

The speed of sound of the late wave shows some variations with time before it stabilises after about $\unit[90]{\mu s}$ for each emitter.
For the temporal development of the speed of sound before that time, a frequency-dependent behaviour is observed: For the emitters with the lowest frequency of $\sim\unit[47]{kHz}$  the speed of sound vs.\ time first shows a small decrease; for $\sim\unit[50]{kHz}$ there is only a very small time-dependence of the speed of sound, and with increasing frequency towards $\sim\unit[60]{kHz}$, the speed of sound rises to a maximum before stabilizing. While differing in various details, this qualitative behaviour is visible for both AMs under investigation.

The most probable cause of this time dependence of the speed of sound is the transient response of the sensors at the beginning of a signal, which shows small differences between sensors. 
In Section~\ref{sec:appendix_sys-errors} of the appendix, the systematic errors will be discussed in detail.

Tables~\ref{tab:v-for-sensors45} and \ref{tab:v-for-sensors23} show the speed of sound as calculated in the manner described above for the two pairs of sensors. 
For the late wave, the speed obtained from the zero crossing of the falling amplitude of maximum 2 (second zero crossing for the respective emitter in Figure~\ref{fig:speed_of_sound_vs_Zero_X})
 is given, as motivated in Section~\ref{sec:propagation-time}. In addition, the difference between the speed of sound obtained from the first zero crossing and from the second zero crossing in Figure~\ref{fig:speed_of_sound_vs_Zero_X} is given as $\Delta_a$.
Furthermore, the speed of sound for the first six zero crossings with $t>\unit[95]{\mu s}$ was averaged for each emitter; the difference between this value and the speed of sound obtained from the second zero crossing is given in the tables as $\Delta_b$.
To account for the changes of the speed of sound observed in Figure~\ref{fig:speed_of_sound_vs_Zero_X}, a systematic error
with absolute value of $ |\Delta_b - \Delta_a|/2 $ will be assigned. This error is assumed to be due to small differences in the transient response of the individual sensors.

\begin{table}
\centering
\caption{Speeds of sound in the glass sphere by line of the emitter for the AM with sensors 4 and 5, ordered by increasing frequency $f$ and 
decreasing  angle $\vartheta_\text{POI}$ between $\vec{n}_\text{POI}$ and the horizontal orthodrome for the same frequency.
This corresponds to an increasing horizontal distance of the emitter to line 12.
Errors on the velocities $v_e$ and $v_\ell$ of the early and late wave, respectively, are statistical only.
The entries $\Delta_a$ and $\Delta_b$ (same units as $v_\ell$) are explained in the text.
}
\begin{tabular}{ r r r| c@{\;\;} r c| c@{\;\;} r c@{\;\;\:} r r}
\toprule
line & \multicolumn{1}{c}{$f\,$[Hz]} & \multicolumn{1}{c|}{\iftoggle{dist}{dist.}{$\vartheta_\text{POI}$} } & \multicolumn{3}{c|}{early wave} & 
 \multicolumn{5}{c}{late wave} \\
\cmidrule{1-11}
 &  & \iftoggle{dist}{[m]}{} & $v_e$\,[mm/$\!\mu$s] & \,ndf\! & $\chi^2\!$/ndf & $v_\ell$\,[mm/$\!\mu$s] & \,ndf\! & 
 $\chi^2\!$/ndf &$\Delta_a$ & $\Delta_b$
 \\ 
\midrule
 3 & 46545 &  \iftoggle{dist}{133}{$71.2^\circ$} & $5.37 \pm 0.08$ &  11 & 0.98 & $1.89 \pm 0.01$ & 23 & 1.68 & $ 0.04$ & $ 0.09$ \\
\cmidrule{1-11}
 8 & 47000 &  \iftoggle{dist}{63}{$80.9^\circ$} & $5.11 \pm 0.07$ &  30 & 0.49 & $1.86 \pm 0.01$ & 30 & 1.03 & $ 0.06$ & $ 0.03$ \\
11 & 47000 &  \iftoggle{dist}{72}{$79.5^\circ$} & $5.35 \pm 0.10$ &  23 & 0.47 & $1.89 \pm 0.01$ & 23 & 0.70 & $ 0.05$ & $ 0.07$ \\
 5 & 47000 & \iftoggle{dist}{180}{$65.3^\circ$} & $-$             & $-$ & $-$  & $1.86 \pm 0.01$ & 16 & 2.19 & $ 0.04$ & $ 0.04$ \\
\cmidrule{1-11}
 1 & 50000 &  \iftoggle{dist}{192}{$63.9^\circ$} & $-$             & $-$ & $-$  & $1.91 \pm 0.01$ & 15 & 1.98 & $ 0.03$ & $ 0.04$ \\
\cmidrule{1-11}
 7 & 53895 &  \iftoggle{dist}{82}{$78.2^\circ$} & $5.08 \pm 0.08$ &  11 & 0.28 & $1.92 \pm 0.02$ & 11 & 0.33 & $ 0.02$ & $ 0.07$ \\
 9 & 53895 &  \iftoggle{dist}{129}{$71.7^\circ$} & $4.99 \pm 0.09$ &   9 & 0.83 & $1.94 \pm 0.01$ & 11 & 0.87 & $ 0.02$ & $ 0.06$ \\
\cmidrule{1-11}
 2 & 56889 & \iftoggle{dist}{179}{$65.4^\circ$} & $-$             & $-$ & $-$  & $1.95 \pm 0.02$ & 10 & 1.08 & $ 0.01$ & $ 0.06$ \\
\cmidrule{1-11}
10 & 60235 &  \iftoggle{dist}{78}{$78.7^\circ$} & $5.17 \pm 0.07$ &  22 & 0.39 & $2.10 \pm 0.01$ & 22 & 2.60 & $-0.03$ & $ 0.10$ \\
 4 & 60235 & \iftoggle{dist}{121}{$72.7^\circ$} & $5.14 \pm 0.05$ &  15 & 0.32 & $2.03 \pm 0.01$ & 15 & 0.89 & $-0.02$ & $ 0.03$ \\
 6 & 60235 & \iftoggle{dist}{129}{$71.7^\circ$} & $5.04 \pm 0.11$ &   8 & 0.83 & $2.00 \pm 0.02$ &  8 & 0.42 & $-0.02$ & $ 0.02$ \\
 2 & 60235 & \iftoggle{dist}{179}{$65.4^\circ$} & $-$             & $-$ & $-$  & $1.98 \pm 0.01$ & 11 & 0.91 & $-0.01$ & $ 0.01$ \\
\botrule
\end{tabular}
\label{tab:v-for-sensors45}
\end{table}

\begin{table}
\centering
\caption{
Speeds of sound in the glass sphere by line of the emitter for the AM with sensors 2 and 3. 
Refer to caption of Table~\ref{tab:v-for-sensors45} for further explanations.
}
\begin{tabular}{ r r r| c@{\;\;} r c| c@{\;\;} r c@{\;\;\:} r r}
\toprule
line & \multicolumn{1}{c}{$f\,$[Hz]} & \multicolumn{1}{c|}{\iftoggle{dist}{dist.}{$\vartheta_\text{POI}$} } & \multicolumn{3}{c|}{early wave} & 
 \multicolumn{5}{c}{late wave} \\
\cmidrule{1-11}
 &  & \iftoggle{dist}{[m]}{} & $v_e$\,[mm/$\!\mu$s] & \,ndf\! & $\chi^2\!$/ndf & $v_\ell$\,[mm/$\!\mu$s] & \,ndf\! & 
 $\chi^2\!$/ndf &$\Delta_a$ & $\Delta_b$
 \\ 
\midrule
 3 & 46545 &  \iftoggle{dist}{133}{$71.2^\circ$} & $5.12 \pm 0.07$ &  18 & 0.79 & $1.98 \pm 0.01$ & 18 & 0.87 & $ 0.01$ & $ 0.09$ \\
\cmidrule{1-11}
 8 & 47000 & \iftoggle{dist}{63}{$80.9^\circ$} & $5.13 \pm 0.09$ &  19 & 0.39 & $1.97 \pm 0.02$ & 19 & 0.43 & $ 0.01$ & $ 0.10$ \\
11 & 47000 & \iftoggle{dist}{72}{$79.5^\circ$} & $5.24 \pm 0.08$ &  19 & 0.49 & $1.98 \pm 0.01$ & 19 & 1.52 & $ 0.01$ & $ 0.12$ \\
 5 & 47000 &\iftoggle{dist}{180}{$65.3^\circ$} & $-$             & $-$ & $-$  & $1.99 \pm 0.01$ & 14 & 2.84 & $ 0.01$ & $ 0.02$ \\
\cmidrule{1-11}
 1 & 50000 &\iftoggle{dist}{192}{$63.9^\circ$} & $5.16 \pm 0.09$ &  13 & 1.40 & $2.04 \pm 0.01$ & 13 & 1.53 & $-0.01$ & $ 0.00$ \\
\cmidrule{1-11}
 7 & 53895 & \iftoggle{dist}{82}{$78.2^\circ$} & $5.21 \pm 0.07$ &  18 & 0.43 & $2.06 \pm 0.01$ & 18 & 1.18 & $-0.02$ & $ 0.08$ \\
 9 & 53895 &\iftoggle{dist}{129}{$71.7^\circ$} & $5.11 \pm 0.07$ &  17 & 1.07 & $2.06 \pm 0.01$ & 17 & 3.46 & $-0.01$ & $ 0.04$ \\
\cmidrule{1-11}
 2 & 56889 &\iftoggle{dist}{179}{$65.4^\circ$} & $5.00 \pm 0.08$ &  16 & 2.10 & $2.08 \pm 0.01$ & 16 & 3.18 & $-0.02$ & $ 0.04$ \\
\cmidrule{1-11}
10 & 60235 & \iftoggle{dist}{78}{$78.7^\circ$} & $5.16 \pm 0.07$ &  18 & 0.47 & $2.13 \pm 0.01$ & 18 & 0.72 & $-0.04$ & $ 0.04$ \\
 4 & 60235 &\iftoggle{dist}{121}{$72.7^\circ$} & $4.97 \pm 0.09$ &  13 & 1.01 & $2.09 \pm 0.01$ & 13 & 0.45 & $-0.02$ & $ 0.07$ \\
 6 & 60235 & \iftoggle{dist}{129}{$71.7^\circ$} & $4.96 \pm 0.10$ &  10 & 1.29 & $2.14 \pm 0.01$ & 10 & 1.35 & $-0.04$ & $ 0.05$ \\
 2 & 60235 & \iftoggle{dist}{179}{$65.4^\circ$} & $5.08 \pm 0.06$ &  12 & 0.73 & $2.13 \pm 0.01$ & 12 & 3.85 & $-0.03$ & $ 0.07$ \\
\botrule
%
\end{tabular}
\label{tab:v-for-sensors23}
\end{table}

\section{Systematic errors}
\label{sec:sys-errors}

A detailed discussion of the systematic errors is given in Section~\ref{sec:appendix_sys-errors} of the appendix.
In Table~\ref{tab:sys-errors} the results are summarised.
Most of the errors associated with the speed of sound measurements within a given AM for the different emitters are correlated.
For the sake of a simplified calculation, a correlation coefficient of 1 is assumed.
Hence, correlated errors will be added linearly when combining errors of signals from several emitters, while uncorrelated errors will be added in quadrature.
\red{For the speed of sound measurements in the two AMs, most errors correlated between measurements for the different emitters within a given AM are also correlated between measurements in different AMs, as they are due to effects on the complete AMS or, as the errors  `thickness and radius of glass sphere' 
and `propagation path (of the wave in the glass sphere)',  
to the principal setup. The systematic error `transient response of sensors' is due to the dependence of the speed of sound on the zero crossing it is calculated from, see Figure~\ref{fig:speed_of_sound_vs_Zero_X}.
As the two sensor pairs 
show a very similar transient behaviour for the time before  $\unit[75]{\mu s}$, this error is assumed to be correlated between the two AMs. Hence, of the correlated errors in Table~\ref{tab:sys-errors}, only the errors `sensor positions relative to the centre of AM' and `sensor displacement' are assumed to be uncorrelated between speed of sound measurements in different AMs. These errors have been marked with an asterisk in the table.
}

\begin{table}
\centering
\caption{Systematic errors for speed of sound measurements of Tables~\ref{tab:v-for-sensors45} and \ref{tab:v-for-sensors23}. 
Errors correlated and uncorrelated between speed of sound measurements in a given AM are added separately.
\red{Errors correlated for measurements with a given AM that are considered uncorrelated between the speed of sound measurements in the two AMs are marked with an asterisk.}
\sout{See text for further explanations.}%
The error due to superposition of early and late wave is contained in the error on the propagation path. 
See Section~\ref{sec:appendix_sys-errors} of the appendix for details on the calculation.
}
\begin{tabular}{l|r@{\ \ \;} c c c@{\ \ \;} c c c c c c r|r}
\toprule
\multicolumn{1}{c|}{error}
& 
\rotatebox{90}{\parbox{\mytableboxsize}{effect of low pass\newline filter}}
& \rotatebox{90}{compass offset} 
& \rotatebox{90}{\parbox{\mytableboxsize}{static positions of \newline emitter and AMS}}
& \rotatebox{90}{\parbox{\mytableboxsize}{dynamic reconstr.\newline of sensor positions}}
& \rotatebox{90}{tilts of AMS}
& \rotatebox{90}{\parbox{\mytableboxsize}{sensor positions rel.\newline to centre of AM$^*$}}
& \rotatebox{90}{sensor displacement$^*$}
& \rotatebox{90}{\parbox{\mytableboxsize}{thickness, radius of\newline glass}}
& \rotatebox{90}{propagation path\ }
& \rotatebox{90}{\parbox{\mytableboxsize}{superpos.\ early,\newline late wave}}
& \rotatebox{90}{\parbox{\mytableboxsize}{transient response\newline of sensors}}
& \multicolumn{1}{c}{
\rotatebox{90}{quadratic sum}
}
\\ \midrule
$v_\text{e}$ unc.  & 3\% &     & 1.5\% & 1\% &     &     &     &       &      & \multicolumn{1}{c}{--} &       & 3.5\% \\  
$v_\ell$ unc.      & $0.5\%$ & & 1.5\% & 1\% &     &     &     &       &      & \multicolumn{1}{c}{--} &       & 2\% \\  
\cmidrule{1-13}
$v_\text{e}$ cor. &     & 2\% & 1.5\% &       & 0.5\% & 0.5\% & 4\% & 3.5\% &4\%   &     & 7.5\% & 10.5\% \\
$v_\ell$ cor.     &     & 2\% & 1.5\% &       & 0.5\% & 0.5\% & 4\% & 3.5\% &4\%   &     & 3\%   & 8\% \\
\botrule
\end{tabular}
\label{tab:sys-errors}
\end{table}

\section{Discussion of the results}
\label{sec:discussion-results}

\begin{figure}
\centering
\includegraphics[width=0.99\columnwidth,angle=0]{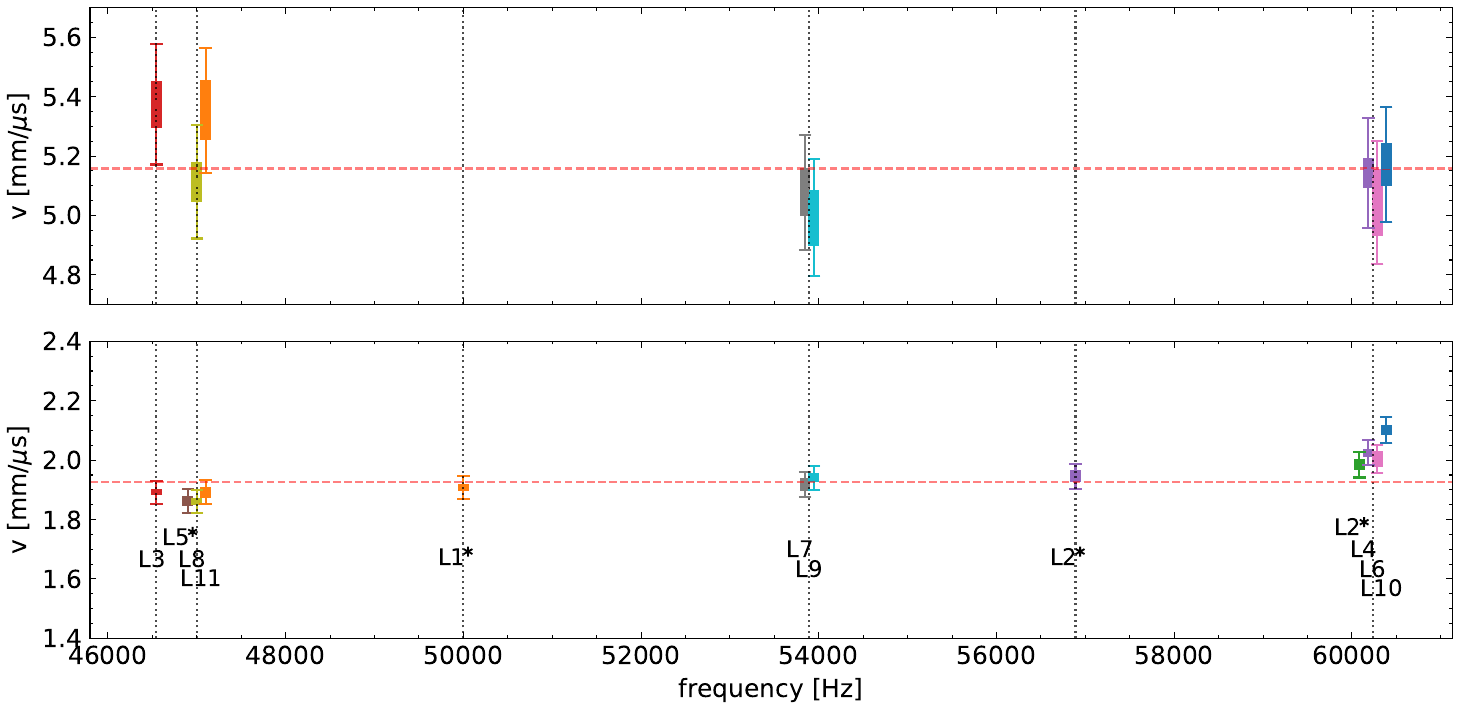}
\includegraphics[width=0.99\columnwidth,angle=0]{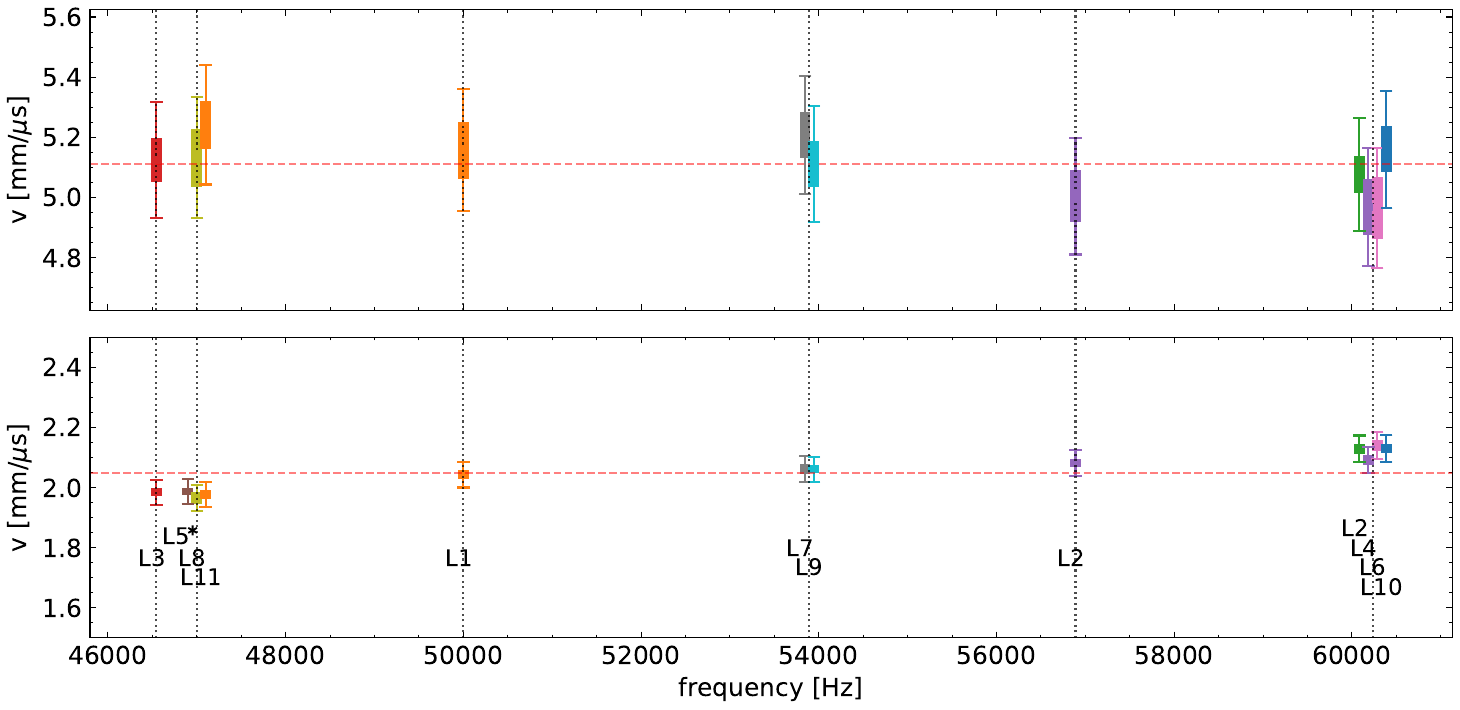}
\caption{Speed of sound vs.\ frequency for 
sensor pair (4, 5) (top) and 
sensor pair (2, 3) (bottom).
For each sub-figure the results for 
the early wave (top) and late wave (bottom) are shown.
Entries correspond to those from Tables~\ref{tab:v-for-sensors45} and \ref{tab:v-for-sensors23}, respectively.
Vertical dotted black lines indicate the nominal frequencies of the emitters, entries with the same frequency are slightly displaced w.r.t.\ each other for visualization.
The entries for the late wave have labels indicating the line on which the respective emitter is located; labels marked with an asterisk have no corresponding entry for the early wave, otherwise the labels also refer to the entry of the early wave directly above. 
For each entry, the solid box indicates the statistical errors from Tables~\ref{tab:v-for-sensors45} and \ref{tab:v-for-sensors23}, for the error bar, the uncorrelated systematic errors from Tab.~\ref{tab:sys-errors} were added in quadrature.
\sout{The dotted and dash-dotted 
lines for the speed of sound of the late wave indicate the values measured for the first zero crossing and the average of the stabilised  values (cf.~Figure~\ref{fig:speed_of_sound_vs_Zero_X_sens_4_5}), indicated by the values $\Delta_a$ and $\Delta_b$ in Tables~\ref{tab:v-for-sensors45} and \ref{tab:v-for-sensors23}, averaged for each frequency.}%
Red broken lines indicate the weighted means of the speed of sound values.
}
\label{fig:Speed_of_sound_vs_frequency}
\end{figure}

Figure~\ref{fig:Speed_of_sound_vs_frequency} shows the speeds of sound of the early and late wave
from Tables~\ref{tab:v-for-sensors45} and \ref{tab:v-for-sensors23} for the respective pairs of sensors.
The weighted means of the speeds of sound have been entered to guide the eye.
While the frequency dependence of the early wave is consistent with a constant value,
there is an indication of an increase of the speed of sound with frequency for the late wave. 
The average speeds of sound for the late (early) wave derived from sensor pairs (4, 5) and (2, 3) are 
$\unit[1.93]{\mmpermus}$ ($\unit[5.16]{\mmpermus}$) 
and 
$\unit[2.05]{\mmpermus}$ ($\unit[5.11]{\mmpermus}$),  
respectively.
An interpretation of the two waves will be given in Section~\ref{sec:interpret-2-waves}.

After propagating over a given distance, 
corresponding phases of the early and late wave show a certain separation in time $\tau$, 
\red{cf.\ Section~\ref{sec:model-superimposing-waves}}. If the maxima coincide at the time of their creation ($\tau=0$), then after propagating a distance $d_\text{prop}$ their time separation is 
\begin{equation*}
\tau = d_\text{prop} {v_\ell}^{-1} - d_\text{prop} {v_e}^{-1}\ .
\end{equation*}
A model in which the early wave is created earlier and closer to the POI will be discussed in Section~\ref{sec:interpret-2-waves}.
  
The time delay $\tau$ of the late wave depends on the heading of the AMS and will now be used to estimate the distance a signal propagates through the glass sphere to a given sensor position. 
The first maxima of the early and late wave in Figure~\ref{fig:signal_vs_time_sens_4_5} 
have a minimal temporal separation of $\tau \approx \unit[15]{\mu s}$.
As discussed above, for sensor pair (4, 5) the emitters on lines 1, 2, 3, 5, and 9 yield waveforms for which 
the first peaks of the early and late wave interfere with each other for small values of $\tau$ 
(cf.\ Figure~\ref{fig:signal_vs_time_sensor_4_5_cycle_2_zoom}). 
For these lines, the temporal separation will not be investigated.

\begin{figure}
\vspace{2ex}
\begin{center}
\includegraphics[width=\textwidth,angle=0]{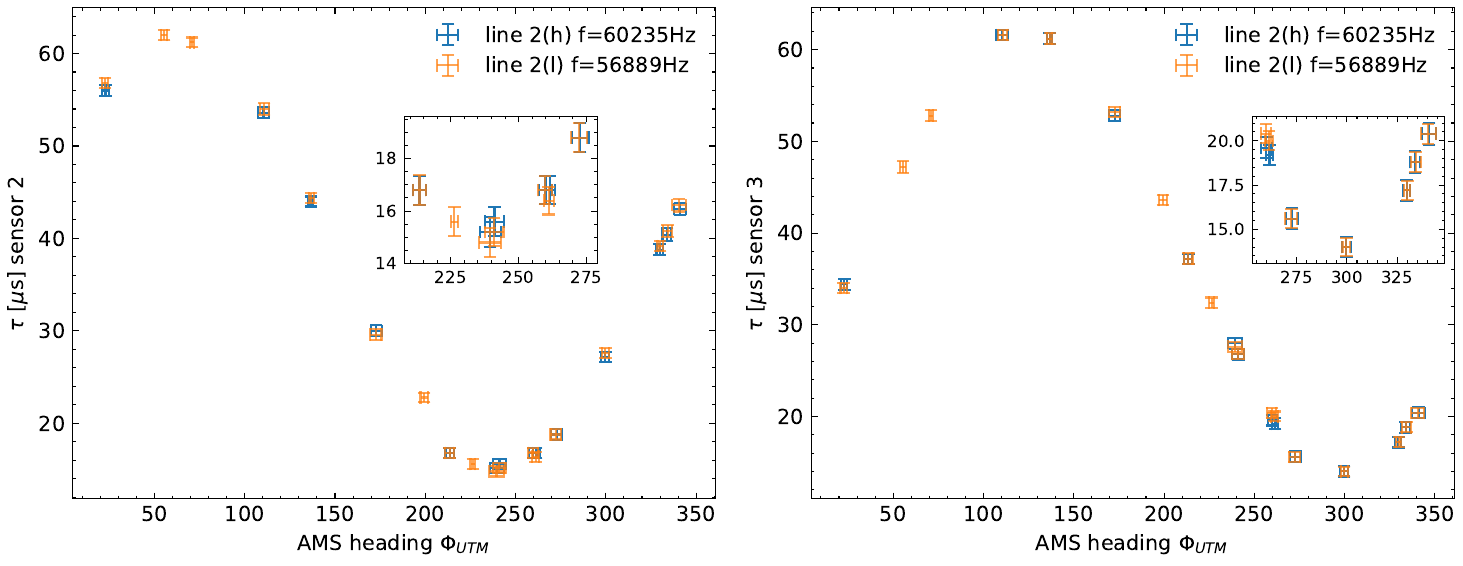}
\caption{
Time difference $\tau$ between first maxima of early and late wave for sensor 2 (left) and 3 (right) as a function of AMS heading $\Phi_\text{UTM}$.
The inserts show a zoom of the region around the minimum of the distribution.
}
\label{fig:late_minus_early_peak_sensors_12942_and_12943}
\end{center}
\end{figure}

Assuming the early and late wave propagate over the same distance $d_\text{prop}$ from a common origin, then this distance is $d_\text{prop} = \tau\,  v_\text{eff}$ where the effective speed\footnote{
The effective speed was introduced for convenience of the notation. For ${v_\ell} = {v_e}$ it is $\tau = 0$ for any $d_\text{prop}$ and hence $v_\text{eff} = \infty$, which has no physical interpretation.}
$ v_\text{eff} = ({v_\ell}^{-1} - {v_e}^{-1})^{-1}$ is 
$v_\text{eff} = \unit[3.42]{\mmpermus}$
for sensors 2 and 3, and
$v_\text{eff} = \unit[3.08]{\mmpermus}$
 for sensors 4 and 5, using the average values of the speeds of the early and late wave from above.
Comparing $\tau$ for the high and low frequency emissions from the emitter on line 2 will allow for insights on how this value -- and hence the point of origin of the early and late wave -- depends on the frequency. For the reasons discussed above, this investigation was done only for the sensor pair (2, 3).
The result is shown in Figure~\ref{fig:late_minus_early_peak_sensors_12942_and_12943}.
No significant difference between $\tau$ for the two different frequencies is observed in either of the sensors. 

\begin{figure}
\begin{center}
\includegraphics[width=\textwidth,angle=0]{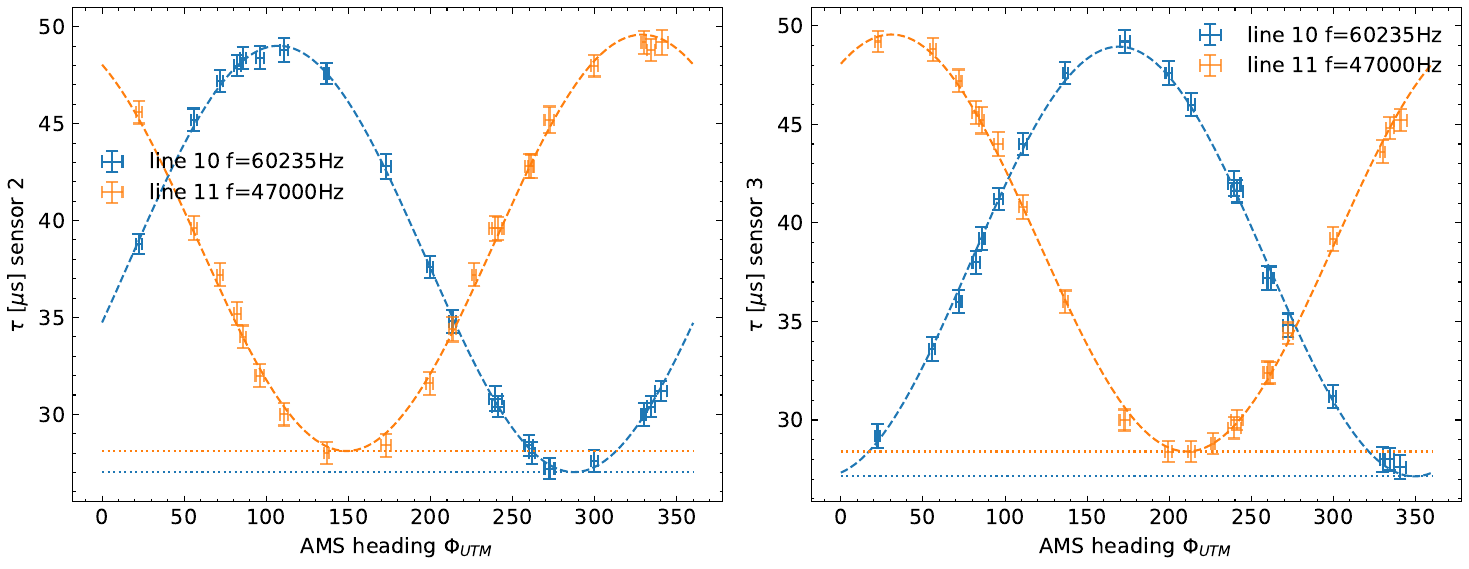}
\includegraphics[width=\textwidth,angle=0]{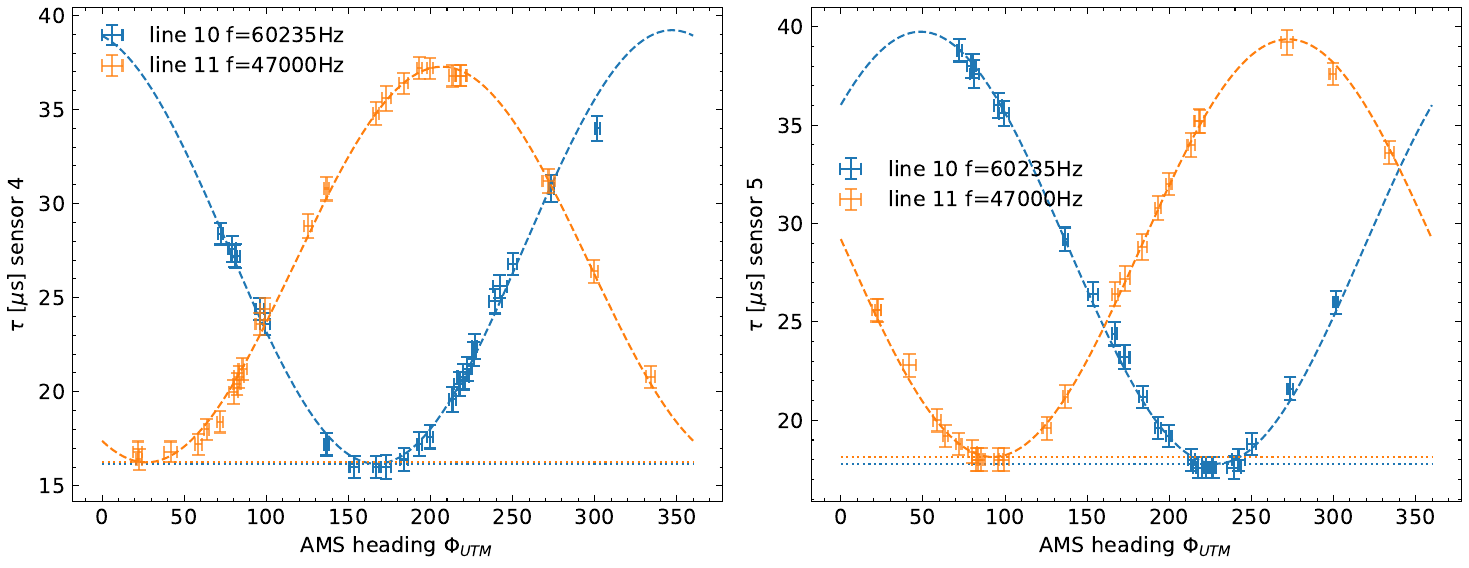}
\caption{
Time difference $\tau$ between first maxima of early and late wave by sensor as a function of AMS heading $\Phi_\text{UTM}$ for emitters on lines 10 and 11. The broken lines show the results of the fits described in the text.
The horizontal dotted lines indicate the minima $\tau_\text{min}$ of the fits with the colours corresponding to those used for the respective sensors.
}
\label{fig:late_minus_early_peak_sensors_l10_11}
\end{center}
\end{figure}

The minima $\tau_\text{min}$ of the $\tau$-values for all emitters were
obtained from a fit of a sine function of the form $\tau_0 \sin(2\pi\cdot\Phi_\text{UTM} /360^\circ + \phi_\text{off}) + \tau_\text{off}$ with the fit parameters amplitude $\tau_0$, offset for the sensor position $\phi_\text{off}$ and offset of the amplitude $\tau_\text{off}$. The minimum of $\tau$ is then given by $\tau_\text{min} =  \tau_\text{off} - \tau_0 $.
This is demonstrated in Figure~\ref{fig:late_minus_early_peak_sensors_l10_11} for the emitters on lines 10 and 11, which have very similar distances from line 12 but very different frequencies.
The values for $\tau_\text{min}$ differ by about $\unit[1]{\mu s}$ between the emitters on the two lines for sensors 2 and 3 and show no significant difference for sensors 4 and 5.

The nominal distance between the POI for a given emitter and the horizontal orthodrome of a given AM shall be denoted by $d_\text{nom}$, measured along the centre of the glass of the sphere and assuming emitters and AMs are at their nominal positions.
As the positions of the centres of the AMs in the local reference system -- as opposed to the positions of the sensors -- are not very well known, the assumption that the sensors are located at the horizontal orthodromes of the AMs is only precise up to a few centimetres. 
Keeping this caveat in mind, the radius of the area, over which the plane wave in the water stimulates the vibration of the AM around the nominal POI, can be estimated as $r_\text{form} \approx d_\text{nom}-d_\text{prop}$. 

The results for the emitters on all lines are summarised in Tab.~\ref{tab:diff-late-early}. Entries are ordered by decreasing angle $\vartheta_\text{POI}$ between $\vec{n}_\text{POI}$ and the horizontal orthodrome, 
which corresponds to an increasing horizontal distance of the emitter to line 12.
The relation to $d_\text{nom}$ is simply given by
$d_\text{nom} = (\vartheta_\text{POI}/180^\circ)\,\pi\, r_c$ with $r_c = \unit[208.5]{mm}$. 
There is a clear dependence of $\tau_\text{min}$ and $d_\text{prop}$ on distance from the emitter for all sensors, but no significant dependence on the frequency. 
The distance $d_\text{prop}$ for sensors 2 and 3 is roughly $\sim\unit[40]{mm}$ shorter than that for sensors 4 and 5.
Presumably this is due to offsets of the sensors from the horizontal orthodrome of their AM.

From the apparent frequency dependence of $v_e$ and $v_\ell$ in Figure~\ref{fig:Speed_of_sound_vs_frequency}, it can be roughly estimated that $v_\text{eff}$ is about $10\%$ lower at $\unit[47000]{Hz}$ than at $\unit[60235]{Hz}$.  For the same propagation distance it follows that $\tau_\text{min}$ is higher than calculated from the average  $v_\text{eff}$ for the lower frequency, and lower for the higher frequency. 
This effect however cannot be isolated in Table~\ref{tab:diff-late-early}.

\begin{table}
\centering
\caption{
For each sensor (s2, s3, s4, s5) and emitter (by line), the minimal temporal separation $\tau_\text{min}$ between early and late wave is shown. The resulting distance $d_\text{prop}$ propagated by the waves, assuming a common starting point, and  $r_\text{form} = d_\text{nom} -d_\text{prop}$ are given for the mean $\tau_\text{min}$ of the sensors in an AM. Here $d_\text{nom}$ is the nominal distance 
between the POI for a given emitter and the horizontal orthodrome of an AM and $\vartheta_\text{POI}$ the corresponding angle.
}
\begin{tabular}{r c r r||c c| c c|| c c| c c}
\toprule
 & & & & s2 & s3 & \multicolumn{2}{c||}{s2, s3 mean} 
       & s4 & s5 & \multicolumn{2}{c}{s4, s5 mean}   
  \\
\cmidrule{5-12}
line & $f$ & \iftoggle{dist}{dist.}{$\vartheta_\text{POI}$} & $d_\text{nom}$ 
& $\tau_\text{min}$ & $\tau_\text{min}$ & $ d_\text{prop} $ & $r_\text{form} $ 
& $\tau_\text{min}$ & $\tau_\text{min}$ & $ d_\text{prop} $ & $r_\text{form} $ 
\\ \cmidrule{1-12}
 & [Hz] & \iftoggle{dist}{[m]}{} & [mm] & [$\mu$s] & [$\mu$s] & [mm] & [mm] & [$\mu$s] & [$\mu$s] & [mm] & [mm] 
 \\ \midrule
 8 & 47000 & \iftoggle{dist}{63}{$80.9^\circ$} & 294.3
& 29.2 & 29.6 & 95.6 & 198.7  
& 18.1 & 19.9 & 61.7 & 232.6  
\\ 
 11 & 47000 & \iftoggle{dist}{72}{$79.5^\circ$} & 289.3
& 28.1 & 28.4 & 91.8 & 197.5  
& 16.2 & 18.1 & 55.9 & 233.5  
\\
 10 & 60235 & \iftoggle{dist}{78}{$78.7^\circ$} & 286.5
& 27.0 & 27.1 & 88.0 & 198.4  
& 16.2 & 17.8 & 55.2 & 231.3  
\\
 7 & 53895 &  \iftoggle{dist}{82}{$78.2^\circ$} & 284.6
& 26.8 & 26.9 & 87.3 & 197.2  
& 15.4 & 16.9 & 52.5 & 232.1  
\\
 4 & 60235 & \iftoggle{dist}{121}{$72.7^\circ$} & 264.7
& 22.2 & 21.6 & 71.1 & 193.6  
& 11.8 & 13.0 & 40.2 & 224.5  
\\
 6 & 60235 & \iftoggle{dist}{129}{$71.7^\circ$} & 260.9
& 20.8 & 20.3 & 66.9 & 194.0  
& 11.1 & 12.3 & 37.9 & 223.0  
\\
 9 & 53895 & \iftoggle{dist}{129}{$71.7^\circ$} & 260.9
& 20.8 & 20.4 & 67.0 & 193.9  
& 10.7 & 11.8 & 36.5 & 224.4  
\\
 3 & 46545 & \iftoggle{dist}{133}{$71.2^\circ$} & 259.0
& 20.2 & 19.9 & 65.2 & 193.8  
& -- & -- & -- & -- 
\\
 2 & 60235 & \iftoggle{dist}{179}{$65.4^\circ$} & 238.1
& 15.3 & 14.2 & 48.0 & 190.2  
& -- & -- & -- & -- 
\\
 2 & 56889 & \iftoggle{dist}{179}{$65.4^\circ$} & 238.1
& 15.3 & 14.3 & 48.1 & 190.0  
& -- & -- & -- & -- 
\\
  5 & 47000 & \iftoggle{dist}{180}{$65.3^\circ$} & 237.8 & -- & -- & -- & -- 
& -- & -- & -- & -- 
\\
 1 & 50000 & \iftoggle{dist}{192}{$63.9^\circ$} & 232.5
& 14.2 & 12.9 & 44.2 & 188.3  
& -- & -- & -- & -- 
\\
\botrule
\end{tabular}
\label{tab:diff-late-early}
\end{table}

The values for $r_\text{form}$ from Table~\ref{tab:diff-late-early} indicate 
that the signals form over a rather large region -- roughly in an area within a radius of $\sim\unit[200]{mm}$ around the POI for a given emitter, measured along the curvature of the sphere.
This corresponds to a cone with opening angle of $\sim 55^\circ$ around the line from the centre of the sphere to the POI.
No significant dependence on the frequency was observed.
The values for $r_\text{form}$ show a small increase with distance. Presumably, this is an artefact due to systematic errors on $v_\text{eff}$ and possibly systematic deviations of the truly propagated distance, as the wave cannot be strictly confined to the centre of the glass of the sphere (cf.\ discussion in Section~\ref{sec:appendix_sys-errors} of the appendix).

If the first discernible maximum of the early wave would indeed be its second maximum (cf.\ 
Section~\ref{sec:sig-shape-vs-orientation}), a frequency dependent term between $\unit[16.6]{\mu s} $ and $\unit[21.5]{\mu s}$ had to be added to $\tau$. This would imply that $\tau_\text{min}$ decreases and hence the formation region increases for larger frequencies.
\red{However, as some characteristic \emph{length} is needed as scale for an area, it seems more likely that the size of the formation region scales with  wavelength rather than frequency.}
\sout{If the formation region were to depend on frequency, 
it seems more likely though that its size scales with wavelength rather than frequency.}%
\red{Furthermore, the values of $\tau_\text{min}$ decrease monotonously with the angle $\vartheta_\text{POI}$ for all sensors in Table~\ref{tab:diff-late-early}. This is expected, and adding the respective frequency dependent terms from above would result in exceptions to that behaviour. This would imply that using the \emph{wrong} maximum of the early wave  results in the expected monotonous behaviour of $\tau_\text{min}$ with $\vartheta_\text{POI}$, whereas using the \emph{correct} one would lead to an unexpected behaviour. This seems like a rather unlikely scenario.} 
Recapitulating, there is no indication that the first discernible maximum of the early wave is not to be identified with the first maximum of that wave.

\section{Interpretation of the two waves}
\label{sec:interpret-2-waves}
In liquids, only longitudinal waves of density fluctuations are possible. The corresponding speed of sound in water at the depth of the ANTARES detector is   $v_\text{water} = \unit[1.54]{\mmpermus}$.
The speeds of sound in the VITROVEX glass sphere (see Section~\ref{sec:appendix_prop-VITROVEX} of the appendix) are 
$v_\text{long} = \unit[5.32]{\mmpermus} $  and
 $v_\text{trans} =   \unit[3.43]{\mmpermus} $  for longitudinal (density) and transverse (shear) waves, respectively.

A theoretical treatment of elastic properties of hollow spherical shells can be found in \cite{bib:Love-elasticity}. More recently, the acoustics of shells submerged in liquids has been investigated for specific experimental settings, see e.g.\ \cite{bib:Talmant-Uberall-Lamb-spheres-1989,bib:Uberall-acou-spheres-2001,bib:kargl-lamb-shere-backscat-JASA-1989,bib:marston-lamb-shere-phase-velo-JASA-1989}.
Hollow spheres have also been discussed as antennas for detection of gravitational waves, see e.g.\
\cite{bib:Coccia-GW-antennas,bib:Bassan-GW-antennas}. 
The treatment however typically concerns the eigenmodes of the vibrating shells or their radiation properties.
The problem at hands is somewhat different: A spherical shell immersed in water is excited by a plane wave,
and of interest is the TOA of the signal at the receiver.
In this case, resonances are not desired. Furthermore, the TOA is determined by the immediate response of the sensor to the first part of the wave reaching the sensor, before the sphere as a whole will start to resonate.

The experimental results indicate that both the early and late wave observed in the AMs are guided in the glass sphere. The conditions are similar to those to which the theory of Lamb waves applies:
Elastic waves which, according to the original theoretical treatment~\cite{bib:lamb-1917}, propagate in  $x$-direction in a plate with infinite extension in the $xy$-plane and some finite thickness $h$ in $z$-direction. 
Lamb waves arise from the superposition of shear and density waves successively reflected at the boundaries of the plate.
Assuming a combination of particle displacements in $z$-direction (shear wave) and $x$-direction (density wave) propagating in form of plane waves, combined with boundary conditions for a plate placed in vacuum, yields two transcendental characteristic equations.
For the product of a given frequency and the thickness of the plate, the two equations specify the allowed values of the phase velocity $v_\text{ph}$ (or of the wave number $k = 2\pi f/v_\text{ph}$) for the symmetric and asymmetric modes, respectively. The solutions to these equations correspond to the Lamb waves. The material properties are most conveniently specified by the  
speed of sound for the longitudinal and transverse wave in that material.

While exact solutions cannot be expressed analytically, numerical solutions yield continuous curves, each corresponding to a mode, in a dispersion diagram of the phase velocity or the group velocity ($2\pi\cdot df/dk$) 
vs.\ $f\times h$ for frequency $f$ and thickness $h$ of the plate. 
New modes set in for $f\times h = n v/2$ where $v$ is either the longitudinal velocity $v_\text{long}$ or the transverse velocity $v_\text{trans}$ and $n$ is a positive integer; even (odd) integers for $v_\text{long}$ correspond to asymmetric (symmetric) modes and even (odd) integers for $v_\text{trans}$ to symmetric (asymmetric) modes.
The \red{symmetric and asymmetric} zeroth order modes\red{, referred to as $S_0$ and $A_0$, respectively,} start at $f=\unit[0]{Hz}$.
All modes extend to infinite frequencies.

Compared to the situation encountered by the waves in the glass spheres of the AMs, there are some differences w.r.t.\ the idealised situation assumed for Lamb waves:
The glass is spherical as opposed to flat or cylindrical and the pressure at the inside (air) and outside (water) is highly asymmetrical with the outside pressure on the sphere exceeding the inside pressure about 200-fold.  
The waves propagate radially from the POI, or from some formation region, cf.\ Section~\ref{sec:discussion-results}, rather than one-dimensionally.
As a generalisation,
waves guided in a plate with a thickness that is small compared to the extend of the plate in the other two dimensions, with wavelengths that are small compared to the extend of the plate, will be referred to as Lamb-like waves -- allowing for curvatures of the plate and in particular the special case of spherical shells.

According to \cite{bib:kargl-lamb-shere-backscat-JASA-1989, bib:williams-lamb-shere-backscat-JASA-1985}, the requirement of phase matching between the wave in water 
and the Lamb-like wave in the sphere along the surface of the sphere leads to an excitation of the Lamb-like wave at an angle
of $\theta_\text{ex} = \arcsin(v_\text{water}/v)$\  w.r.t.\ $\vec{n}_\text{POI}$.
Here $v$ is either the speed of sound of the early ($v_e$) or late ($v_\ell$) wave, yielding excitation angles of 
$\theta_\text{ex}^{e} \approx 18^\circ$ and $\theta_\text{ex}^\ell \approx 50^\circ$, respectively. This would correspond to an additional distance propagated by the early wave along the centre of the glass of $\sim \unit[115]{mm}$, while the late wave would be excited $r_o (\cos\theta_\text{ex}^{e} - \cos\theta_\text{ex}^\ell)/c_\text{water} \approx \unit[43]{\mu s}$ later than the early wave.

With these numbers, the expected value of $\tau_\text{min}$ for each emitter can be calculated.
The results range from  $\sim 45$  to $\sim\unit[55]{\mu s}$ for increasing values of $\vartheta_\text{POI}$, assuming in the calculation that the sensor is located at the horizontal orthodrome.
This is $\unit[20\sim 25]{\mu s}$ larger than the corresponding values for sensors 2 and 3 in  Table~\ref{tab:diff-late-early}.
Note that in this model, a sensor located at $\theta_\text{ex}^\ell$ would measure a value of $\tau \approx \unit[20]{\mu s}$, constituting the smallest possible time separation between corresponding peaks of the early and late wave. 
As most measured values of $\tau_\text{min}$ are smaller than this value,
 it is likely that the formation of the early and late wave cannot be treated independently in the fashion described above.
Nonetheless, the opening angle of the formation region of $\sim 55^\circ$ from Section~\ref{sec:discussion-results} agrees well with 
$\theta_\text{ex}^\ell \approx 50^\circ$, and  the predicted independence of the size of the formation region from the wavelength of the signal is indeed observed.

No effort will be made for a detailed theoretical treatment of the formation or propagation of the Lamb-like waves. 
Instead, the expected modes for Lamb waves in an infinitely extended plate will be compared to the experimental results. 
\begin{figure}[htb]
\centering
\includegraphics[width=\columnwidth,angle=0]{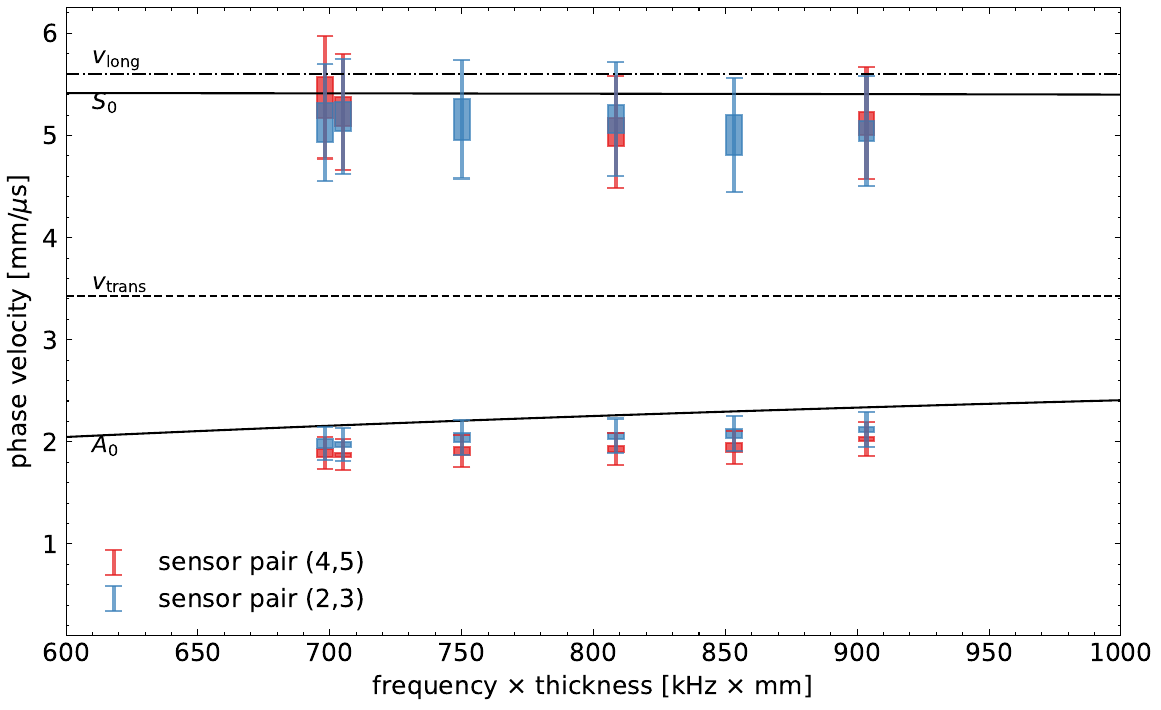}
\caption{
Speed of early and late wave vs.\ product of frequency and thickness of the glass sphere
for sensor pairs (2, 3) and (4, 5), compared to 
zeroth order symmetric ($S_0$) and asymmetric ($A_0$) Lamb wave modes. 
The velocities of the early and late waves have been averaged for frequencies with multiple measurements.
For each entry, the solid box indicates the combined statistical and systematic uncorrelated error from Figure~\ref{fig:Speed_of_sound_vs_frequency}, added in quadrature in case of multiple measurements. 
The error bars indicate the quadratic sum of uncorrelated and correlated (cf.~Tab.~\ref{tab:sys-errors}) errors.
The longitudinal and transverse phase velocities of sound $v_\text{long}$ and $v_\text{trans}$  in VITROVEX glass are shown for reference.
}
\label{fig:Velocity_lamb}
\end{figure}

As the acoustic emitters send narrowband signals and the speed of sound of the early and late wave were obtained by tracking a particular maximum/zero crossing, it appears reasonable to associate the measured speeds of sound with the phase velocity of the nominal emitter frequencies. 
 
The frequency thickness product for the AMs ranges from $\unit[700]{kHz\times mm}$ to $\unit[900]{kHz\times mm}$. The first higher order mode is excited at 
$v_\text{trans}/2 = \unit[1715]{kHz \times mm}$, i.e.\ only zeroth order modes are present for the investigated experimental setup. 

In Figure~\ref{fig:Velocity_lamb}, the phase velocities of the two waves observed in the AMs are compared to those expected for Lamb waves\footnote{from
\url{https://github.com/franciscorotea/Lamb-Wave-Dispersion}, installed 25 June 2023 }. 
\sout{The measured speeds of sound lie roughly 10\% below the phase velocities of the zeroth order symmetric and asymmetric Lamb wave modes.}%
The frequency dependence and size of the measured velocities 
are strong indications that these two waves can indeed be associated with zeroth order modes of Lamb-like waves in the glass sphere.
\red{A linear $\chi^2$-minimisation fit to the phase velocity vs.\ frequency thickness product of the form $v_\ell^\text{fit}(f\times h) = v_\ell^\text{ctr} + (f\times h - \{f\times h\}_\text{ctr} )\cdot b $ with fit parameters $v_\ell^\text{ctr}$ (phase velocity at central frequency thickness product value $\{f\times h\}_\text{ctr} = \unit[800]{kHz \times mm}$) and $b$ (unitless slope) was done for the late wave. As errors, the quadratic sum of the statistical uncertainties and the uncorrelated errors from Table~\ref{tab:sys-errors} were used.  
The results are 
$v^\text{ctr}_\ell = 1.942 \pm  0.012\, \unit{\mmpermus}$ and $b= 0.72 \pm  0.14$
for sensor pair (4, 5), and  
$v^\text{ctr}_\ell = 2.051 \pm  0.012\, \unit{\mmpermus}$ and $b= 0.70 \pm  0.15$
for sensor pair (2, 3). The corresponding reduced $\chi^2$-values for 4 degrees of freedom are 0.435 and 0.137, respectively, suggesting that the errors used in the fit have been overestimated.

As over the validity of the fit from $\unit[700\sim 900]{kHz\times mm}$ the change of $v_\ell^\text{fit}$ due to the slope $b$ does not exceed $10\%$ of $v_\ell^\text{ctr}$, 
the correlated errors do not significantly affect the error on the respective slopes but essentially move the offset $v_\ell^\text{ctr}$ up or down. The difference between the offsets $v^\text{ctr}_\ell$ for the two sensor pairs is $5\sim6\%$. The error \emph{uncorrelated} between the measurements for the sensor pairs in the two AMs is the quadratic sum of the two errors marked with an asterisk in 
Table~\ref{tab:sys-errors}, corresponding to $4\%$ for each sensor pair. Hence, the  measurements for  $v^\text{ctr}_\ell$ in the respective AMs are consistent within the systematic uncertainties.
The combined error of  $4\%/\sqrt{2}$ on the average $v^\text{ctr}_\ell$ is added in quadrature with the error correlated between measurements in the two AMs from Table~\ref{tab:sys-errors}, yielding $7\%$, which renders the error on $v^\text{ctr}_\ell$ from the fit above of less than $1\%$ negligible.
The combined result for the speed of sound of the late wave hence is
\begin{equation*}
v_\ell^\text{fit}(f\times h) = \unit[(2.00\pm0.14)]{\mmpermus} + (0.71 \pm 0.10) \times (f\times h - \unit[800]{kHz \times mm})  
\end{equation*}
A linear regression of the phase velocity of the $A_0$ mode in the range $\unit[700\sim 900]{kHz\times mm}$ agrees with the actual values to 0.2\%  or better for any value of the frequency thickness product and yields
\begin{equation*}
v_{A_0}(f\times h) = \unit[2.25]{\mmpermus} + 0.88 \times (f\times h - \unit[800]{kHz \times mm})  
\end{equation*}

The phase velocity of the early wave is consistent with a constant value over the frequency range of the measurement. Hence, rather than a linear fit, the weighted average of the measurements was calculated. The errors were treated in exactly the same fashion as those of the late wave.
The procedure yields 
$v_e^\text{ave} =  \unit[(5.155 \pm 0.069)]{\mmpermus}$
for sensor pair (4, 5) and
$v_e^\text{ave} =  \unit[(5.103 \pm 0.059)]{\mmpermus}$
for sensor pair (2, 3). The errors are the combined 
uncorrelated errors (statistical and systematic) in each case.
Note that basically all systematic errors should affect the phase velocity of the early and late wave in the same fashion. Hence, it would have been expected that 
the same  difference of $5\sim6\%$ is observed between the values $v_e^\text{ave}$ for the two sensor pairs as for the offsets $v^\text{ctr}_\ell$.
However, the uncorrelated error is larger for the phase velocity of the early wave and in particular for the systematic error from the `transient response of sensors', it is presumably a simplification to assume that the error is 100\% correlated, which cannot be experimentally tested for the early wave.

In analogy to the procedure for the late wave,  errors of $4\%/\sqrt{2}$ (uncorrelated between the measurements in the two AMs) and $9.5\%$ (correlated between the measurements in the two AMs) are added in quadrature, yielding 
a combined speed of sound of the early wave for the two AMs of
\begin{equation*}
v_e^\text{ave} = \unit[(5.12 \pm 0.51)]{\mmpermus}  
\end{equation*}

The average phase velocity of the $S_0$ mode in the range $\unit[700\sim 900]{kHz\times mm}$ is given by 
\begin{equation*}
v_{S_0}(f\times h) = \unit[5.41]{\mmpermus}
\end{equation*}
where the largest deviation of a phase velocity in the given range of the frequency thickness product from the average is $<0.1\%$.

} 
\sout{The measurements are not conclusive as to whether the deviations from the phase velocities expected for the Lamb wave modes are due to the systematic errors of the measurements or the approximation of the glass sphere as an infinitely extended flat plate in vacuum.}
\red{Even though due to the relatively large systematic errors (in particular for the early wave)  caution  must be exercised in the interpretation, the measurements indicate that the phase velocities in the AM spheres are lower than those calculated for Lamb waves in an infinitely extended flat plate in vacuum.}
According to \cite{bib:kargl-lamb-shere-backscat-JASA-1989}, the phase velocities of zeroth order Lamb-like waves in a hollow sphere in water at low pressure are about 10\% \emph{higher} than those of the corresponding Lamb waves in an infinitely extended flat plate.
The sphere described in that reference, however, is made of steel, with a radius of only $\unit[19.05]{mm}$ and a larger ratio of outer to inner radius.

\red{
\section{Aspects of Lamb wave propagation}
\label{sec:lamb-wave-prop}
Lamb waves are used for non destructive testing and health monitoring of materials or structures in a wide range of industrial applications. See e.g.\ \cite{bib:Alleyne-Cawley-lamb-waves-defects-1992, bib:su-ye-lu-guided-lamb-waves-review-2006, bib:lamb-waves-aircraft-appl-2022} for reviews. Lamb waves are scattered at geometrical discontinuities, or defects. Detection of the scattered waves allows for an assessment of the discontinuities.
As a rule of thumb, discontinuities should have dimensions of the order of half a wavelength of the Lamb wave or larger in order to
be detected through their scattering. Applied to the AMs, this implies that structures for a phase velocity of $\sim\unit[2]{\mmpermus}$ and frequency of $\sim\unit[60]{kHz}$ would need to have sizes of at least $\sim\unit[15]{mm}$, i.e.\ roughly the thickness of the glass sphere. 
However, as discussed in \cite{bib:lee-staszewski-lamb-wave-interactions-2003}, defects much smaller than that have been reported to scatter Lamb waves. 

In \cite{bib:lee-staszewski-lamb-wave-interactions-2003}, a slot of $\unit[10]{mm}$ length and varying width from $1$ to $\unit[5]{mm}$ in the direction of the propagation of the Lamb waves in an aluminium plate with $\unit[2]{mm}$ thickness has been used as defect. Lamb wave were excited with $\unit[325]{kHz}$, and the $S_0$ mode with a corresponding wavelength of about $\unit[15]{mm}$ was investigated.
Despite of the severe nature of the discontinuity, the amplitude of the transmitted Lamb wave is reduced by only about $10\sim15\%$.

In \cite{bib:maio-ren-memmolo-lamb-change-thickness-2023} the effect of a discontinuity in form of an abrupt thickness change from $\unit[3]{mm}$ to $\unit[6]{mm}$ in an aluminium plate on  $A_0$ mode waves is investigated. 
Again, the severe discontinuity results in amplitudes of the reflected waves of only about 10\%.

In the light of these findings, it is difficult to imagine inhomogeneities of the glass of the AMs that could lead to any noticeable scattering. The titanium structures could induce scattering and the reflected waves could overlay with the direct wave in the AM from the POI to a sensor. Given the small dimensions of the structures w.r.t.\ the surface area of the glass sphere, however, any effect of scattering is expected to be smaller than in the two cases referenced above. And the superposition of waves with a 1:10 amplitude ratio was discussed in Section~\ref{sec:model-superimposing-waves} and shown to have a very small effect. Any scattered wave would reach the sensor later than the direct wave, with a delay that differs for the two sensors, depending on the heading of the AMS.
A scattered wave with significant amplitude reaching the sensor with some delay would superimpose with the direct wave and shift the zero crossings. This in turn would result in an apparent abrupt change of the velocity of the waves in the glass, as the delays would be different for the two sensors in an AM. There is however no indication of any such abrupt change in Figure~\ref{fig:dist_vs_time_cycle10} or any of the related figures (e.g.\ Figure~\ref{fig:line8_dist_and_rot_ang} in the appendix), in Figure~\ref{fig:speed_of_sound_vs_Zero_X} or in the time separation between the first maxima of the early or late wave shown in Figures~\ref{fig:late_minus_early_peak_sensors_12942_and_12943} and \ref{fig:late_minus_early_peak_sensors_l10_11}. It is therefore concluded that the measurements of the phase velocity of the Lamb-like waves in the AMs are not  affected by reflections or other effects of geometric discontinuities beyond the systematic errors that result from the propagation path and the transient response of the sensors, cf.\ Section~\ref{sec:appendix_sys-errors} of the appendix.
 
Note that radiation of power of Lamb-like waves in a solid plate with contact to a fluid (also referred to as `leaky waves') into the fluid is a known effect, see e.g.~\cite{bib:watkins-etal-lamb-wave-attenuation-1982}. 
A reduction of the signal amplitude with increasing distance from the POI to a given sensor is also observed for the late waves in the AMs, cf.\ Section~\ref{sec:model-superimposing-waves}, which is the result of power radiated into the water and geometric effects.
While the phase velocity of the waves in the solid material is affected by the presence of the liquid, it is not affected by the attenuation itself. Note that according to \cite{bib:kiefer-etal-lamb-spectrum-fluid-interaction-2019}, attenuation for the $A_0$ and $S_0$ modes differs for a given frequency thickness product. 
Furthermore,
in \cite{bib:maio-ren-memmolo-lamb-change-thickness-2023} it is pointed out that the $S_0$
mode is characterized mostly by in-plane displacements, to which  piezo sensors are not sensitive. This is consistent with the observation that the signal from the early wave is much weaker than that of the late wave. 

In \cite{bib:maio-ren-memmolo-lamb-change-thickness-2023}, the effect of mode conversion is observed, where the incident $S_0$ wave is not observed directly, but converted into a $A_0$ mode wave at the thickness change of the aluminium plate, which consequently is detected. The amplitude of the converted $S_0$ mode is again about 10\% of the amplitude of the incident $A_0$ mode.
It is unlikely though that the observed early waves can be associated with the $S_0$ modes created in a process of mode conversion: this would imply that independent of the heading of the AMS, the defect leading to such a conversion would have to be reached by the late wave some time before the direct wave reaches the sensor. Furthermore, the linear behaviour observed for $\Delta d$ vs.\ $\Delta t_e$ (cf.\ e.g.\ Figure~\ref{fig:dist_vs_time_cycle10}) would then be purely coincidental, as $\Delta d$  would be determined by the points of the conversion, rather than the POI.
} 

\section{Position calibration revisited}
\label{sec:pos-reco-revisit}
\begin{figure}[htbp]
\centering
\includegraphics[width=\myscale\columnwidth]{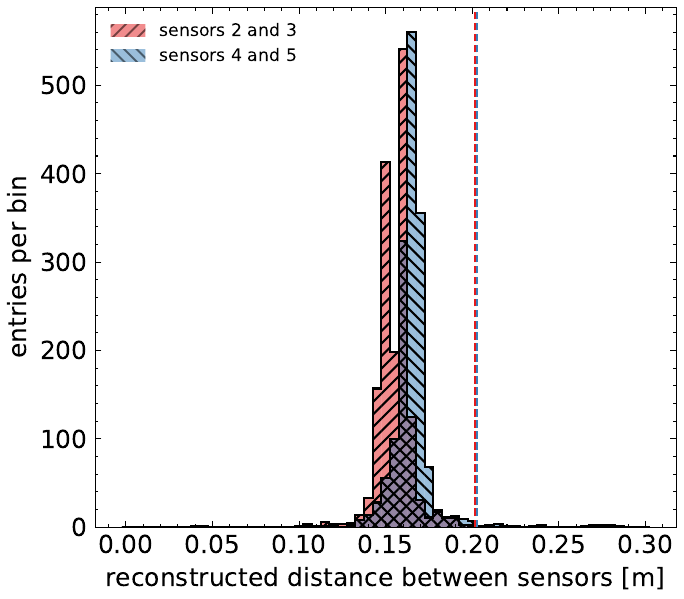}
\caption{
Reconstructed distances between the positions of sensors 2, 3 and 4, 5. The broken lines show the nominal values of the distances calculated from the sensor positions measured before deployment, scaled to the inside of the glass sphere. The colours correspond to those used for the respective sensor pairs. The cross hatched area indicates the overlap of the two distributions. The TOAs for the signals in each sensor were set to the time of the  zero crossing after the second maximum of the late wave, minus 1.5 times the period of the given frequency. 
}
\label{fig:Sensor_distance_TOA_latewave}
\end{figure}

With the findings about the presence of an early and late wave in the signal recorded by the sensors, the determination of the TOF can be refined.
As a first step, the fits of the sensor positions from Section~\ref{sec:pos-calibration} are repeated, but defining the TOA for a given emission cycle in a given sensor as the absolute time of the 
zero crossing after the second maximum of the late wave (i.e. the absolute times used to calculate $\Delta t_\ell$), 
minus 1.5 times the period of the signal frequency. The latter correction moves the TOA to the nominal start of the first maximum of the late wave.
In this fashion, the distances between the two sensors in the AMs under consideration that were shown in Figure~\ref{fig:Sensor_distance} were recalculated. The result is shown in Figure~\ref{fig:Sensor_distance_TOA_latewave}.
The reconstructed distances are 
$156 \pm \unit[10]{mm}$ 
for sensors 2 and 3,
and
$166 \pm  \unit[17]{mm}$
for sensors 4 and 5.
This is closer to the expected value w.r.t.\ the results from Section~\ref{sec:pos-calibration}, but still $\sim\unit[5]{cm}$ too short.

Given the notion that the signal from an emitter excites the glass sphere, in which it then propagates to the receiver, the calculation of the sensor positions can be further adjusted. 
\sout{Assuming an average speed of sound in the glass sphere of the late wave of $\unit[2.05]{\mmpermus}$ and $\unit[1.93]{\mmpermus}$}%
\red{Using the values for $v_\ell^\text{fit}(f\times h)$ from Sec.~\ref{sec:interpret-2-waves}} for the sensor pairs (2, 3) and (4, 5), respectively, the time $t_\text{g}$ that the late wave travels in the glass from the 
\sout{POI}%
\red{edge of the formation region} 
to a given sensor can be calculated for the distance given by Eq.~\ref{eq:dist_in_glass}, 
\red{minus $\unit[200]{mm}$. Here the subtrahend corresponds to the estimated edge of the formation region, see Section~\ref{sec:discussion-results}.
Modifying Equation~\ref{eq:dist-in-water} in a similar fashion, the time $t_\text{w}$ that a plane wave takes to propagate in water from the edge of the formation region to a given sensor can be calculated.}
\sout{In the same fashion, using Eq.~\ref{eq:dist-in-water}, and the speed of sound in water,  the time $t_\text{w}$ that a signal would have propagated in a straight line in water to the sensor can be calculated.}%
Then the TOF, used above to calculate the sensor positions for Figure~\ref{fig:Sensor_distance_TOA_latewave}, is adjusted by adding the term  $-t_\text{g}+t_\text{w}$. This corrected TOF corresponds to the time that the signal would have propagated in a straight line in water from the emitter to the sensor. This TOF multiplied with the speed of sound in water then corresponds to the true distance between emitter and receiver.

\begin{figure}[htbp]
\centering
\includegraphics[width=\myscale\columnwidth]{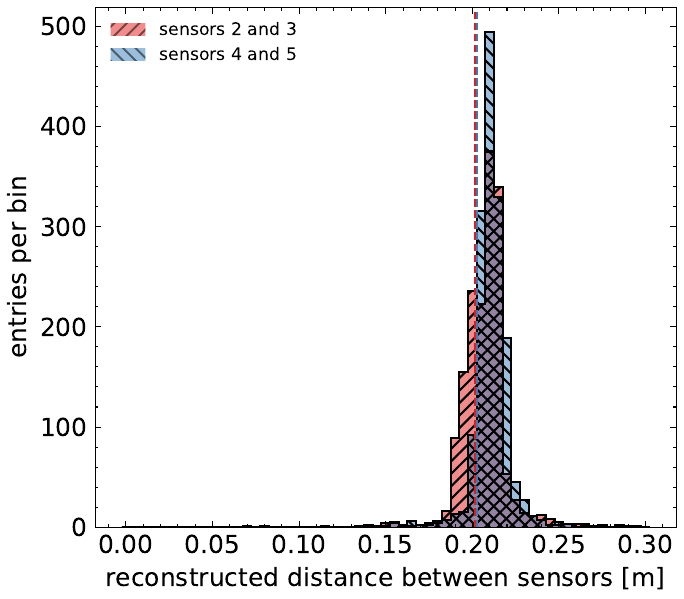}
\caption{
Reconstructed distances between the positions of sensors 2, 3 and 4, 5. The broken lines show the nominal values of the distances calculated from the sensor positions measured before deployment, scaled to the inside of the glass sphere. The respective colours correspond to those of the sensor pairs. The cross hatched area indicates the overlap of the two distributions. The TOAs for the signals in each sensor were set to the time of the  zero crossing after the second maximum of the late wave, minus 1.5 times the period of the given frequency, and corrected for the propagation through the AM as described in the text. 
}
\label{fig:Sensor_distance_TOA_latewave_withcorrection}
\end{figure}

The results for the reconstruction of the distances between the two sensors in the two AMs are shown in Figure~\ref{fig:Sensor_distance_TOA_latewave_withcorrection}.
The reconstructed distances are 
$207 \pm  \unit[12]{mm}$ 
for sensors 2 and 3, 
and
$213 \pm  \unit[19]{mm}$ 
for sensors 4 and 5.
These are rather close to the distances of $\unit[201]{mm}$ for sensors 2 and 3, and $\unit[203]{mm}$ for sensors 4 and 5 measured before the deployment.

Note that the results are not model-independent, as the nominal positions of the sensors enter the calculation of the speed of sound of the late wave. Nonetheless, the increase of the reconstructed distance between the two sensors when correcting for the effect of the signal propagating in the glass indicates that the model is a valid description of the signal propagation.

Furthermore, the distance $d$ is more strongly affected by systematic errors than the difference $\Delta d$ according to Eq.~\ref{eq:delta_d}. Reasons for this are among others offsets of the distance $d$, e.g.\ from a shift of the sensor positions in $z$-direction w.r.t.\ the centre of the glass sphere inside the AM, and the effect of the 
formation region, cf.\ Section~\ref{sec:discussion-results}. 
The reconstructed distance between the sensors however is again robust against changes of the POI. Offsets common to both sensors have mostly the effect of coherently moving the reconstructed positions up or down.

\section{Summary and conclusions}
\label{sec:conclusions}

Acoustic position calibration in the deep sea, using piezo sensors glued to the inside of submerged air filled glass spheres, was investigated with the AMADEUS system of the ANTARES deep sea neutrino telescope.
The ANTARES neutrino telescope was located in the Mediterranean Sea, 
roughly 40 km South of the town of Toulon at the French coast at the geographic position of 
$42^\circ 48'$ N, $6^\circ 10'$ E.
The acoustic neutrino detection test system AMADEUS 
 was equipped with three so-called acoustic modules, each comprising a glass sphere with outer diameter of 432 mm and thickness of at least 15 mm, with two acoustics sensors glued to the inside.
With all six sensors of the three acoustic modules, time series of the external pressure were recorded with $\unit[250]{ksps}$ digitisation rate and with precise synchronisation. 

For the analysis presented in this paper, 
the acoustic emitters of the ANTARES position calibration system were used, which were installed at fixed locations at the sea floor and emitted sinusoidal oscillations at six monochromatic frequencies between $\unit[46545]{Hz}$ and $\unit[60235]{Hz}$ of $\unit[5]{ms}$ duration at intervals of 2 minutes. 

Upon inspection of the signals from the acoustic emitters recorded by the sensors, two waves could be identified to reach each sensor,
a fast one \red{with a speed of sound of $v_e^\text{ave} = \unit[(5.12 \pm 0.51)]{\mmpermus}$} (early wave) and a slow one with \red{$v_\ell^\text{fit}(f\times h) = \unit[(2.00\pm0.14)]{\mmpermus} + (0.71 \pm 0.10) \times (f\times h - \unit[800]{kHz \times mm}) $} (late wave). 
The late wave is dominant with an amplitude $\gtrsim 10$ times higher than that of the early wave. 
\red{The speeds of sound of the early and late wave were compared to the phase velocities of the symmetric and asymmetric zeroth order Lamb wave modes, respectively, calculated for a flat plate with finite thickness and infinite extend in the other two dimensions. The agreement in the range of a frequency thickness product from $700$ to $\unit[900]{kHz\times mm}$ is $5\sim6\%$ between the early wave and the symmetric mode and $12\sim13\%$ between the late wave and the asymmetric mode.} 
\sout{The two waves are interpreted as symmetric and asymmetric modes of Lamb-like guided waves.
The measured phase velocities of the two waves agree with the phase velocities of Lamb waves -- calculated for a flat plate with finite thickness and infinite extend in the other two dimensions -- on a level of 10\% for the zeroth order symmetric and asymmetric mode in the range of a frequency thickness product between $700$ and $\unit[900]{kHz\times mm}$.}%
This indicates that
\red{the two waves observed in the AMs can be interpreted as `Lamb-like' guided waves, for which
deviations from the phase velocities of the corresponding Lamb wave modes due to the effects of the curvature, the different media inside and outside of the sphere, and the pressure exceeding $\unit[200]{bar}$ on the outside of the sphere, are relatively low.}
\sout{the effect of the curvature and of the different media inside and outside of the sphere, as well as of the depth that implies a pressure exceeding $\unit[200]{bar}$ on the outside of the sphere, is relatively low.}%
 Hence it seems viable to transfer the results \red{qualitatively} to different setups, such as the digital optical modules of the KM3NeT neutrino telescope~\cite{bib:KM3NeT-DOMs} with its integrated acoustic sensors. 
\red{For a quantitative understanding of the wave propagation and the resulting phase velocities, applying finite element methods for a given setup would be an appropriate tool.}

The notion of the faster wave propagating through the glass and the slower one through the water, exciting the glass sphere ``along the way'' is not supported by the observations. Reconstructing the positions of the individual sensors by means of multilateration using the time of flight of signals from at least three emitters, multiplied by the average speed of sound in water, yields systematic offsets. Taking into account that the signals propagate the very last part of their path in glass with a speed of sound different from that in water reduces the offset significantly. 
Using piezo sensors glued into glass spheres or other containers existing in a given deep sea experimental setup constitute a viable method for position calibration when taking the propagation of the sound wave in the container into account. It appears recommendable to use containers in which higher order modes of Lamb waves cannot form,
so to avoid additional interferences that may result in additional complexities in the calculation of the time of arrival.

In the setup of  AMADEUS, the shortest distance between the point of impact, i.e.\ the geometrically calculated point where the plane wave of an emitter signal hits an acoustic module, to a sensor near the horizontal orthodrome of the glass sphere is about $\unit[230]{mm}$. 
From propagation time differences between the early and late wave it was derived that the signals propagating through the glass sphere form over a region with radius $\sim\unit[200]{mm}$, measured along the curvature of the glass sphere, or $\sim\unit[170]{mm}$ radius for the projected area. Hence this formation region does not extend to the positions of the piezo sensors.
For a sensor placed in this region, the recorded time traces would require further scrutiny before applying the results of this paper, in particular when relating propagation times of signals to the distances propagated in the glass sphere.

\backmatter

\bmhead{Acknowledgments and funding}

The authors acknowledge the financial support of the funding agencies:
Centre National de la Recherche Scientifique (CNRS), Commissariat \`a
l'\'ener\-gie atomique et aux \'energies alternatives (CEA),
Commission Europ\'eenne (FEDER fund and Marie Curie Program),
LabEx UnivEarthS (ANR-10-LABX-0023 and ANR-18-IDEX-0001),
R\'egion Alsace (contrat CPER), R\'egion Provence-Alpes-C\^ote d'Azur,
D\'e\-par\-tement du Var and Ville de La
Seyne-sur-Mer, France;
Bundesministerium f\"ur Bildung und Forschung
(BMBF), Germany; 
Istituto Nazionale di Fisica Nucleare (INFN), Italy;
Nederlandse organisatie voor Wetenschappelijk Onderzoek (NWO), the Netherlands;
Ministry of Research, Innovation and Digitalisation (MCID), Romania;
MCIN for PID2021-124591NB-C41, -C42, -C43, funded by MCIN/AEI/10.13039/501100011033 and by ``ERDF A way of making Europe'', for ASFAE/2022/014, ASFAE/2022 /023, with funding from the EU NextGenerationEU (PRTR-C17.I01), Generalitat Valenciana,
for Grant AST22\_6.2 with funding from Consejer\'{\i}a de Universidad, Investigaci\'on e Innovaci\'on and Gobierno de Espa\~na and European Union - NextGenerationEU,
and for CSIC-INFRA23013, Generalitat Valenciana for PROMETEO/2020/019, for CIDEGENT/2018/034, /2019/043, /2020/049, /2021/23 and for GRISOLIAP/2021/192 and EU for MSC/101025085, Spain;
Ministry of Higher Education, Scientific Research and Innovation, Morocco, and the Arab Fund for Economic and Social Development, Kuwait.
We also acknowledge the technical support of Ifremer, AIM and Foselev Marine
for the sea operation and the CC-IN2P3 for the computing facilities.

\appendix

\section{Speed of sound for assumption of late wave propagating in water}
\label{sec:appendix-late-wave-in-water}

Tables~\ref{tab:v-in-water-45} and \ref{tab:v-in-water-23} show the speed of sound of the late wave as calculated for the AM with sensor pair (4, 5) and (2, 3), respectively,  obtained for the assumption of the wave propagating through the water, using Eq.~\ref{eq:dist-in-water} to calculate the distance. 
This assumption does not yield a speed of sound consistent with the known speed of sound in water at the depth of the AMS, $v_\text{water} = \unit[1.54]{\mmpermus}$ and is not accounted for by the systematic error from Table~\ref{tab:sys-errors} of less then 10\% for the late wave. 
The results are compared to those for the assumption of the late wave propagating through the glass of the AMs from Tables~\ref{tab:v-for-sensors45} and \ref{tab:v-for-sensors23}. The speed of sound for the latter assumption is lower because it is assumed that the waves travel in the centre of the glass.
The quality of the fit as indicated by the reduced $\chi^2$ is worse for the assumption of the wave propagating through the water for all emitters in the case of sensor pair (4, 5). 
For sensor pair (2, 3), for some emitters the reduced $\chi^2$ improves for the assumption of the wave propagating through the water.
The fit for emitter 1, which shows the largest decrease of the reduced $\chi^2$, was investigated in detail. 
The reduced $\chi^2$ fit is strongly affected by one run in which the sensors have a very high value of the relative orientation, i.e.\ are rotated such that they are pointing away from the emitter.
Removing this one run yields a reduced $\chi^2$ of 0.825 (1.138) for the assumption of the wave propagating in the glass (water).

\begin{table}[bthp]
\centering
\caption{Speed of sound of the late wave for the assumption of propagation in water for sensor pair (4, 5), compared to the result of the fit for the assumption of propagation in the glass sphere from Table~\ref{tab:v-for-sensors45}.
}
\begin{tabular}{rrr|r|rc|rc}
\toprule
line & \multicolumn{1}{c}{$f\,$[Hz]} & \multicolumn{1}{c|}{\iftoggle{dist}{dist.}{$\vartheta_\text{POI}$} }
 & ndf & \multicolumn{2}{c|}{late wave in glass} & \multicolumn{2}{c}{late wave in water} \\ 
 \cmidrule{1-8}
 &  & \multicolumn{1}{c|}{\iftoggle{dist}{[m]}{}} &  &  $v_\ell$ [mm/$\mu$s]  & $\chi^2$/ndf & $v_{\ell w}$ [mm/$\mu$s] & $\chi^2$/ndf
  \\ \midrule
 3 & 46545 & \iftoggle{dist}{133}{$71.2^\circ$} &  23 & $1.89 \pm 0.01$ & 1.68 &
 $1.94 \pm 0.01$ & 2.36 \\ 
\cmidrule{1-8}
 8 & 47000 & \iftoggle{dist}{63}{$80.9^\circ$} &  30 & $1.86 \pm 0.01$ & 1.03 & 
 $1.92 \pm 0.01$ & 1.04 \\ 
11 & 47000 &  \iftoggle{dist}{72}{$79.5^\circ$} &  23 & $1.89 \pm 0.01$ & 0.70 &
$1.95 \pm 0.01$  & 0.71 \\ 
 5 & 47000 & \iftoggle{dist}{180}{$65.3^\circ$} &  16 & $1.86 \pm 0.01$ & 2.19 & 
$1.89 \pm 0.01$  & 2.67 \\ 
\cmidrule{1-8}
 1 & 50000 & \iftoggle{dist}{192}{$63.9^\circ$} &  15 & $1.91 \pm 0.01$ & 1.98 & 
 $1.94 \pm 0.01$  & 3.83 \\ 
\cmidrule{1-8}
 7 & 53895 & \iftoggle{dist}{82}{$78.2^\circ$} &  11 & $1.92 \pm 0.02$ & 0.33 & 
$1.97 \pm 0.02$  & 0.38 \\ 
 9 & 53895 & \iftoggle{dist}{129}{$71.7^\circ$} &  11 & $1.94 \pm 0.01$ & 0.87 & 
 $1.99 \pm 0.01$ & 1.28 \\ 
\cmidrule{1-8}
2 & 56889 & \iftoggle{dist}{179}{$65.4^\circ$} &  10 & $1.95 \pm 0.02$ & 1.08 &
$1.96 \pm 0.02$ & 2.07 \\ 
\cmidrule{1-8}
10 & 60235 & \iftoggle{dist}{78}{$78.7^\circ$} &  22 & $2.10 \pm 0.01$ & 2.60 & 
$2.17 \pm 0.01$ &  2.84 \\ 
 4 & 60235 & \iftoggle{dist}{121}{$72.7^\circ$} &  15 & $2.03 \pm 0.01$ & 0.89 & 
$2.08 \pm 0.01$  & 1.43 \\ 
 6 & 60235 & \iftoggle{dist}{129}{$71.7^\circ$} &   8 & $2.00 \pm 0.02$ & 0.42 & 
 $2.05 \pm 0.02$ & 0.97 \\ 
 2 & 60235 & \iftoggle{dist}{179}{$65.4^\circ$} &  11 & $1.98 \pm 0.01$ & 0.91 & 
 $2.01 \pm 0.02$ & 2.15 \\ 
\botrule
\end{tabular}
\label{tab:v-in-water-45}
\end{table}

\begin{table}
\centering
\caption{Speed of sound of the late wave for the assumption of propagation in water for sensor pair (2, 3), compared to the result of the fit for the assumption of propagation in the glass sphere from Table~\ref{tab:v-for-sensors23}.
}
\begin{tabular}{rrr|r|rc|rc}
\toprule
line & \multicolumn{1}{c}{$f\,$[Hz]} & \multicolumn{1}{c|}{\iftoggle{dist}{dist.}{$\vartheta_\text{POI}$} }
& ndf & \multicolumn{2}{c|}{late wave in glass} & \multicolumn{2}{c}{late wave in water} \\ 
 \cmidrule{1-8}
 & & \multicolumn{1}{c|}{\iftoggle{dist}{[m]}{}} &  &  $v_\ell$ [mm/$\mu$s]  & $\chi^2$/ndf & $v_{\ell w}$ [mm/$\mu$s] & $\chi^2$/ndf
  \\ \midrule
 3 & 46545 & \iftoggle{dist}{133}{$71.2^\circ$} &  18 & $1.98 \pm 0.01$ & 0.87 &
 $2.04 \pm 0.01$ & 1.05 \\ 
 \cmidrule{1-8}
 8 & 47000 & \iftoggle{dist}{63}{$80.9^\circ$} &  19 & $1.97 \pm 0.02$ & 0.43 & 
$2.03 \pm 0.02$  & 0.43 \\ 
11 & 47000 & \iftoggle{dist}{72}{$79.5^\circ$} &  19 & $1.98 \pm 0.01$ & 1.52 & 
 $2.04 \pm 0.01$ & 1.45 \\ 
 5 & 47000 & \iftoggle{dist}{180}{$65.3^\circ$} &  14 & $1.99 \pm 0.01$ & 2.84 &
 $2.03 \pm 0.01$ & 2.80 \\ 
 \cmidrule{1-8}
 1 & 50000 & \iftoggle{dist}{192}{$63.9^\circ$} &  13 & $2.04 \pm 0.01$ & 1.53 & 
 $2.08 \pm 0.01$ & 1.20 \\ 
 \cmidrule{1-8}
 7 & 53895 & \iftoggle{dist}{82}{$78.2^\circ$} &  18 & $2.06 \pm 0.01$ & 1.18 &
 $2.13 \pm 0.01$ & 1.16 \\ 
 9 & 53895 & \iftoggle{dist}{129}{$71.7^\circ$} &  17 & $2.06 \pm 0.01$ & 3.46 &
 $2.12 \pm 0.01$ & 3.50 \\ 
 \cmidrule{1-8}
 2 & 56889 & \iftoggle{dist}{179}{$65.4^\circ$} &  16 & $2.08 \pm 0.01$ & 3.18 & 
$2.12 \pm 0.01$  & 3.56 \\ 
 \cmidrule{1-8}
10 & 60235 & \iftoggle{dist}{78}{$78.7^\circ$} &  18 & $2.13 \pm 0.01$ & 0.72 & 
 $2.20 \pm 0.01$ & 0.68 \\ 
 4 & 60235 & \iftoggle{dist}{121}{$72.7^\circ$} &  13 & $2.09 \pm 0.01$ & 0.45 & 
 $2.15 \pm 0.01$ & 0.54 \\ 
 6 & 60235 & \iftoggle{dist}{129}{$71.7^\circ$} &  10 & $2.14 \pm 0.01$ & 1.35 & 
  $2.20 \pm 0.02$ & 1.59 \\ 
 2 & 60235 & \iftoggle{dist}{179}{$65.4^\circ$} &  12 & $2.13 \pm 0.01$ & 3.85 & 
 $2.17 \pm 0.01$ & 4.78 \\ 
\botrule
\end{tabular}
\label{tab:v-in-water-23}
\end{table}

\section{Discussion of systematic errors}
\label{sec:appendix_sys-errors}
The systematic errors discussed below will be assumed to be correlated between the speed of sound measurements in a given AM for the different emitters, unless noted otherwise.

\paragraph{Effect of low pass filter on early wave:}
To mitigate effects from artificial high frequency components introduced by the upsampling of the signal, a low pass filter (128th order FIR filter with Hamming window and $\unit[100]{kHz}$ corner frequency) was applied throughout the analysis. The linear phase shift was compensated for. The filter has no significant effect on the late wave, but given the small amplitude of the early wave, it affects the shape of the first maximum and consequently the corresponding time of flight. 
To estimate the effect of the low pass filter, it was removed and the fit of the speed of sound of the early wave repeated.
The 19 measurements for the speed of sound of the late wave from the two sensor pairs obtained in this way were divided by the speed of sound measurements with low pass filter applied from Tables~\ref{tab:v-for-sensors45} and \ref{tab:v-for-sensors23}. This procedure yielded mean and standard deviation of $1.00 \pm 0.03$ with minimum value of $0.94\ (-0.06)$ and maximum value of $1.05\ (+0.05)$. Seven values are outside the one standard deviation interval, in good agreement with the expectation from a normal distribution. 
Correspondingly, an error of $\pm 3\%$ was assumed to result from the upsampling for the early wave.
For the late wave, an upper bound of the error of $\pm 0.5\%$ is assumed.
The error is assumed to be uncorrelated between the speed of sound measurements in a given AM for the different emitters.

\paragraph{Compass offset:}
Compass offsets of  $+3^\circ$ and $-3^\circ$ were subsequently applied and the rotated positions of the sensors calculated based on a rotation of the AMS according to the modified compass readings. 
The size of the variations correspond to the statistical error on the compass offset $\delta_\text{off}$ discussed in Section~\ref{sec:orientation-AM}.
The speed of sound fit was repeated with the resulting altered values of $\Delta d$. 
The resulting deviations did not exceed $\pm 1.5\%$ for sensor pair (2, 3) for the signal from any emitter. For sensor pair (4, 5), six out of the 20 measurements of the speed of sound (early and late waves) showed deviations between $\pm 2\%$ and $\pm 3\%$; in addition one outlier (late wave recorded for the signal from the emitter on line 6) had a deviation of $\pm 3.7\%$ for a compass offset of $\pm 3^\circ$. The overall error was estimated as $\pm 2\%$.

\paragraph{Static positions of emitter and AMS:}
In what follows, the AMS is considered fixed at its nominal position and effects of the position uncertainties of the emitters on the speed of sound measurement are investigated. 
As only relative distances are relevant, the same error then results for the situation of fixed emitter positions and an AMS position with uncertainties. Errors on the positions of the emitters and the AMS will be assumed to be the same.

The nominal position of the AMS is defined as the position of the centre of gravity of the six sensors for a perfectly vertical alignment of the line. Differences of this nominal position from the positions of the actual sensors and from movements of the line due to sea currents will be considered separately.

The position error of the emitter has a component in vertical direction ($z$-direction) and in the horizontal plane. The latter can be decomposed into a radial position uncertainty along the line between an emitter and the base of line 12 carrying the AMS, and a lateral position uncertainty along a circle around the base of line 12 with a radius corresponding to the distance to the emitter.
The lateral error has the same effect as a compass offset, only that it is different for each emitter.
The error on the nominal position of the emitter will be -- conservatively -- estimated at $\pm \unit[1]{m}$ in all directions ($z$, radial, lateral). 

The effect of a lateral displacement by a given distance 
has the strongest effect for the closest line, i.e.\ for line 8 at a distance of about $\unit[63]{m}$.
For a movement of $\pm \unit[1]{m}$ along the corresponding circle with circumference approximated by $2\pi\times\unit[60]{m}$ for computational convenience, the corresponding angular error is $\pm\unit[1/60]{rad} $ or $\pm3/\pi \approx 1$ degree. 
This error for a distance of
 $\unit[60]{m}$ scales linearly with the distance and is only $\pm 0.3^\circ$ for the lines at the furthest distance.
For the effect of a compass offsets of $\pm 3^\circ$, an error of $\pm 2\%$ was estimated. Hence, the error from a lateral offset is about $\pm 0.7\%$ for the closest lines and only about one third of this for the furthest lines. For simplicity, and since the error is very small, the same error of $\pm 0.7\%$ will be assumed for all emitters as the error resulting from a lateral offset of their position. 

As only the POI for a given emitter is relevant, the errors on the radial and $z$-position of the emitter can be combined into a ``polar error''. To this end, the calculation of the speed of sound was redone,
where $\Delta d$ was recalculated for each run with the $z$-position of the AMS and the radial distance to the emitter varied each by $\pm \unit[1]{m}$. The changes of the speed of sound were then added in quadrature to get the polar error.  
The effect is largest for the closest lines, where it results in an error of about $\pm 1.5\%$. For line 1, at the greatest distance, it is less than $\pm 0.5\%$.

The worst case errors for the lateral and polar error are added in quadrature yielding $0.7\% \oplus 1.5\% \approx 1.7\%$ for the error on the static positions of the emitters for a fixed AMS. 
As average for all lines, an error of $\pm 1.5\%$ will be assumed, uncorrelated between measurements for different emitters.
An error of the same size, correlated between measurements for different emitters, will be assumed for the case of fixed emitter positions and an uncertainty on the AMS position.

\paragraph{Dynamic reconstruction of sensor positions:}
Movements of the ANTARES lines and consequently of the AMS  with submarine sea currents and the dynamic reconstruction of its position were discussed in Section~\ref{sec:position-reco}.
The mean reconstructed positions in $x$ and $y$ of the AMS for each run were used
for the calculation of $\Delta d$, used subsequently in the fit of the speed of sound. 
The deviations of these mean positions from the nominal position of the AMS were visualised in Figure~\ref{fig:AM_z-position}. The $z$-coordinate was assumed to be constant at $\unit[-2086.700]{m}$ as it changes only on the order of $\unit[1]{cm}$ for the various runs.
To estimate the effect on the speed of sound resulting from potential errors in the dynamic position calibration, the AMS was assumed to be fixed at its nominal position for a perfectly vertical line 12, and the calculation of $\Delta d$ and the fit to the speed of sound were repeated.
The resulting deviations of the speed of sound are small and can be conservatively estimated as $\pm 1\%$.
For simplicity, the errors for the speed of sound measurements from different emitters will be assumed to be uncorrelated -- this is not perfectly true as the runs used for the determination of the speed of sound are not mutually exclusive for the emitters.

\paragraph{Tilts of AMS:}
Tilts of the AMS due to the bending of the line are small; for a height of $\sim\unit[390]{m}$ of the AMS above the sea floor and a maximal displacement from the nominal position by  $\unit[3]{m}$, an angle of less than $0.5^\circ$ follows if the complete line were to be tilted as a straight line. In practice, the tilt is smaller at the top than at the bottom of the line due to the combination of drag (flow resistance) and vertical forces (pull from buoyancy)~\cite{bib:antares-pos-2012}.

For a conservative estimate of the effect, a tilt was applied to the AMS in the direction of the displacement as it would result if the line remained completely straight as the AMS is displaced by the distance observed for a given run and shown in Fig.~\ref{fig:AM_z-position}.
The resulting speed of sound did not differ significantly from the speed of sound calculated for an AMS in upright position. 

It is a reasonable assumption that the centre of gravity of the AMS is not exactly located along the $z$-axis of the local coordinate system of the AMS, i.e.\ the AMS, when looked at from straight above, will not have a perfect symmetry for rotations of multiples of $120^\circ$. The three titanium bars of the frame (cf.\ Figure~\ref{fig:line12_AM_HTI_gray}) can be expected to show differences from the $120^\circ$-symmetry and the centres of the AMs may have different distances from the $z$-axis of the local coordinate system of the AMS. This would result in a tilt of the AMS that has a specific direction in the local coordinate system of the AMS, irrespective of the direction of the displacement of the AMS. 

For the $x$- and $y$-axis of the local coordinate system and the three axes through back-to-back sensors (cf.~Figure~\ref{subfig:AM_schematic_gray}),
a tilt of $\pm 5^\circ$ was applied to the AMS. Depending on the axis, the fit to $\Delta d$ vs.\ $\Delta t$ is moved up or down, but the slope, i.e.\ the speed of sound, was not affected significantly.

A small but still conservative systematic error of $\pm0.5\%$ was assigned as being due to tilts of the AMS.

\paragraph{Sensor positions relative to centre of AM:} 
The positions of the sensors were determined before deployment within the local coordinate system of the AMS, 
but a precise measurement of the positions w.r.t.\ the centre of the respective AM was not done. 
It was assumed for the calculation of the distance between the POI and the sensors that a plane perpendicular to the local $z$-axis (cf.\ Figure~\ref{subfig:AM_schematic_gray}) through the centre of gravity of all sensors contains the centres of all AMs. 
This cannot be exactly true, as is indicated by the different delays between arrival times of the early and late wave as discussed in Section~\ref{sec:discussion-results}.

The position of the centre of the AM was moved $\pm\unit[5]{cm}$ w.r.t.\ its assumed position and 
the differences between the propagation distances $\Delta d$ were recalculated.
This changes the vector $\vec{n}_{s}$ in Eq.~\ref{eq:dist_in_glass}, whereas the effect on $\vec{n}_\text{POI}$ is negligible.
No significant effect was observed. Again, a systematic error of $\pm 0.5\%$ was assigned as upper bound.

\paragraph{Displacement of sensors from measured position:} 
The precision of the measurement of the sensor positions in the local coordinate system of the AMS is estimated as $\unit[5]{mm}$ in any direction along the inner surface of the sphere.
When both sensors move by the same distance in the same direction in lateral direction, the effect on the speed of sound is comparable to an offset of the compass reading. With a distance of a given sensor from the centre of the AMS of about $\unit[60]{cm}$,  a shift of $\unit[5]{mm}$ corresponds to an angular offset of only about $0.5^\circ$, which in the light of the discussion of angular offsets from above is negligible. A common shift of $\unit[5]{mm}$ of both sensor positions in clockwise and counter-clockwise direction was also implemented for the calculation of the values of $\Delta d$ and the effect on the fit to  the speed of sound was confirmed to be negligible.  

A worst case scenario for the lateral displacement of the sensors was considered, in which the positions of each of the two sensors was moved by $\unit[5]{mm}$ towards the other one (distance reduced by $\unit[10]{mm}$) and away from each other (distance increased by $\unit[10]{mm}$). Repeating the fit of the speed of sound with the altered positions yields an error on the speed of sound measurement of $\pm 4\%$.

For a displacement of the sensors in polar direction (up- or downwards), a common movement up or down has basically the same effect as a displacement of the sensor positions relative to the centre of the AM, which was shown above to be negligible.
For a polar displacement, the worst case scenario is an upward displacement of one sensor, combined with a downward displacement of the other sensor. 
The fit was repeated with one sensor moved upwards by $\unit[5]{mm}$  and the other downwards by $\unit[5]{mm}$  and vice versa. 
These displacements move the fitted line up or down, without changing the slope, i.e.\ the speed of sound, significantly. This does not come as a surprise, as the displacements have an effect that is similar to that of a tilt of the AMS. 

In total, the error resulting from a displacement of the sensor positions is estimated as $\pm 4\%$.  

\paragraph{Thickness and radius of glass sphere:}
It was assumed for the calculation of the propagation distance that the wave travels in the centre of the glass. The speed of sound scales linearly with the distance over which the wave propagates. Assuming that the wave travels at the outer or inner edge of the sphere leads to a change of  
the speed of sound of $3.5\%$ according to the corresponding change in radius
 $r = 216 \pm  \unit[7.5]{mm} \rightarrow \Delta r / r = 3.5\% $. 
The diameter shrinkage of the glass sphere per $\unit[1000]{m}$ depth is $\unit[0.30]{mm}$~\cite{bib:km3net-tdr}, which can be neglected. 

\paragraph{Propagation path of the wave in the glass sphere:}
For most emitters, signals were also recorded for both sensors when the AMS is rotated such that 
at least one sensor is located in the shadow zone of the emitter\footnote{Refer to Section~\ref{sec:late-wave-in-water} for a definition of the shadow zone.}.
The corresponding signals have a reduced amplitude and the angles over which they are recorded depend on the combination of emitters and receivers. Recall from Section~\ref{sec:analysis-strategy} that for some emitters and some orientations of the AMS only few or no suitable runs may exist.

\begin{figure}[tbh]
\centering
\includegraphics[width=\columnwidth,angle=0]{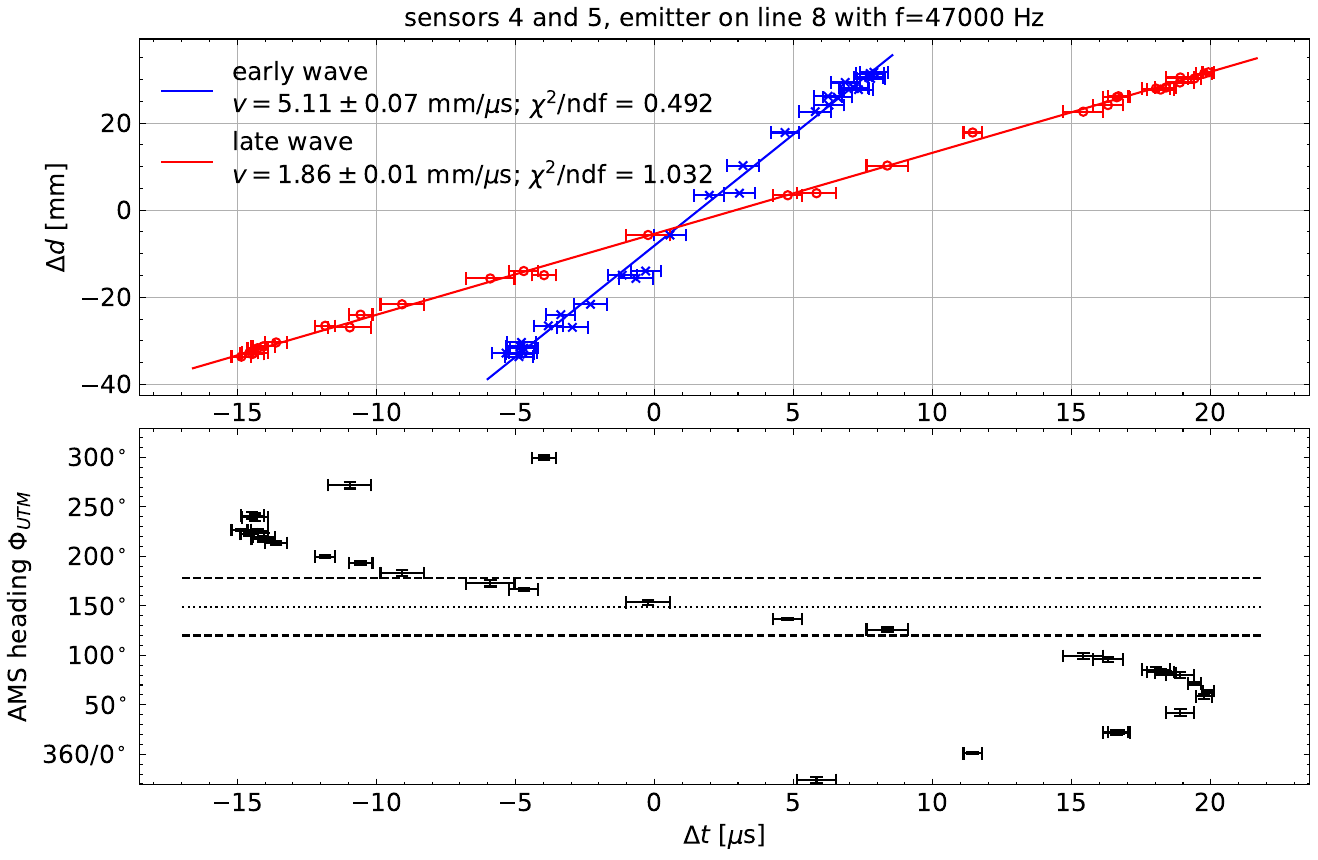}
\caption{
Top: $\Delta d$ vs.\ $\Delta t$ and fit of speed of sound for the early and late wave;
Bottom: 
AMS heading vs.\ $\Delta t$ of the late wave.
The broken lines indicate the AMS headings for which sensor 4 and 5, respectively, point towards the emitter on line 8 ($\phi_{8\rightarrow 4} = 0^\circ $, $\phi_{8\rightarrow 5} = 0^\circ$, respectively). The dotted line at a  $\Phi_\text{UTM} = 150^\circ$ is in the middle between the two broken lines. 
The $y$-axis is centred at this value of the AMS heading. 
For this angle, $\Delta t =0$ when the sensors are located at the same value of the local $z$-coordinate of the AMS and in absence of tilts of the AMS.   
}
\label{fig:line8_dist_and_rot_ang}
\end{figure}

For the emitter on line 8, signals were recorded for almost the complete $360^\circ$ range of AMS headings as shown in Figure~\ref{fig:line8_dist_and_rot_ang} for the sensor pair (4, 5). 
Two orientations of the AMS correspond to each value of $\Delta t$ on the $x$-axis (bottom plot of the figure). 
When the AM is rotated $180^\circ$ away from the emitter ($\Phi_\text{UTM} = 330^\circ$ for the case of the emitter on line 8 and receiving sensors 4 and 5 in Figure~\ref{fig:line8_dist_and_rot_ang}) $\Delta t$ and $\Delta d$ again approach zero, albeit with longer propagation distances to the respective sensors when compared to a AMS heading towards the emitter ($\Phi_\text{UTM} = 150^\circ$).

Upon closer inspection it is observed that for a given value of $\Delta t$, the orientation with the larger value of $d$, i.e.\ larger rotation of the sensors away from the POI, tends to have a larger corresponding value of $\Delta d$. 
This corresponds to an apparent increase of the speed of sound with increasing distance of the sensors from the POI.

\begin{figure}[htbp]
\centering
\includegraphics[width=0.7\columnwidth,angle=0]{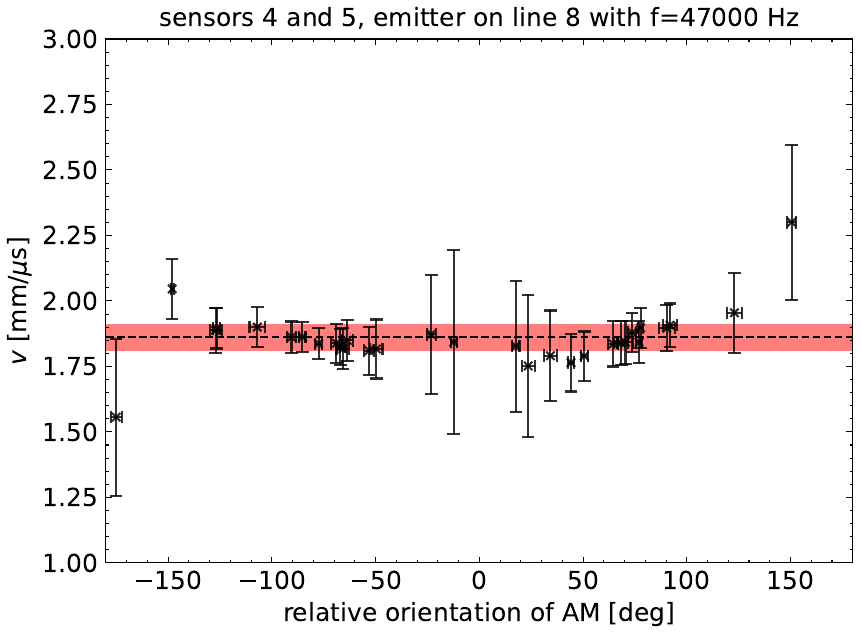}
\caption{
Speed of sound vs.\ relative orientation of the AM with sensor pair (4, 5) for emitter on line 8.
The black dotted line is the speed of sound obtained from the fit of $\Delta d$ vs.\ $\Delta t$ as shown in Figure~\ref{fig:line8_dist_and_rot_ang}, the red bar corresponds the uncertainty as explained in the text.
Speeds of sound for relative orientations of the AMS w.r.t.\ the emitter with an absolute value smaller than $10^\circ$ were ignored.
}
\label{fig:Speed_of_sound_vs_angle_for_cycle_8}
\end{figure}

The effect is demonstrated in Figure~\ref{fig:Speed_of_sound_vs_angle_for_cycle_8} where the speed of sound was calculated as a function of the relative orientation of the AM, i.e.\ rotation w.r.t.\ the position where the centre between the sensors is pointing towards the emitter.
The speed of sound was calculated by assuming that the intersection of the fits to the early and late wave, $(\Delta t)_\text{off} = \unit[0.85]{\mu s}$ and $(\Delta d)_\text{off} = \unit[-4]{mm}$ for the emitter on line 8 (see Figure~\ref{fig:line8_dist_and_rot_ang}), constitute the offsets of the corresponding quantities. Hence the speed of sound for a given angular rotation with measured values  $\Delta t$ and $\Delta d$ was calculated according to $v = (\Delta d - (\Delta d)_\text{off} )/( \Delta t - (\Delta t)_\text{off} )$.
Figure~\ref{fig:Speed_of_sound_vs_angle_for_cycle_8} shows the result for the late wave. 

An increase of the speed of sound for increasing absolute values of the relative rotation can be observed.
Assuming this is an artefact, it implies that for a large rotation with a correspondingly large measured value of $\Delta t$,  the calculated value of $\Delta d$ 
is overestimated and should be smaller -- or for a low rotation, $\Delta d$ 
is underestimated and should be larger.
Varying the sensor positions
from their measured position within their corresponding AM according to the precision of the measurement, and repeating the fit on the speed of sound  has no significant effect on the observed dependence of the speed of sound on the relative orientation.

The observation can be explained if it is taken into account that the wave inside the glass sphere
clearly cannot be confined to the centre but
has to travel somehow from the outside of the sphere -- where the emitter signal ``hits'' -- to the inside of the sphere, where the sensor is attached. This additional distance would increase short distances along the centre of the glass by a larger factor than long distances.

In the presence of a trend of the speed of sound as shown in Figure~\ref{fig:Speed_of_sound_vs_angle_for_cycle_8}, calculating a statistical uncertainty, as done to estimate the error on the slope from the $\Delta d$ vs.\ $\Delta t$ distribution, does not seem sufficient.
The red band shown in the figure is an estimate of the systematic error $(\Delta v)_{\Phi_\text{UTM}}$ due to the apparent dependence of the speed of sound on the relative orientation of the AMS. It was obtained by squaring the difference between the speed of sound $v$ for a given angle  and the speed of sound obtained from the fit (see values in Tables~\ref{tab:v-for-sensors45} and \ref{tab:v-for-sensors23}); for all of these terms the weighted sum was calculated and the square root taken. As weight, the squared inverse of the statistical uncertainties of the values of $v$ as shown as error bars in Figure~\ref{fig:Speed_of_sound_vs_angle_for_cycle_8} were used. 
Speeds of sound for relative orientations of the AMS w.r.t.\ the emitter with an absolute value smaller than $10^\circ$ were ignored as the errors for the correspondingly small time differences get excessively large.

This procedure was done only for the late wave -- for the early wave, the short time differences lead to imprecise measurements. As the error is presumed to be due to an incorrect estimation of the distance propagated by the wave, it is expected to have the same relative effect on the early and late wave.

For emitters on lines 8, 10, and 11 signals were recorded in sensors 4 and 5 over almost the complete $360^\circ$ range of orientations of the AMS. For the emitters on the remaining lines with a smaller range of orientations, $(\Delta v)_{\Phi_\text{UTM}}$ is expected to be smaller, which is indeed observed:
The absolute errors on the speed of sound of the early wave from sensor pair (4, 5) are calculated as 
$\unit[0.056]{\mmpermus}$ (line 8),
$\unit[0.10]{\mmpermus}$ (line 10), and
$\unit[0.060]{\mmpermus}$ (line 11).
The errors for all other lines are smaller than the lowest of these values.

The signal from the emitter on  line 10 shows some unexpected behaviour as demonstrated in Figure~\ref{fig:speed_of_sound_vs_Zero_X_sens_4_5}. For this emitter, the systematic error of $\unit[0.10]{\mmpermus}$ or 5\% is found, whereas for the other emitters, an error of 3\% is calculated. 

For sensor pair (2, 3), more lines recorded signals over almost the complete $360^\circ$ range of orientations of the AMS. 
The largest value of $\unit[0.085]{\mmpermus}$  or 4\% is observed for line 9. 
As an overall value, the systematic error is estimated  as 4\%.

\paragraph{Effect of superposition of early and late wave:}
In Section~\ref{sec:model-superimposing-waves}, a worst case
phase shift of $6^\circ$, corresponding to a shift of the zero crossing of the late wave
of  $\unit[0.35]{\mu s}$ for the lowest frequency of $\unit[46545]{Hz}$
 was estimated as result of the superposition with the early wave. 

The time delay
$\tau = d_\text{prop} {v_\text{eff}}^{-1}$ with $v_\text{eff} = ({v_\ell}^{-1} - {v_e}^{-1})^{-1}$
between the arrival times of corresponding maxima of the late and early wave  at a given sensor after propagating the distance $d_\text{prop} $ was introduced in Section~\ref{sec:discussion-results}. 
If now the distance to the receiving sensor is changed by $\Delta d_\text{prop}$,  the time difference between corresponding maxima of the early and late wave changes by  $\Delta \tau$ according to
$ \Delta d_\text{prop} = \Delta \tau\,v_\text{eff} $. 
Approximating $v_\ell \approx \unit[2]{\mmpermus}$ and $v_e \approx \unit[5]{\mmpermus}$ yields $\Delta d_\text{prop} = \Delta \tau \times  \unit[10/3]{\mmpermus}$. 

At a certain distance of the sensor from the POI, the 
phase shift between the early and late wave will be $\pi/2 + n\pi\ (n = 0,1,2,\dots)$ so that their
superposition will have a maximal effect on the times of the zero crossings of the late wave. 
If then this distance is changed by $\Delta d_\text{prop}$ such that the resulting $\Delta\tau$ 
introduces an additional phase shift of $\pi/2$, corresponding to a time delay between early and late wave of a quarter of a period of the frequency of the emitter, the effect of the superposition for this new propagation distance is zero.

Taking a frequency of $\unit[50]{kHz}$ as average value,
the corresponding period is $\unit[20]{\mu s}$ and 
a change by a quarter of a period corresponds to a change in distance by $\Delta d_\text{prop} = \unit[5]{\mu s} \cdot  \unit[10/3]{\mmpermus} = \unit[17]{mm} $.
As the orientation of a given AM changes, the distances propagated by the waves to the individual sensors change accordingly. 
The nominal distances propagated by the signals from the POI to a sensor, depending on the heading of the AMS, is in the range $\unit[295\sim360]{mm}$ for line 1 at the furthest distance from line 12, and  $\unit[230\sim420]{mm}$ for line 8 at the closest distance to line 12.
Over this  variation of the distance of $\unit[\sim70]{mm}$ and more, effects of the superposition of the early and late wave on the  time of the zero crossing should vary between minimal and maximal several times. 
This should lead to a variation of the speed of sound with the relative orientation of an AM, as the maximal and minimal effects on the zero crossing appear at different orientations for the two sensors. 

In the context of the systematic error from the ``propagation path of the wave in the glass sphere'', see above, the effect of the relative orientation of an AM on the speed of sound measurement was investigated.
Only an \emph{increase} of the speed of sound with increasing relative orientation, and no variation, is observed. 
Hence it is concluded that the superposition of early and late wave has no observable effect on the speed of sound of the late wave. And any potential effect would already be included in the systematic error
assigned to the ``propagation path of the wave in the glass sphere''.

\begin{figure}[thbp]
\centering
\includegraphics[width=0.75\columnwidth,angle=0]{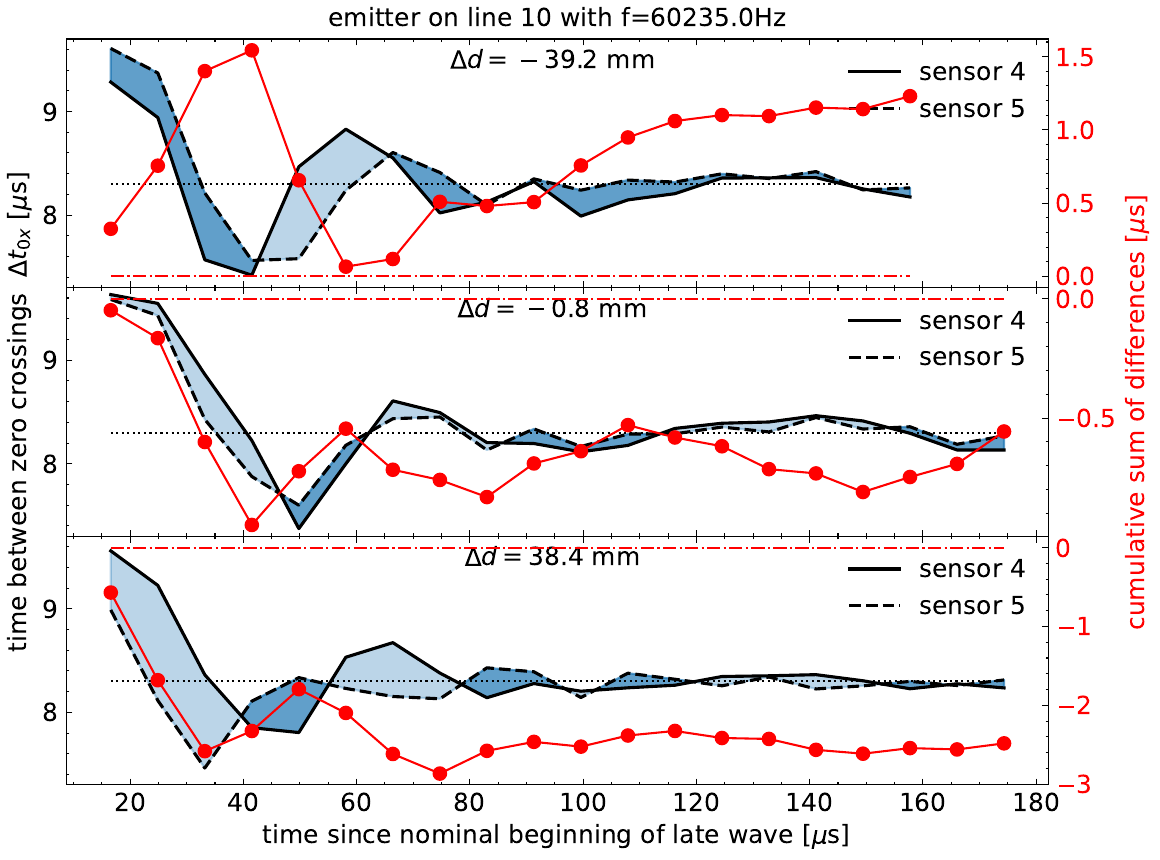}
\includegraphics[width=0.75\columnwidth,angle=0]{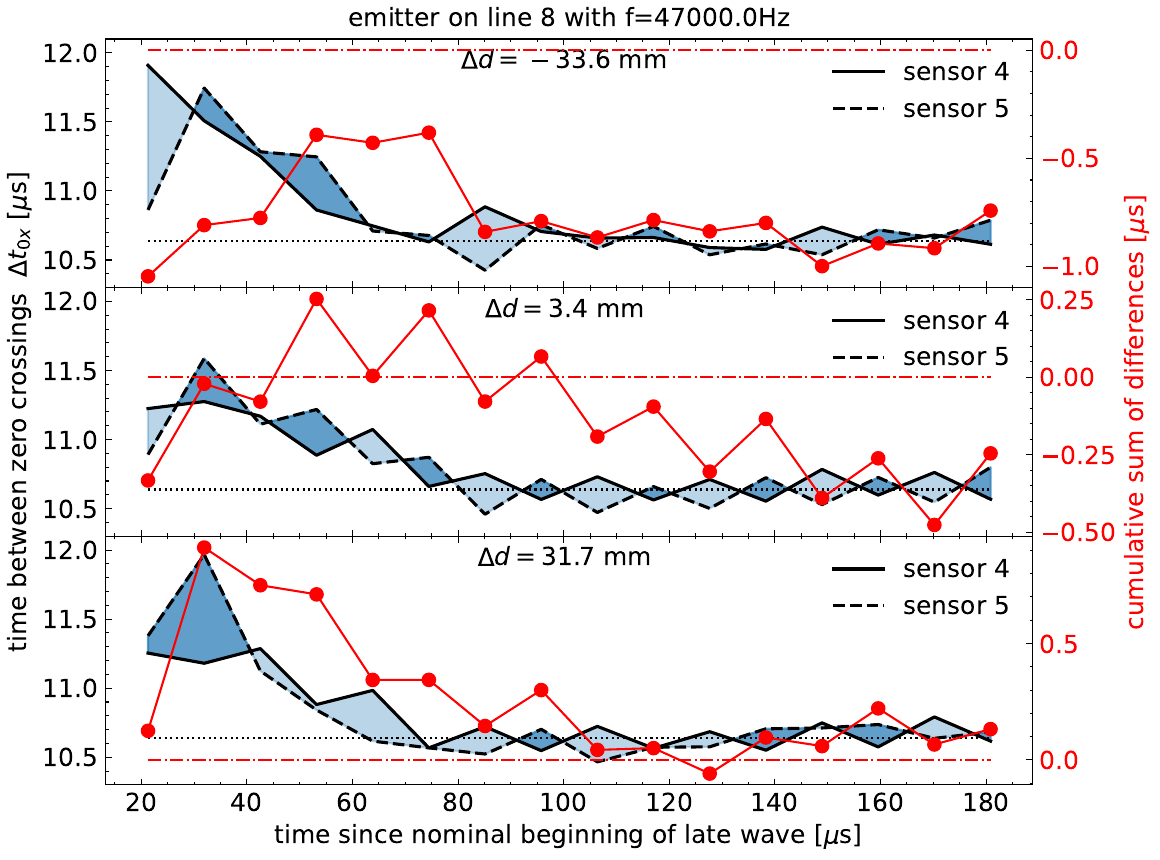}
\caption{
Time between zero crossings $\Delta t_{0x\,s4}$ for sensor 4 (left axis, solid line) and $\Delta t_{0x\,s5}$ for sensor 5 (left axis, broken line) and cumulated sum of the differences $\Delta t_{0x\,s5} - \Delta t_{0x\,s4}$ (right axis, line with dots.)
The area between the lines for  $\Delta t_{0x\,s5}$ and $\Delta t_{0x\,s4}$ is dark when $\Delta t_{0x\,s5} > \Delta t_{0x\,s4}$ (positive term entering cumulative sum of differences) and light when $\Delta t_{0x\,s5} < \Delta t_{0x\,s4}$ (negative term entering cumulative sum of differences). The dotted line shows the expected value for $\Delta t_{0x\,s5}$ and $\Delta t_{0x\,s4}$ ($0.5/f$ for signal frequency $f$), the broken dotted line indicates zero cumulative time difference (right axis).
The three plots for each of the two emitters correspond to orientations of the AM that produce a large negative value, a value near zero, and a large positive value of $\Delta d$, as indicated in the plots. 
The $x$-axis corresponds to that of Figure~\ref{fig:speed_of_sound_vs_Zero_X}, where the time difference between consecutive zero crossing has been assigned to the nominal time of the first zero crossing.
}
\label{fig:transient-response-sens-4-5}
\end{figure}

\paragraph{Transient response of sensors:}
Figure~\ref{fig:transient-response-sens-4-5} shows the time between two consecutive zero crossings for sensors 4 and 5, $\Delta t_{0x\,s4}$  and $\Delta t_{0x\,s5}$, respectively,  as a function of the time since the nominal beginning of the wave, i.e.\ the extrapolated start of the rising of the amplitude to maximum 1, cf.\ Section~\ref{sec:sound-speed-vs-freq}. 
This corresponds to the $x$-axis of Figure~\ref{fig:speed_of_sound_vs_Zero_X}. The values for $\Delta t_{0x\,s4}$  and $\Delta t_{0x\,s5}$ at early times differ from the value expected for a given frequency of the emitter, before settling to the expected value after about $\unit[80\sim 90]{\mu s}$. This behaviour could in principal be a feature of the emitter, but $\Delta t_{0x\,s5}$ and $\Delta t_{0x\,s4}$ differ from each other for the beginning of the signal, where the difference varies with the orientation of the AM. 

For each of the two emitters from Figure~\ref{fig:transient-response-sens-4-5}, three plots are shown that correspond to an orientation of the AM that produces a large negative value, a value near zero, and a large positive value of $\Delta d$, respectively. In Figures~\ref{fig:dist_vs_time_cycle10} and \ref{fig:line8_dist_and_rot_ang}, these values of $\Delta d$ correspond to values on the $y$-axis at the respective ends and near the centre of the fit shown in the figures.
In Figure~\ref{fig:transient-response-sens-4-5}, the cumulative sum of  $\Delta t_{0x\,s5} - \Delta t_{0x\,s4}$ for the emitter on line 10 starts at about $\unit[0.3]{\mu s}$  and settles at about $\unit[1.2]{\mu s}$ for $\Delta d=\unit[-39.3]{mm}$ and varies from about $\unit[-0.6]{\mu s}$ to $\unit[-2.5]{\mu s}$ for $\Delta d=\unit[38.4]{mm}$. Over the time covered for the fit in Figure~\ref{fig:dist_vs_time_cycle10} of $\unit[\sim 40]{\mu s}$, this corresponds to an increase of the speed of sound by about $7\sim 8\%$, if the slope of the line were to be determined from these two points alone. This demonstrates qualitatively how the changing of the cumulative sum of the differences $\Delta t_{0x\,s5} - \Delta t_{0x\,s4}$ leads to the varying speeds of sound observed in Figure~\ref{fig:speed_of_sound_vs_Zero_X}.

It is not necessarily more precise to use the speed of sound for the later times of that plot, where the speed of sound stabilises, as delays from the beginning of the signal have a cumulative effect. 
Hence, the speed of sound for the late wave was determined from an early zero crossing (the second one in Figure~\ref{fig:speed_of_sound_vs_Zero_X}, listed in Tables~\ref{tab:v-for-sensors45} and \ref{tab:v-for-sensors23}) and the error on $v_\ell$ was estimated from the largest observed value of $|\Delta_b - \Delta_a|/2$ as $\sim$3\%. 
As the cause of the error is assumed to be a time offset, and the time differences used to calculate the speed of sound for  the early wave scales with $v_\ell/v_\text{e} \approx 2/5$, the relative error on $v_\text{e}$ is estimated as $5/2\cdot 3\% = 7.5\%$.

From the measurement of the speed of sound vs.\ time in Figure~\ref{fig:speed_of_sound_vs_Zero_X}, a strong correlation between the error for emitters with the same frequency can be deduced, which may get weaker as differences between emission frequencies increase. For simplicity, it will be conservatively assumed that the errors for all emitters are correlated.

Note that there are no indications that reflections,  e.g.\ from the boundaries between hemispheres, affect the measurements of the speed of sound. 
Such reflections would lead to abrupt changes in the times between zero crossings for either sensor, which would in turn result in a  change of the speed of sound for the later parts of the wave. No such effects are observed in Figures~\ref{fig:speed_of_sound_vs_Zero_X} or \ref{fig:transient-response-sens-4-5}.

\section{Elastic properties of VITROVEX glass}
\label{sec:appendix_prop-VITROVEX}
The relevant elastic properties of VITROVEX glass, used for both ANTARES and KM3NeT glass housings, 
are as follows~\cite{bib:vitrovex}:

\begin{tabular}{lc@{\,=\,}l}
Specific gravity at $25^\circ$C & $\rho$ & $\unit[2.23]{g/cm^3}$ \\
Young's modulus & $E$ & $\unit[63]{GPa}$ \\
Poisson's ratio & $\nu$ & $0.20$
\end{tabular}
\\

From these properties, the following elastic parameters are furthermore derived:
\\

\begin{tabular}{lc@{\,=\,}l}
{Lam\a'{e}}\ first parameter & $\lambda$ & $\tfrac{E\nu }{(1+\nu )(1-2\nu )} $ \\ 
Shear modulus\footnotemark
& $G$ & $\tfrac{E}{2(1+\nu )} = \unit[26.25]{GPa}$ 
\\
P-wave modulus &  $M$ & $\lambda + 2G = \frac{E(1-\nu)}{(1+\nu)(1-2\nu)} = \unit[70.00]{GPa}$
\end{tabular}
\\
\footnotetext{The shear modulus is also frequently called Lam\'e second parameter and denoted by $\mu$.}

With the numerical value for $G$ and $M$, 
 the longitudinal and transverse speed of sound in VITROVEX glass, $v_\text{long}$ and $v_\text{trans}$, respectively, are readily calculated:
\begin{align*}
v_\text{long} &= \sqrt{M/\rho} 
= \unit[5.60]{\mmpermus} 
\\
v_\text{trans} &= \sqrt{G/\rho} 
= \unit[3.43]{\mmpermus}  
\end{align*}

%

%
%

\end{document}